\documentclass[usegraphicx,usenatbib]{mn2e}
\usepackage{verbatim}
\usepackage{color}
\usepackage{aas_macros}

\usepackage[T1]{fontenc}
\usepackage{ae,aecompl}

\usepackage{hyperref}  
\usepackage{amsmath} 
\usepackage{breqn}  

\usepackage{pdflscape} 

\newcommand{\beq}{\begin{equation}}
\newcommand{\eeq}{\end{equation}}

\newcommand{\hmpc}{\,$h^{-1}$Mpc }
\newcommand{\hmpcii}{\,$h^{-1}$Mpc}

\newcommand{\hgpcii}{\,\ensuremath{h^{-3}\mathrm{Gpc}^3}}

\newcommand{\dvrs}{\,$D_{\mathrm V}/r_{\mathrm s}$ }
\newcommand{\dvrsii}{\,$D_{\mathrm V}/r_{\mathrm s}$}
\newcommand{\iidvrsii}{$D_{\mathrm V}/r_{\mathrm s}$}

\newcommand{\dvrsrsfid}{\,$D_{\mathrm V}\left(r_{\mathrm s}^{\mathrm{fid}}/r_{\mathrm s}\right)$ }
\newcommand{\dvrsrsfidii}{\,$D_{\mathrm V}\left(r_{\mathrm s}^{\mathrm{fid}}/r_{\mathrm s}\right)$}

\newcommand{\baf}{\,baryonic acoustic feature }
\newcommand{\bafii}{\,baryonic acoustic feature}

\newcommand{\wizcola}{\,WiZ-COLA }
\newcommand{\wizcolaii}{\,WiZ-COLA}

\newcommand{\Dznear}{\,$\Delta z^{\mathrm{Near}}$ }
\newcommand{\Dznearii}{\,$\Delta z^{\mathrm{Near}}$}
\newcommand{\Dzmid}{\,$\Delta z^{\mathrm{Mid}}$ }
\newcommand{\Dzmidii}{\,$\Delta z^{\mathrm{Mid}}$}
\newcommand{\Dzfar}{\,$\Delta z^{\mathrm{Far}}$ }
\newcommand{\Dzfarii}{\,$\Delta z^{\mathrm{Far}}$}

\newcommand{\scb}{}

\newcommand{\dvrsrsfidN}{\,$1716$ }
\newcommand{\dvrsrsfidNii}{\,$1716$}
\newcommand{\upmdvrsrsfidN}{\,$83$ }
\newcommand{\dvrsrsfidM}{\,$2221$ }
\newcommand{\dvrsrsfidMii}{\,$2221$}
\newcommand{\upmdvrsrsfidM}{\,$101$ }
\newcommand{\dvrsrsfidF}{\,$2516$ }
\newcommand{\dvrsrsfidFii}{\,$2516$}
\newcommand{\upmdvrsrsfidF}{\,$86$ }

\begin{document}




\title[Improved Distance Measures with Reconstructed WiggleZ]{The WiggleZ Dark Energy Survey: Improved Distance Measurements to $z=1$ with Reconstruction of the Baryonic Acoustic Feature}


\author[Kazin E., Koda J., Blake C. et al.]
{\parbox[t]{\textwidth}{
Eyal A. Kazin$^{1 \ 2}$, 
Jun Koda$^{1 \ 2}$,
Chris Blake$^{1}$, 
Nikhil Padmanabhan$^{3}$,  
    Sarah Brough$^4$,
    Matthew Colless$^4$,
    Carlos Contreras$^{1,5}$,
    Warrick Couch$^{1,4}$,
    Scott Croom$^6$,
    Darren J. Croton$^1$,
    Tamara M. Davis$^7$,
    Michael J.\ Drinkwater$^7$,
    Karl Forster$^8$,
    David Gilbank$^9$,
    Mike Gladders$^{10}$,
    Karl Glazebrook$^1$,
    Ben Jelliffe$^6$,
    Russell J.\ Jurek$^{11}$,
    I-hui Li$^{12}$,
    Barry Madore$^{13}$,
    D.\ Christopher Martin$^8$,
    Kevin Pimbblet$^{14,15}$,
    Gregory B.\ Poole$^{1,16}$,
    Michael Pracy$^{1,6}$,
    Rob Sharp$^{16,17}$,
    Emily Wisnioski$^{1,18}$,
    David Woods$^{19}$,
    Ted K.\ Wyder$^8$ and H.K.C. Yee$^{12}$
}
\vspace*{6pt} \\ 
$^{1}$ Centre for Astrophysics $\&$ Supercomputing, Swinburne University of Technology, PO Box 218, Hawthorn, VIC 3122, Australia.\\
$^{2}$ ARC Centre of Excellence for All-sky Astrophysics (CAASTRO). \\
$^{3}$Department of Physics, Yale University, 260 Whitney Ave, New Haven, CT 06520, USA.\\
  $^4$ Australian Astronomical Observatory, P.O. Box 915, North Ryde, NSW 1670, Australia \\
  $^5$ Carnegie Institution of Washington, Las Campanas Observatory, Colina el Pino s/n, Casilla 601, Chile \\
  $^6$ Sydney Institute for Astronomy, School of Physics, University of Sydney, NSW 2006, Australia \\
  $^7$ School of Mathematics and Physics, University of Queensland, Brisbane, QLD 4072, Australia \\
  $^8$ California Institute of Technology, MC 278-17, 1200 East California Boulevard, Pasadena, CA 91125, United States \\
  $^9$ South African Astronomical Observatory, PO Box 9 Observatory, 7935 South Africa \\
  $^{10}$ Department of Astronomy and Astrophysics, University of Chicago, 5640 South Ellis Avenue, Chicago, IL 60637, United States \\
  $^{11}$ Australia Telescope National Facility, CSIRO, Epping, NSW 1710, Australia \\
  $^{12}$ Department of Astronomy and Astrophysics, University of Toronto, 50 St.\ George Street, Toronto, ON M5S 3H4, Canada \\
  $^{13}$ Observatories of the Carnegie Institute of Washington, 813 Santa Barbara St., Pasadena, CA 91101, United States \\
  $^{14}$ School of Physics, Monash University, Clayton, VIC 3800, Australia \\
  $^{15}$ Department of Physics and Mathematics, University of Hull, Cottingham Road, Hull, HU6 7RX, UK \\
  $^{16}$ School of Physics, University of Melbourne, Parksville, VIC 3010, Australia \\
  $^{17}$ Research School of Astronomy \& Astrophysics, Australian National University, Weston Creek, ACT 2600, Australia \\
  $^{18}$ Max Planck Institut f\"{u}r extraterrestrische Physik, Giessenbachstra$\beta$e, D-85748 Garching, Germany\\
  $^{19}$ Department of Physics \& Astronomy, University of British Columbia, 6224 Agricultural Road, Vancouver, BC V6T 1Z1, Canada
}
%

\maketitle

\begin{abstract}
We present significant improvements in cosmic distance 
measurements from the WiggleZ Dark Energy Survey, 
achieved by applying  
the reconstruction of the \baf technique. 
We show using both data and simulations 
that the reconstruction 
technique can often be   
effective despite patchiness of the survey, significant edge effects 
and shot-noise.
We investigate three redshift bins in the redshift range  
$0.2<z<1$, and in all three find improvement 
after reconstruction 
in the detection 
of the \baf and its usage as a standard ruler.  
We measure model independent distance measures \dvrsrsfid
of \dvrsrsfidNii$\pm$\upmdvrsrsfidN Mpc,  \dvrsrsfidMii$\pm$\upmdvrsrsfidM Mpc, \dvrsrsfidFii$\pm$\upmdvrsrsfidF Mpc ($68\%$ CL) at 
effective redshifts $z=0.44,0.6,0.73$, respectively, 
where  $D_{\rm V}$ is the volume-average-distance, 
and $r_{\rm s}$ is the sound horizon at the end of the baryon drag epoch. 
These significantly improved 
4.8, 4.5 and 3.4 per-cent accuracy measurements 
are equivalent to those expected from 
surveys with up to 2.5 times the volume of WiggleZ without reconstruction applied. 
These measurements are fully 
consistent with cosmologies allowed by the analyses of the Planck Collaboration  
and the Sloan Digital Sky Survey.  
We provide the \dvrsrsfid posterior probability distributions and their covariances. 
When combining these measurements with temperature fluctuations measurements 
of Planck, the polarization of WMAP9, and 
the 6dF Galaxy Survey \bafii, 
we do not detect deviations from a flat $\Lambda$CDM model. 
Assuming this model we constrain 
the current expansion rate to $H_{0}=67.15 \pm 0.98$ kms$^{-1}$Mpc$^{-1}$. 
Allowing the equation of state of dark energy to vary we obtain $w_{\rm DE}=-1.080 \pm 0.135$. 
When assuming a curved $\Lambda$CDM model we obtain a curvature value of $\Omega_{\rm K}=-0.0043\pm 0.0047$.


\end{abstract}


\thanks{E-mail: eyalkazin@gmail.com}
\section{Introduction}\label{section:intro}
The baryonic acoustic feature is regarded 
as a reliable tool for measuring distances, which 
can be used to probe cosmic expansion rates and 
hence assist in understanding the  
mysterious nature of the recent cosmic acceleration (\citealt{riess98, perlmutter99a, seo03, blake03}). 
Early plasma-photon acoustic waves that came to a near-stop 
at a redshift $z\sim 1100$ left these baryonic signatures imprinted at 
a co-moving radius of $\sim 150$ Mpc 
in both the cosmic microwave background temperature fluctuations 
and in the distribution of matter,  
as an enhancement in the clustering amplitude of overdensities 
at this ``standard ruler" distance  
(\citealt{peebles70a}). 

However, the signature in the distribution of matter, 
and hence in galaxies,   
experienced a damping due to long-range coherent bulk motions 
generated by tidal gravitational forces.  
In the linear density field, 
galaxies coherently move by $\sim 5$ Mpc from their original positions, 
which causes smoothing of the otherwise sharp feature 
at the scale of $150$ Mpc in 
the clustering correlation function. 
Although this damping is well understood and modeled  
(\citealt{meiksin99a,seo07,seo08a,smith08,angulo08,crocce08,sanchez08,kim09a}), 
it decreases the accuracy with which the feature may be used as a standard ruler. 

To correct for the effects 
of large-scale motions, 
\cite{eisenstein07} suggested the method of reconstruction of the \bafii.  
By using  
the density field to infer the displacements caused by these bulk flows in linear theory,  
one can 
retract the galaxies to their near-original positions, 
and hence sharpen the baryonic acoustic signature.
They concluded that this method improves the usage 
of the \baf as a standard ruler. 
The technique has since been further developed, 
showing that this procedure minimizes the systematic 
errors in the bias 
of geometric information obtained 
from matter and galaxies (\citealt{padmanabhan09a,noh09a,seo10a,mehta11a}). 
\cite{mehta11a} concluded that distance 
measurements made when using galaxies 
with a low galaxy to matter density bias of $b=\delta_{\rm gal}/\delta_{m}\sim 1$, 
such as those analyzed here, have a low 
systematic error of $\sim 0.2-0.25\%$ which is reduced to $0.1-0.15\%$ 
when applying reconstruction (see their Figure 5).

The first successful application of the technique to galaxy data was 
reported by \cite{padmanabhan12a}, who improved the distance constraint 
to $z=0.35$ by sharpening the \baf of the luminous red galaxy sample 
(\citealt{eisenstein01a}) of the SDSS-II (\citealt{york00a}). Testing 
realistic mock catalogs, they showed that the technique yields 
unbiased improved results. A further application 
of the technique was performed by the SDSS-III Baryon Oscillation Spectroscopic Survey (BOSS) 
using a massive galaxy sample at $z=0.57$ (\citealt{anderson12a, anderson13a}). 
The inability of the technique to improve constraints 
in this particular case may be attributed 
to sample variance 
in the sense that the pre-reconstruction measurement was 
on the fortunate side of expectations  
(\citealt{kazin13a}). 
Recently the BOSS collaboration have shown this to be the mostly likely 
explanation, by showing improvement of distance measures   
when probing galaxy samples two and three times as large (\citealt{anderson13b}, see their Figure 4.)


In this study, we apply the reconstruction technique to galaxies 
mapped by the WiggleZ Dark Energy Survey (\citealt{drinkwater10a}). 
The $0.2<z<1$ range of WiggleZ enables the survey to probe dark energy 
at a unique effective redshift of $z=0.73$, 
which is close to the beginning 
of the acceleration phase, 
according to the dark energy cold dark matter paradigm.  
We have previously reported measurements using the \baf  
in this redshift range with accuracies of $\sim 4.5-7.5\%$ (\citealt{blake11c}). 
In this analysis we show that 
reconstruction improves the detectability of the \baf 
and yields {\it substantially} tighter distance constraints.  

When applying reconstruction to WiggleZ  
we are confronted by various challenges 
compared to other galaxy surveys. 
The WiggleZ volumes are patchy 
with substantial edge effects,  
because each survey region is only $\sim 500$\hmpc 
in dimension with additional incompleteness due to the input catalogues. 
In addition, 
clustering measurements using 
the highest redshifts of the volume also 
contain fairly high shot-noise
corresponding to $nP\sim 1$, 
where $n$ is the number density and 
$P$ is the characteristic power spectrum amplitude 
at wave number $k\sim 0.15 \ h$Mpc$^{-1}$.   
Hence we are required to test if reconstruction 
of the density fields of such volumes 
could potentially cause possible 
biases when displacing the galaxies. 

To test for this, 
we apply the algorithm to a myriad of realistic 
simulated realizations. 
Constructing mock catalogues from 
$N-$body simulations for WiggleZ is a challenging 
problem because the galaxies trace dark matter haloes with masses 
$\sim 10^{12}h^{-1}$M$_\odot$, 
an order of magnitude lower than those populated by luminous red galaxies.  
For this reason in 
past analyses of WiggleZ (e.g, \citealt{blake11c})
we used log-normal realizations 
to support the data analysis 
(e.g, to determine the covariance of the measurement). 
These, however, 
do not contain realistic displacement information.  
Hence, to support this study we generated $600$ mock realizations based on 
a more accurate Lagrangian co-moving scheme, as described in 
\S\ref{section:wizcola}. 

Another difference between the past and current WiggleZ analyses is 
the manner in which we model the correlation function $\xi$. 
In past analyses, we modeled the full shape, resulting in 
model-dependent measurements. 
The reason for this is that when assuming a  
theoretical model for $\xi$ its full shape 
may be used as a standard ruler (e.g, see \citealt{eisenstein05b, sanchez12a,sanchez13a}). 
As reconstruction involves smoothing of the density field, 
it is difficult to model the overall broadband shape of the post-reconstruction 
power spectrum. 
For this reason, in this analysis we are only interested in the baryonic acoustic peak position, 
and hence focus on the geometric information. 
This means that 
we are required to marginalize over the shape 
information, which makes 
the distance 
measurements reported here model-independent.


This study is presented as follows. 
In \S\ref{section:data} we present the data, simulated data, the reconstruction technique 
and the construction of the two-point correlation functions.  
In \S\ref{section:method} we describe the method 
used to calculate the geometric constraints, including the construction of the fitting model. 
In \S\ref{section:results} we present distance measurements from 
the data and compare with those obtained with the simulations. 
This section is concluded by cosmological implications. 
We summarize in \S\ref{section:summary}. 

Unless otherwise stated, 
we assume a flat $\Lambda$CDM fiducial cosmology as defined in 
\cite{komatsu09a}: a dark matter density of $\Omega_{\rm m}=0.27$, 
a baryon density of $\Omega_{\rm b}=0.0226$, 
a spectral index of $n_{\rm s}=0.963$, a rms of density fluctuations 
averaged in spheres of radii at $8$\hmpc of $\sigma_{8}=0.8$, 
and $h=0.71$, where the local expansion rate is defined as 
$H_0=100h$ kms$^{-1}$Mpc$^{-1}$.

\section{Data}\label{section:data}
\subsection{Galaxy sample}\label{section:galaxy_sample}
The WiggleZ Dark Energy Survey (\citealt{drinkwater10a}) is a 
large-scale galaxy redshift survey of bright emission-line galaxies over 
the redshift range $z < 1$, which was carried out at the 
Anglo-Australian Telescope between August 2006 and January 2011.  In 
total, of order $200{,}000$ redshifts of UV-selected galaxies were 
obtained, covering of order 1000 deg$^2$ of equatorial sky.  In this 
study we analyze the same final WiggleZ galaxy sample as utilized by 
\cite{blake11c} for the measurements of BAOs in the galaxy 
clustering pattern.  After cuts to maximize the contiguity of the 
observations, the sample contains $158{,}741$ galaxies divided into six 
survey regions -- the 9-hr, 11-hr, 15-hr, 22-hr, 1-hr and 3-hr regions.  
The survey selection function within each region was determined using 
the methods described by \cite{blake10a}.

For purposes of this study, 
following the analysis of \cite{blake11c}, 
we divided the galaxies into three redshift bins of width  
$\Delta z=0.4$, defined here as: 
\Dznear ($0.2<z<0.6$), \Dzmid ($0.4<z<0.8$) 
and \Dzfar ($0.6<z<1.0$). 
Notice that the second bin fully overlaps 
with the other two, which are independent from each other. 

\cite{blake11c} calculated  
the effective redshift $z_{\rm eff}$ 
of $\xi$ in each slice as the weighted
mean redshift of the galaxy pairs in the separation bin 
$100<s<110$\hmpcii, where the $z$ 
of a pair is the average ($z_1+z_2$)/2. 
For \Dznearii, \Dzmidii and \Dzfar this results in 
$z_{\rm eff}=0.44, 0.6, 0.73$, respectively. 

\subsection{The WiZ-COLA simulation}\label{section:wizcola}

Simulated galaxy catalogs are a key 
tool for interpretation of large-scale structure measurements 
which are used to determine covariances, 
and test methodologies for potential biases. 
In this section we briefly describe the construction of the 
mock catalogs used in this analysis. 
For full details, the reader is referred to 
Koda et al. (in prep).

Constructing hundreds of mock catalogues from 
$N-$body simulations for WiggleZ is a challenging 
problem because the galaxies trace dark matter haloes with masses 
$\sim 10^{12}h^{-1}$M$_\odot$, 
an order of magnitude lower than those populated by luminous red galaxies.  
For this reason we employed cheaper 
methods of production of mocks that yield  
a good approximation to $N-$body simulations. 

In our first attempt to build mock catalogs, 
we tried implementing the second order Lagrangian Perturbation Theory 
method (2LPT; \citealt{bernardeau02a}), as described in \cite{manera12a}. 
However, we found  
that because of poor resolution, 
this method failed  
to identify correctly  
low-mass haloes 
such as those in which the low-bias WiggleZ galaxies reside. 

For this reason 
we developed a parallel version of the COmoving Lagrangian Acceleration simulation
\citep*[COLA,][]{tassev13a} which we used to generate 3600 realisations of
10-time step simulations --- 600 realisations for each of the 6 observational
regions in the WiggleZ survey.

Each of the WiggleZ COLA (\wizcolaii)
simulations consists of $1296^3$ $N$-body particles in a box of
$600h^{-1}\mathrm{Mpc}$ on a side, which gives a particle mass of
$7.5\times10^{9}h^{-1}M_\odot$. 
We use $3\times1296$ grids per dimension
to calculate the gravitational force with enough spatial resolution
\citep{tassev13a}. This 
simulation configuration has sufficient 
volume to contain one region of the WiggleZ survey for each redshift
range $z=0.2-0.6$, $0.4-0.8$, or $0.6-1.0$, and simultaneously
resolves dark matter haloes down to $10^{12}h^{-1}M_\odot$, which host
emission-line galaxies observed in the WiggleZ survey. 
Each simulation takes 15 minutes with 216 computation
cores, including halo finding. 

As fully described in Koda et al. (in prep), 
we populate the haloes using a Gaussian halo occupation 
distribution, such that the resulting projected correlation 
functions $w_{\rm p}(r_{\rm p})$ match those of the observations.
 
We then apply the WiggleZ selection function to the mock galaxies to
make simulated catalogues with correct survey geometry. When we apply the
mask, we rotate the simulation box to fit the survey volume into the 
box with minimum overlap, using the remapping algorithm by
\cite{carlson10a} to find the best rotation.
We output three snapshots at z =0.44, 0.6, and 0.73, for the three
redshift bins, \Dznearii, \Dzmid and \Dzfarii, respectively. 
In each redshift bin we use the appropriate independent 600
mocks to generate covariance matrices, as described in \S\ref{section:xi} 
and \S\ref{section:stat_methods}, 
and analyse each redshift bin
separately to measure \dvrs (as defined below).

Our simulation box is large enough for one redshift bin, but not for 
the full range $0.2<z<1$. This is not a problem when we treat
different redshifts separately 
(\S\ref{section:significance_detection} and \S\ref{section:distance_constraints}), 
but does not give
the correct correlation between \Dzmid and the other two redshift bins. 
For this reason, 
we also create 300 additional mock catalogues for each of the 6
regions to evaluate the correlation coefficient between the \dvrs 
measurements in  
the overlapping
redshift regions (as presented in \S\ref{section:cosmo_implications}).  
We combine, or {\it stitch},  two mock catalogues from different
realizations of $z=0.2-0.6$ and $z=0.6-1.0$, by joining them together 
at their sharp edges of $z=0.6$ and cut out the redshift
region $z=0.4-0.8$ appropriately from each. 
This mock does not have accurate clustering across
the boundary at $z=0.6$, but contains the the same mock galaxies that
exist in the other two redshift regions $0.2-0.6$ and $0.6-1.0$, which
is necessary to compute the correlation between the overlapping redshift
data. 
Because for each of the 600 realizations we use 
different snapshots to create the three original $\Delta z$ volumes, 
by stitching \Dznear and \Dzfar from different realizations  
we end up with 300 stitched versions.


\subsection{Reconstruction of the density field}\label{section:reconstruction}

In order to reduce effects of large-scale coherent motions 
on the \bafii,  
the reconstruction of the density field method  
is applied by shifting 
the galaxies to their near-original positions 
in the linear density field. 
Here we describe the calculation of the displacement vectors 
from the density fields, 
including the survey selection effects. 

We determine the displacement field ${\bmath \Psi}$ within the Zel'dovich 
approximation (\citealt{zel'dovich70a}) following the method described by \cite{padmanabhan12a}.  
Given that large-scale structure outside the survey regions 
contributes gravitationally to displacements within, it is necessary to 
enclose the observed volume within a larger ``embedded'' volume, into 
which we must extrapolate the density field in a statistically 
consistent manner.  The extrapolation is over any unobserved regions 
inside the survey cone, and into a ``padding'' volume which extends $200 
\, h^{-1}$ Mpc beyond each edge of a cuboid enclosing the survey region.  
For each of the eighteen volumes analyzed (6 angular regions and 3 redshift slices), 
we apply the reconstruction technique described here independently, 
because we do not expect volumes to affect each other due to the large distances between them. 

We summarize the steps of the method as follows, distinguishing between 
quantities evaluated over the observed and embedded volumes:

\begin{itemize}

\item We evaluate the smoothed, observed galaxy overdensity field, 
${\delta}({\bmath x})$, in each survey region.  We carry out this 
calculation by binning the galaxy distribution and normalized selection 
function in a 3D co-moving co-ordinate grid with a cell size of $5$\hmpc on the side, 
denoting these gridded 
distributions as \scb{$D_{\rm c}$} and \scb{$R_{\rm c}$} (where c is the cell number), 
and then determining ${\delta}$ by smoothing 
these distributions with a Gaussian kernel $G({\bmath x}) = 
e^{-({\bmath x}\cdot{\bmath x})/2\lambda^2}$ such that \scb{$\delta_{\rm c} = {\rm 
smooth}(D_{\rm c})/{\rm smooth}(R_{\rm c}) - 1$ and $\langle\delta_{\rm c}\rangle = 0$}.  We choose an r.m.s.\ 
smoothing scaling $\lambda = 15$ \hmpc for our analysis, noting 
that our results are not sensitive to this choice. \scb{From here on we drop 
the `c' notation from $\delta$, for  convenience.} 

\item We generate a realization of an ``unconstrained'' Gaussian random 
field across the embedded volume, $\tilde{\delta}_U$, using an assumed 
galaxy power spectrum $P(k)$ consistent with fits to the data in the 
observed region.  We smooth the unconstrained overdensity field in the 
same manner as the observed overdensity field.

\item We use the Hoffman-Ribak algorithm (\citealt{hoffman91a}; Equation 3 in \citealt{padmanabhan12a}), 
as our best estimate of the overdensity field in the 
embedded volume:
\beq\label{equation:delta_constrained}
 \tilde{\delta} = \tilde{\delta}_U + {\bf \tilde{\textbfss C} }\, {\textbfss C}^{-1} \left( {\delta} -  {\textbfss P} 
\tilde{\delta}_U \right)
\eeq
where ${\textbfss P}$ is a matrix of zeros and ones which projects a vector from the 
embedded volume to the observed volume, and ${\textbfss C}$ and ${\tilde{\textbfss  C}}$ are the 
covariance matrices of pixels in the observed and embedded volumes, 
respectively, which are just the correlation functions $\xi$:
\begin{equation}
C_{ij} = \langle {\delta}({\bmath x}_i) \, {\delta}({\bmath x}_j) \rangle = 
\xi(|{\bmath x}_i-{\bmath x}_j|)
\end{equation}

Following \cite{padmanabhan12a}, we solve Equation \ref{equation:delta_constrained} 
in a number of steps. (i) We evaluate ${ u} = {\delta} - {\textbfss P} \tilde{\delta}_U$ 
by simple projection of $\tilde{\delta}_U$ from the embedded to the 
observed volumes.  (ii) We solve ${ v} = {\textbfss C}^{-1} {u}$ using a preconditioned 
conjugate gradient algorithm to determine the solution of ${\textbfss C}{v} = {u}$, 
using a modified version of the Numerical Recipes subroutine \texttt{linbcg}.  
For each iteration, the expression ${\textbfss C}{v}$ is evaluated by Fast Fourier 
Transforms, using the fact that multiplication by ${\textbfss C}$ is equivalent to 
convolution by $\xi(r)$.  Therefore, ${\rm FT}({\textbfss C} {v})$ is equal to the product 
of $P(k)$ and ${\rm FT}({v})$, where we note that the power spectra contain the 
galaxy shot noise contribution $1/n$ in terms of mean galaxy density 
$n$.  (iii) We project ${v}$ into the embedded space, $\tilde{v} = 
{\textbfss P}^{-1} {v}$, and calculate $\tilde{w} = \tilde{\textbfss C} \tilde{v}$ as above.  
(iv) The final overdensity field in the embedded volume is given by 
$\tilde{\delta} = \tilde{\delta}_U + \tilde{w}$.

\item Finally, we estimate the displacement field ${\bmath \Psi}$ in the 
 Zel’dovich approximation as
\begin{equation}
{\bmath \nabla}\cdot{\bmath \Psi} + \left(f/b\right) {\bmath \nabla} \cdot (\Psi_s \hat{\bmath s}) = - \tilde{\delta}/b 
\label{eqpsi}
\end{equation}
 where $f$ is the growth rate of structure at the survey redshift, $b$ is 
the galaxy bias factor, and $\Psi_s = {\bmath \Psi} \cdot \hat{\bmath s}$ is the 
displacement in the line-of-sight direction.  We assume values $f = 0.70  \ (z=0.44), 0.76 \ (z=0.6), 0.79 \ (z=0.73)$ 
and $b = 1, 1.1, 1.2$ (for \Dznearii, \Dzmid and \Dzfarii, respectively), noting that our results are not sensitive to these 
choices.  The flat-sky approximation is valid for the WiggleZ survey 
regions, and we can therefore take the line-of-sight direction as 
parallel to a single Cartesian axis, which we take as the $x$-direction, 
such that $\Psi_s = {\bmath \Psi}\cdot \hat{\bmath x}$.  We then solve Equation 
\ref{eqpsi} by substituting ${\bmath \Psi} = {\bmath \nabla} \phi$ and taking the Fourier 
transform of the equation to obtain
\begin{equation}
\left[ (1+f/b) k_x^2 + k_y^2 + k_z^2 \right] {\rm FT}[\phi](k_x,k_y,k_z) = 
\frac{{\rm FT}[\tilde{\delta}](k_x, k_y, k_z)}{b}, 
\end{equation}
\scb{where FT is the Fourier Transform.}

The inverse Fourier transform then yields the displacement field 
${\bmath \Psi(x,y,z)} = {\bmath \nabla} \phi$. 

\item We then shift each galaxy and random point by $-{\bmath \Psi}$.  
To subtract the Kaiser effect in redshift space, 
the galaxies are also shifted an additional $-f\Psi_x$ 
in the $x$ dimension. 
This additional shift is not applied to the random points. 

\end{itemize}

At the end of this procedure, 
for each of the eighteen volumes we obtain a 
shifted data catalog and a shifted random point catalog. 


\subsection{Correlation Functions}\label{section:xi}

To estimate the correlation function, 
we compare pair counts of the data 
to those of a sample of random points.
The random points are distributed 
in a Poisson-like manner, 
such that they trace the mask of the survey, 
as described in \cite{blake10a}. 
To reduce shot-noise effects of the mask, 
we use a ratio of 100 random points per data point. 

Before calculating pairs, 
we first convert the data and randoms from 
the R.A, Dec, $z$ coordinate system 
to a co-moving Euclidian system assuming a 
flat $\Lambda$CDM fiducial cosmology as defined in 
\cite{komatsu09a}: $\Omega_{\rm m}=0.27$. 
When calculating the pairs, 
each galaxy and random point is assigned  
a weight according to the \cite{feldman94} 
minimum variance weighting, which takes into 
account the number density at a given redshift $n(z)$: 
\beq
w(z) = \frac{1}{1+P\cdot n(z)},
\eeq
where we assume $P=5000h^{-3}$Mpc$^3$ 
as the characteristic power spectrum amplitude 
at the physical scales of interest. 

We calculate the \cite{landy93a} correlation function estimator $\xi$  for 
each of the eighteen volumes. This is done first by calculating:  
\beq\label{equation:xi(mu,s) estimator}
\xi(\mu,s)=\frac{DD-2DR+RR^{\rm num}}{RR^{\rm denom}}, 
\eeq
where the line-of-sight direction $\mu=1$ is 
defined as the direction which bisects the separation vector $\bmath{s}$ 
between each pair, and $s\equiv |\bmath{s}|$. 
The normalized galaxy-galaxy pair count is $DD(\mu,s)$ 
and similarly for the normalized galaxy-random $DR$ and normalized random-random $RR$ counts. 

The reconstruction procedure described in \S\ref{section:reconstruction} results in various 
data and random sets which we use as follows. 
For the pre-reconstruction case we use the original data and random point counts 
where both $RR$ terms in Equation \ref{equation:xi(mu,s) estimator} are the same. 
In the reconstruction case we use the shifted data for $DD$ and $DR$, and shifted randoms for 
$DR$ and $RR^{\rm num}$. 
Finally, for the $RR^{\rm denom}$ term we use the original non-shifted randoms. 
In this study we examine results  
using two different separation bin widths $\Delta s$, 
of 3.3\hmpc and 6.7\hmpcii.

To account for the volume limitation of each region, the integral constraint correction 
is calculated as:  
\beq\label{equation:ic}
I.C = \frac{\sum_{s_i}{\xi^{\rm theory}(s_i)RR^{\rm num}(s_i)}}{\sum_{s_i}RR^{\rm num}(s_i)}
\eeq 
and added to $\xi(\mu,s)$. For this purpose the $RR$ terms used are calculated 
in each region to a large separation $s$ at which $RR$ is zero. In the largest region 
this is just over $1h^{-1}$Gpc. 
The theoretical model used, $\xi^{\rm theory}$, is a combination 
of the template used in the analysis for $s>50$\hmpc (see \S\ref{section:modeling}), and  
a linear model for lower separation bins $s_i$. For the reconstruction 
case we use the shifted random point count $RR$, and do not include the 
Kaiser boost term in $\xi^{\rm theory}$. 
We verify that the resulting values of $I.C$ are not sensitive to details of this procedure. 

We then obtain the angle-averaged correlation function $\xi_0$ and quadrupole $\xi_2$ of each of the 
18 volumes by 
integrating each $\xi(\mu,s)$ using the appropriate Legendre polynomials. 
We follow this procedure for both the data and the 600 mock catalogues,  
performing measurements before and after reconstruction. 

To calculate the three redshift slice correlation functions $\xi^{\Delta z}$ 
we combine the correlation functions 
of six angular regions $\Omega$ in the following manner. 
To account for the correlations between the multipoles (\citealt{taruya11a,kazin11a}), 
we define the vector ${\bmath \xi_{[0,2]}^\Omega}$ 
that contains $\xi_0^\Omega$ and $\xi_2^\Omega$ and 
therefore has a length equal to double the number of bins. 
We emphasize that we use the $\xi_2^\Omega$ 
information to construct the $\xi_0^{\Delta z}$ because the multipoles  
are not independent, as shown below. 

The resulting covariance matrix ${\ \textbfss C_{[0,2]}}$ is defined as: 
\begin{dmath}\label{equation:cij_equation}
C^{\Omega}_{[0,2]_{ij}}= \frac{1}{N_{\rm mocks}-1} \sum_{m=1}^{N_{\rm mocks}} \left(   {\overline{\xi^{\Omega}_{\rm [0,2]}}}_i - {\xi_{\rm [0,2]}^{\Omega \ m}}_i \right) \left(   {\overline{\xi^{\Omega}_{\rm [0,2]}}}_j - {\xi_{\rm [0,2]}^{\Omega \ m}}_j \right), 
\end{dmath}
where the over-line denotes the mean value of $N_{\rm mocks}=600$.

Following \cite{white10a}, we then combine these to obtain: 
\beq\label{equation:xi02_combined}
{\bmath \xi_{[0,2]}^{\Delta z}} ={\textbfss C_{[0,2]}^{\Delta z}}\sum_{\Omega}
\left({\textbfss C^{\Omega}_{[0,2]}}\right)^{-1}{\bmath \xi^{\Omega}_{[0,2]}}, 
\eeq
where 
\beq\label{equation:Cij02_combined}
\left({\textbfss C_{[0,2]}^{\Delta z}}\right)^{-1}=\sum_{\Omega} \left({\textbfss C^{\Omega}_{[0,2]}}\right)^{-1}. 
\eeq

\begin{figure*}
\begin{center}
\includegraphics[width=0.32\textwidth]{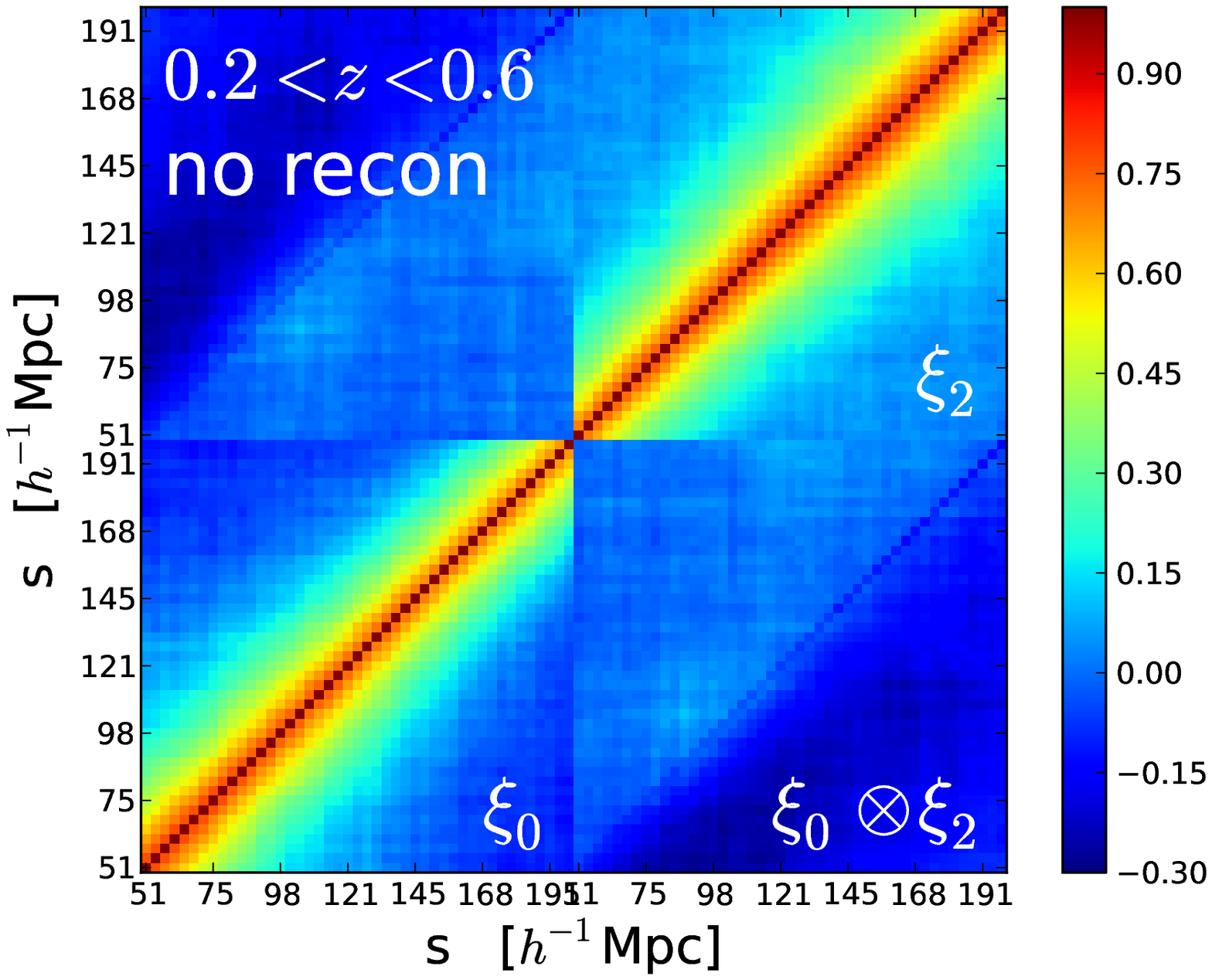}
\includegraphics[width=0.32\textwidth]{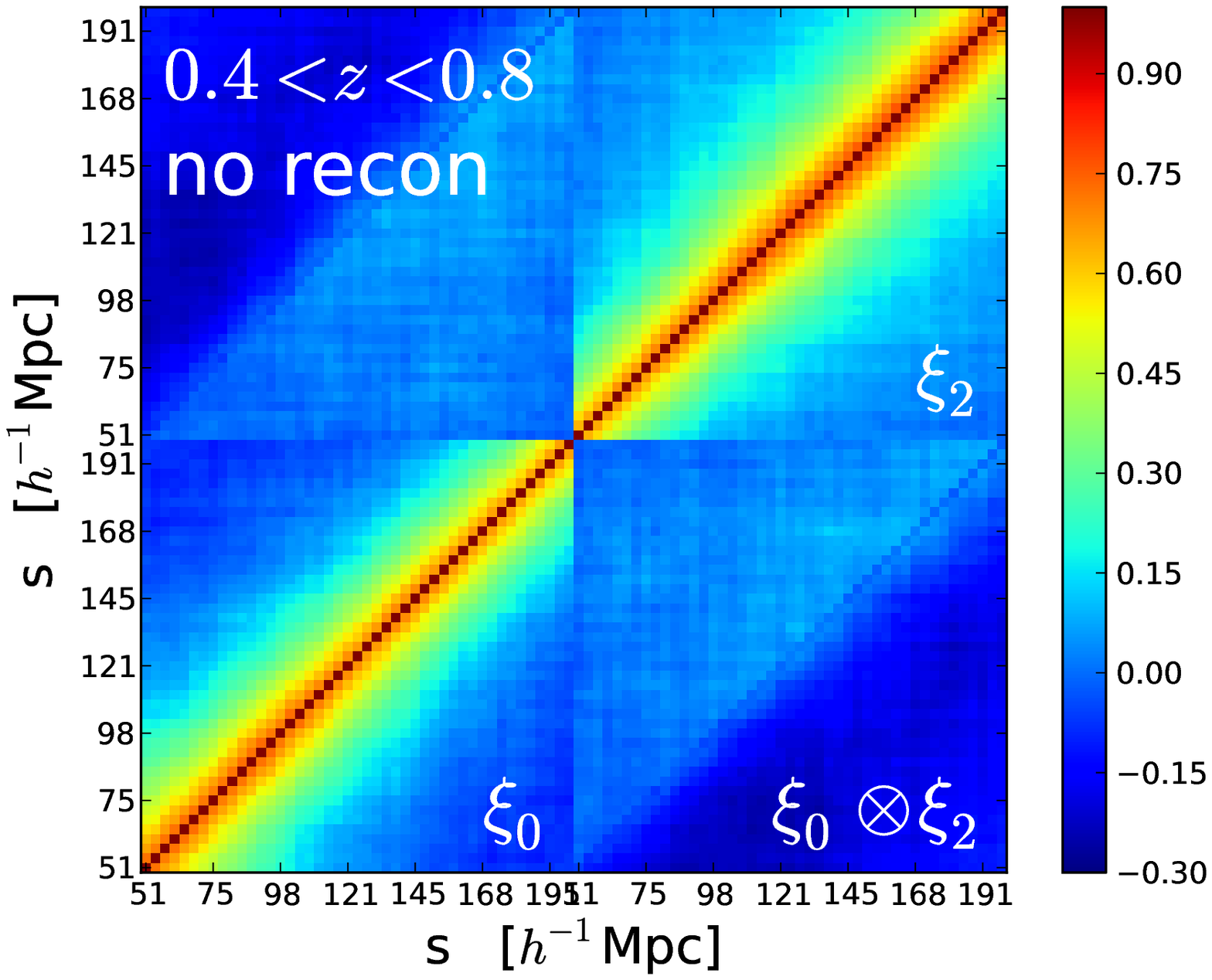} 
\includegraphics[width=0.32\textwidth]{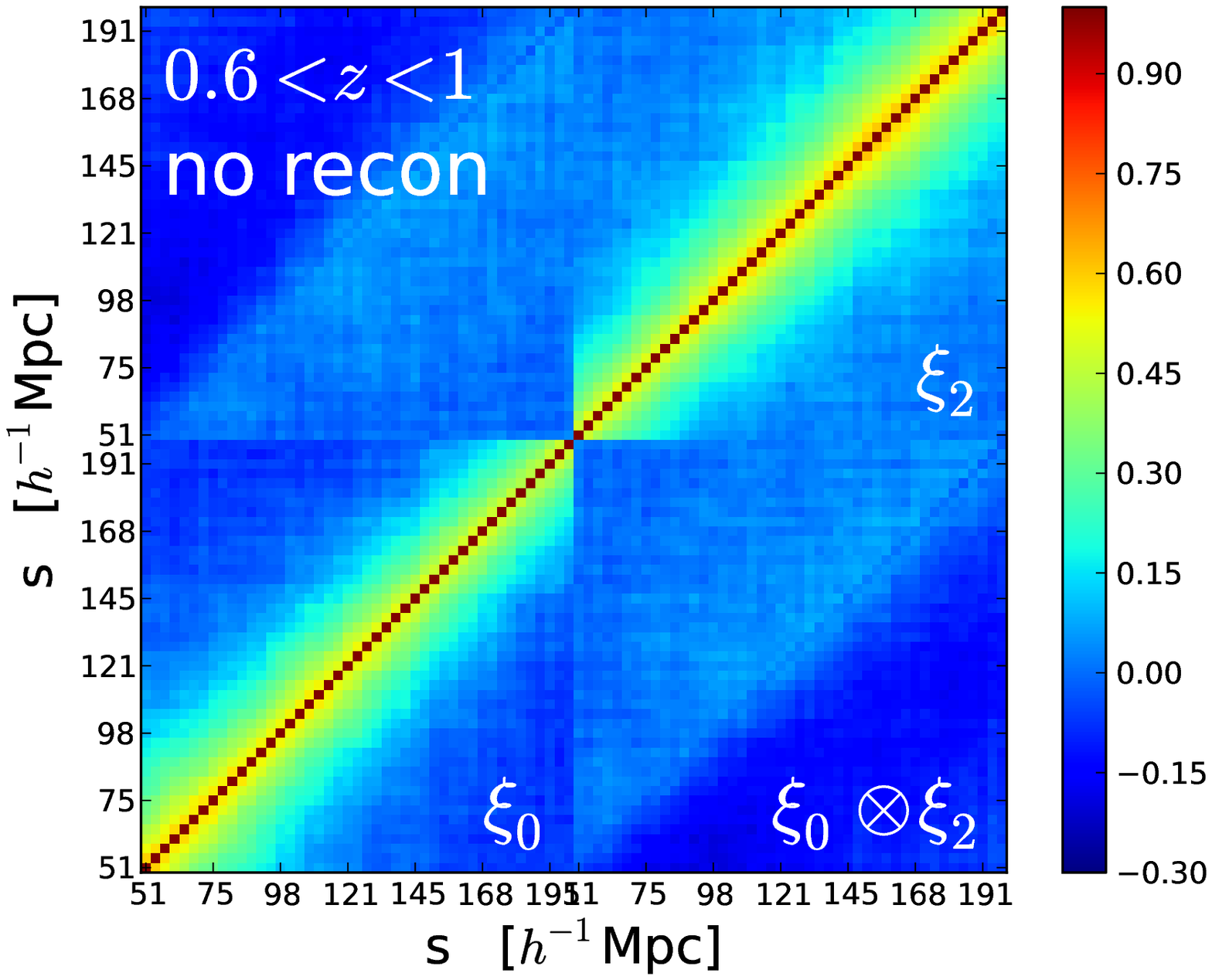} 
\includegraphics[width=0.32\textwidth]{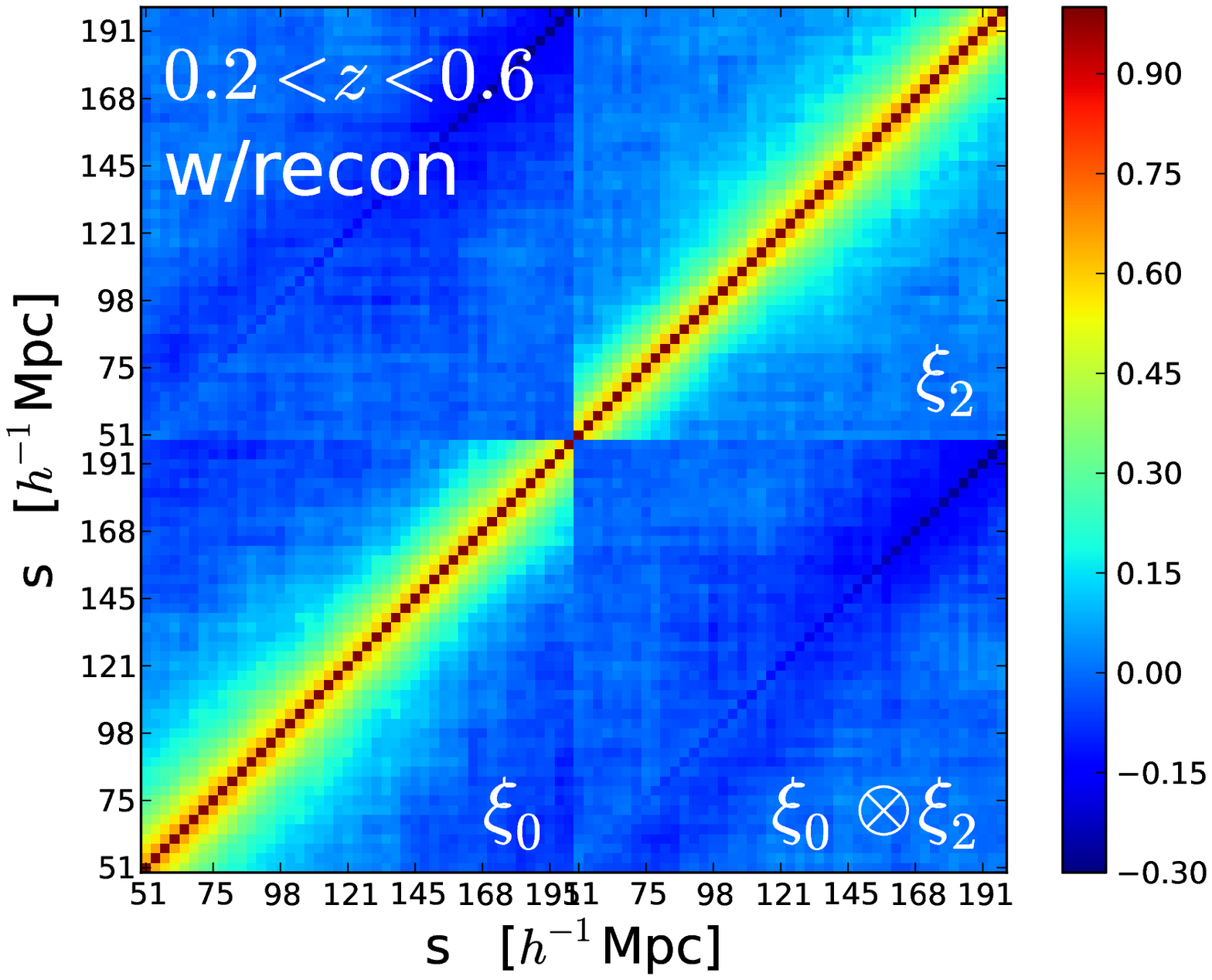} 
\includegraphics[width=0.32\textwidth]{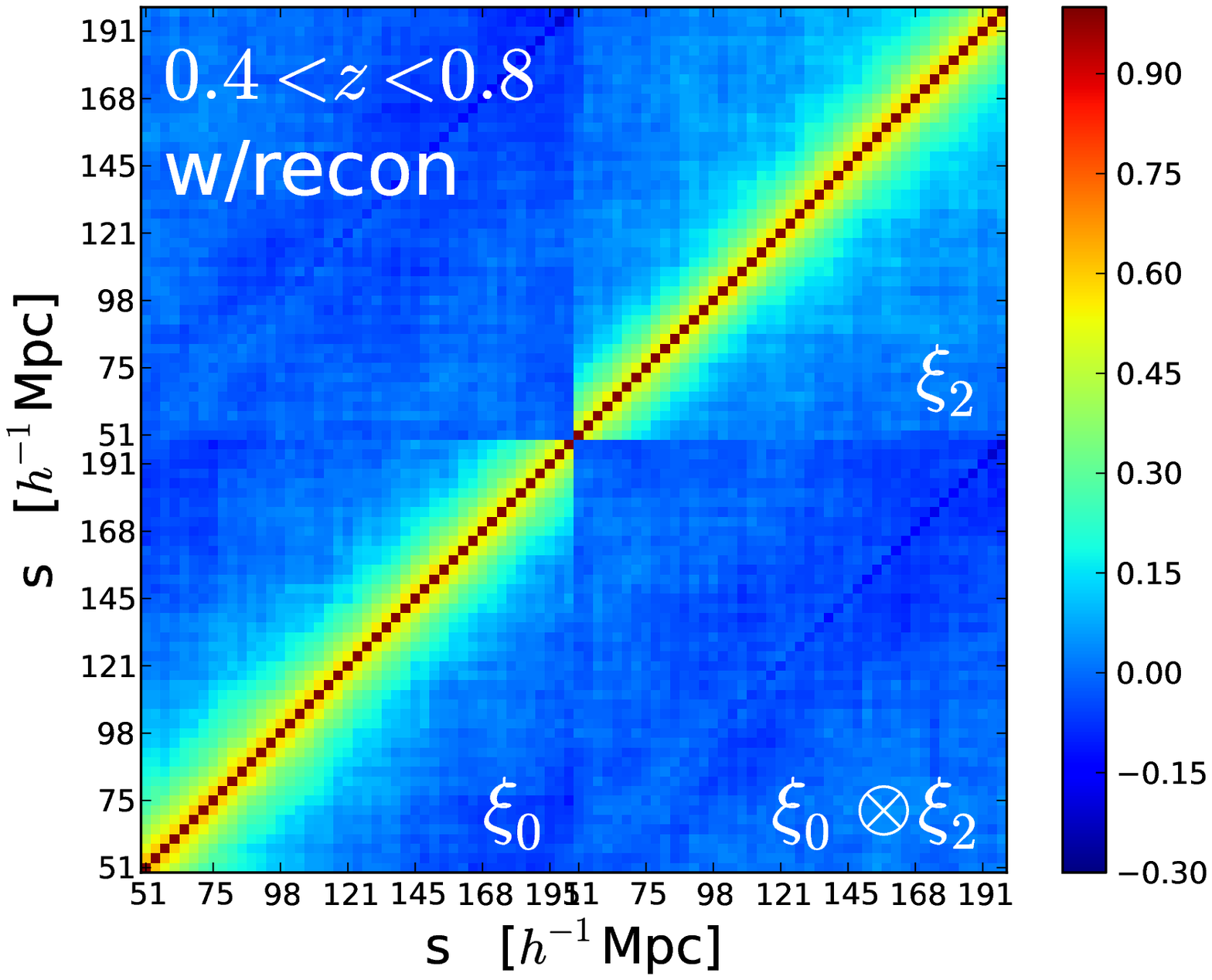} 
\includegraphics[width=0.32\textwidth]{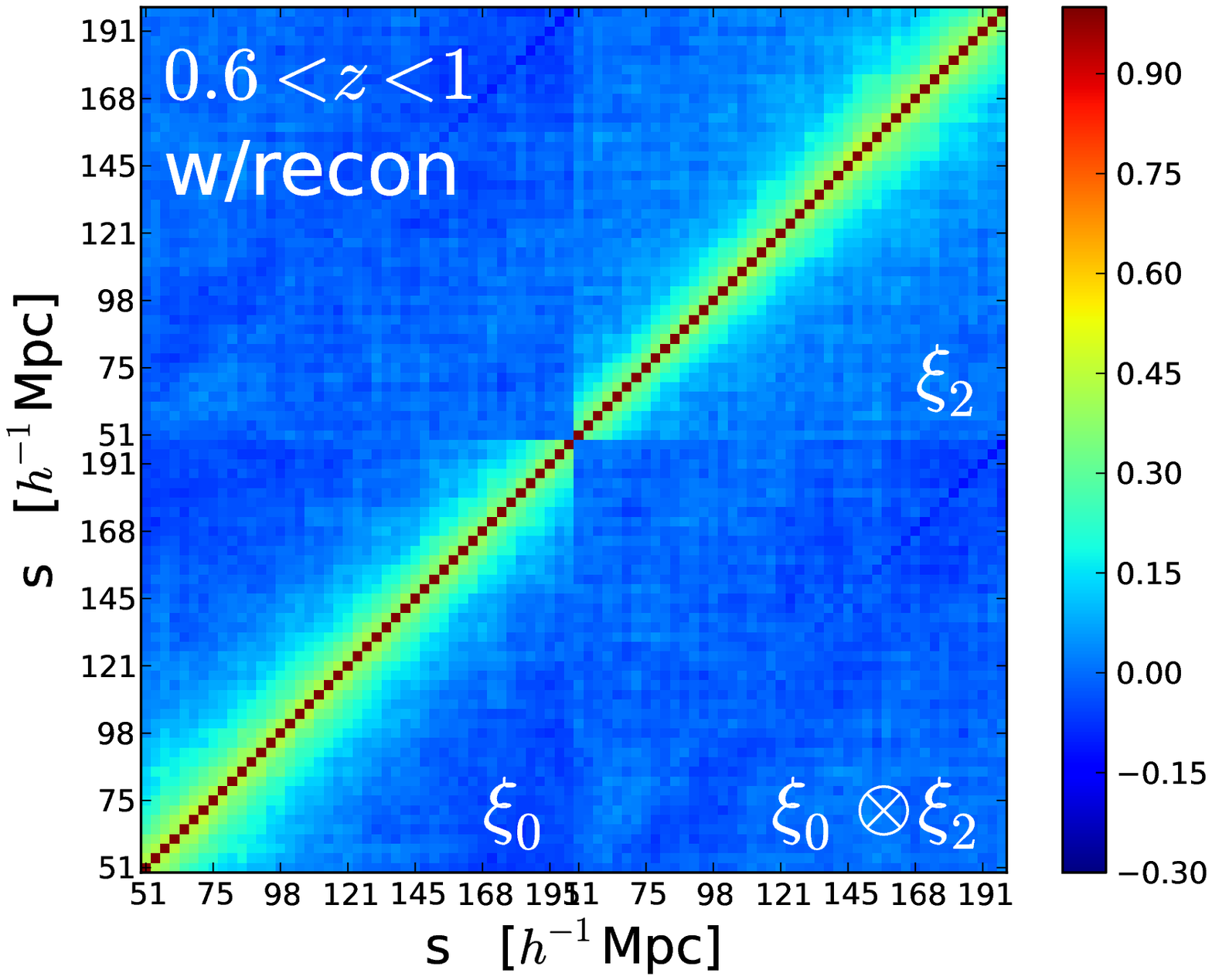} 
\includegraphics[width=0.32\textwidth]{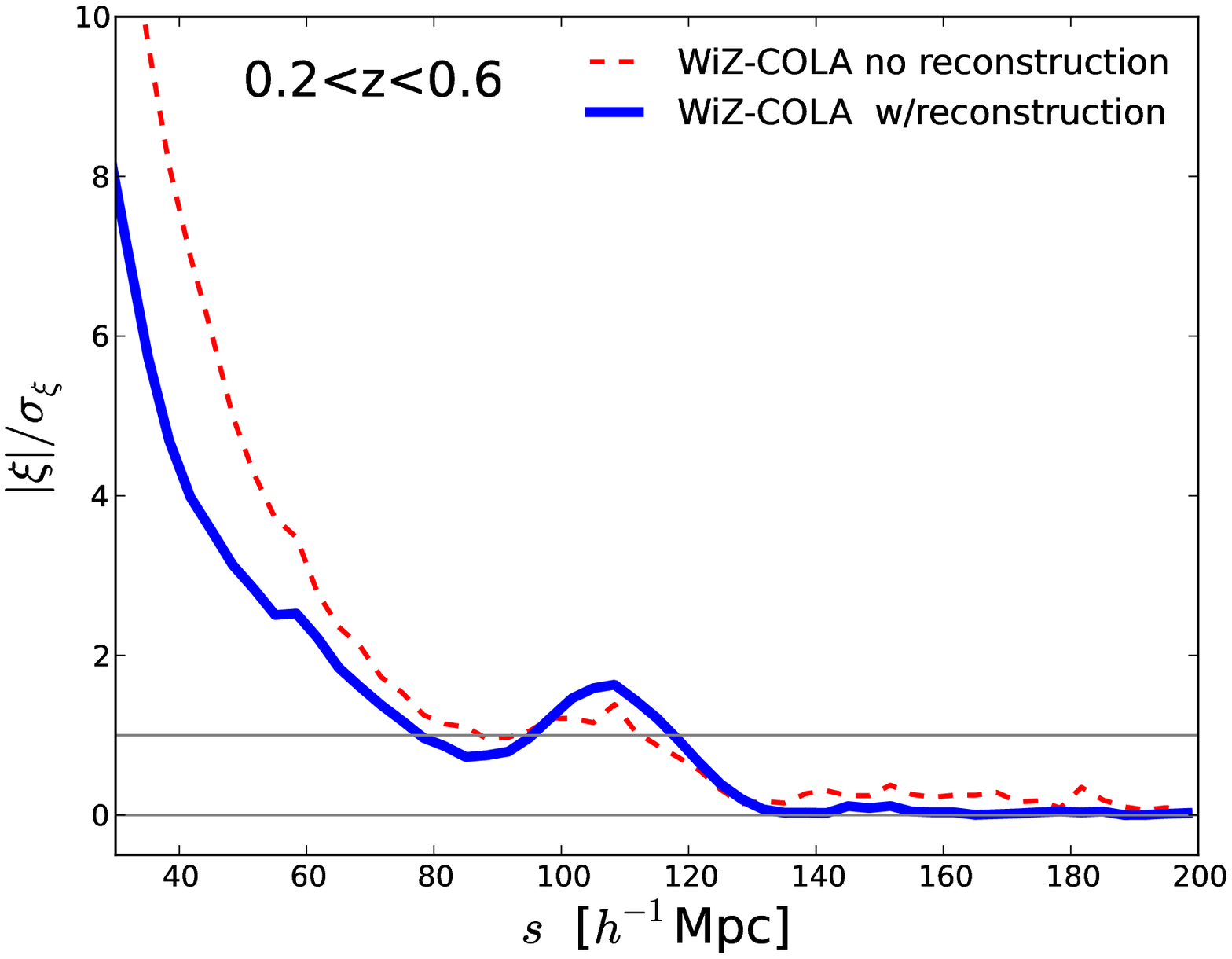}
\includegraphics[width=0.32\textwidth]{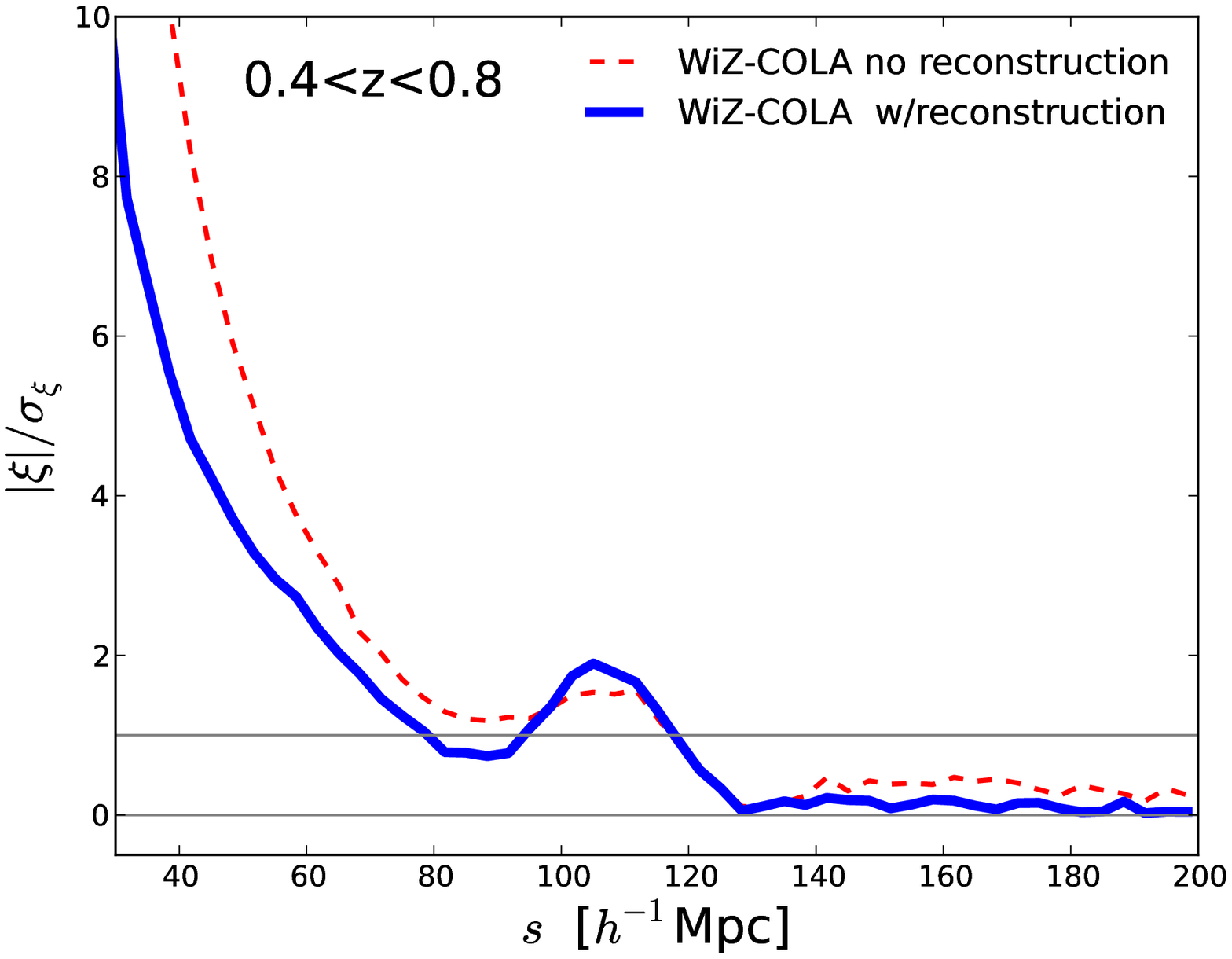}
\includegraphics[width=0.32\textwidth]{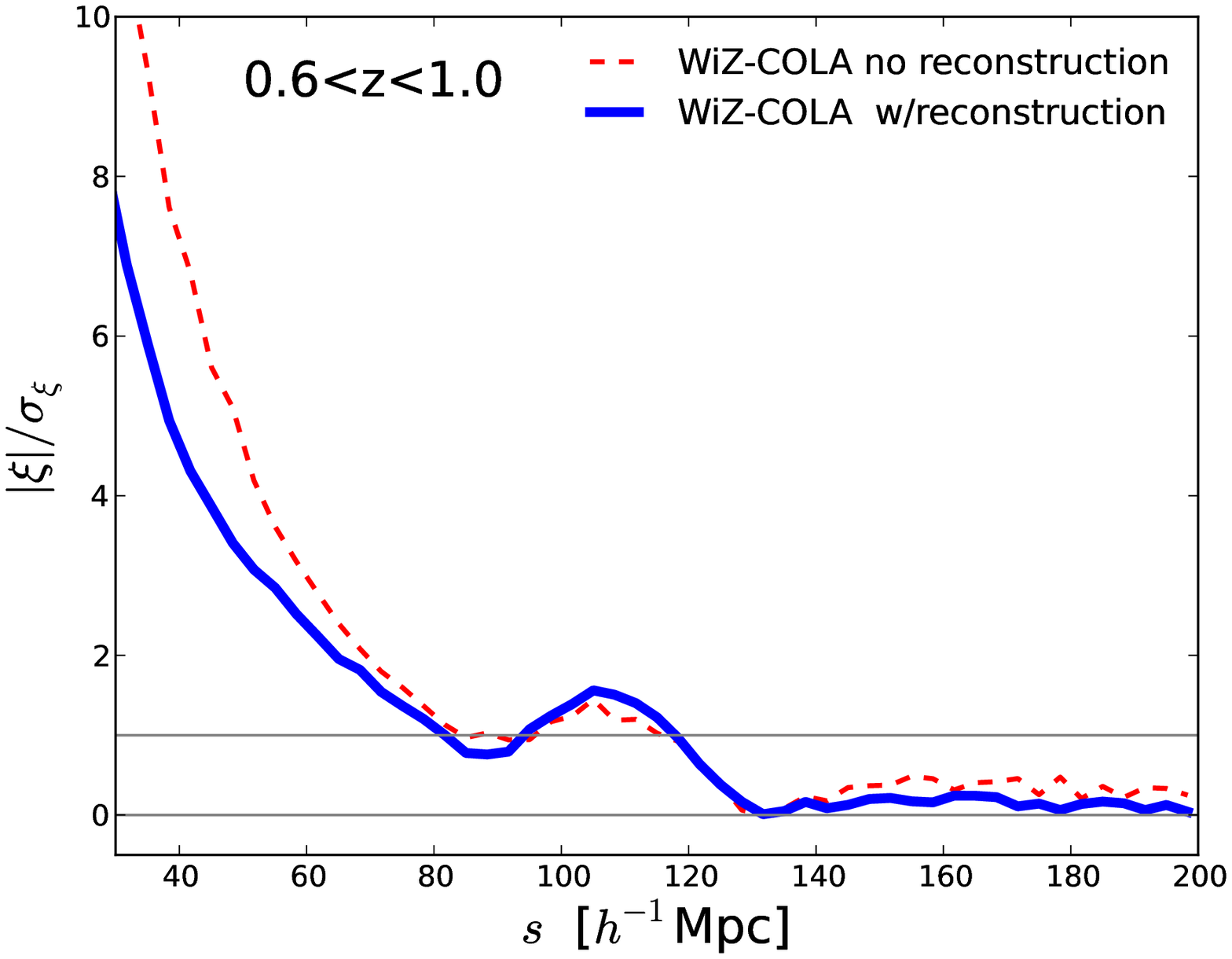}
\caption{
The top and center panels show the normalized covariance matrix 
$C^{ij}_{[0,2]}/\sqrt{C^{ii}_{[0,2]}C^{jj}_{[0,2]}}$ before and after reconstruction, respectively, 
for each of the $\Delta z$ volumes, as indicated.   
The bottom panels show comparisons of S/N ratios of the monopole $|\xi_0|/\sigma_{\xi_0}$, 
before (dashed red) and after reconstruction (solid blue), 
where we define 
the uncertainties $\sigma_{\xi_0} = \sqrt{C_{ii}}$ of the ``0" component. 
}
\label{figure:Cij_pre_post_wizcola} 
\end{center}
\end{figure*}

Figure \ref{figure:Cij_pre_post_wizcola} displays the 
resulting ${\textbfss C^{\Delta z}_{[0,2]}}$ for all 
three redshift volumes. 
The top and center row of panels show the normalized values 
pre- and post-reconstruction, respectively. 
The bottom row of panels displays the 
signal-to-noise (S/N) of the monopole defined as $|\xi_0|/\sigma_{\xi_0}$, 
where the uncertainty $\sigma_{\xi_0}$ is the 
square root of the diagonal elements 
of the monopole component of ${\textbfss C_{[0,2]}}$. 


We notice that the off-diagonal normalized terms in the $\xi_0$ and $\xi_2$ quadrants  
are suppressed in the post-reconstruction case compared to pre-reconstruction. 
This can be explained by the restoration of the linear density field 
and removal of the galaxy displacements.

The bottom panels of  Figure \ref{figure:Cij_pre_post_wizcola}  show clear improvement 
in the S/N of $\xi_0$ at the scale of the 
\baf in all $\Delta z$. 
The improvement with reconstruction is 
$40\%$ for \Dznearii, 
$25\%$ for \Dzmid and $15-25\%$ for \Dzfarii. 
This is the case for both separation widths of 
 $\Delta s=3.3$\hmpc and $6.7$\hmpcii. 
The S/N is lower at other scales ($s<90$\hmpc and $s>130$\hmpcii) 
because of the suppression of the 
redshift-space clustering power. 

We defer investigation of the cosmological content of $\xi_2$ 
to future studies, 
and from hereon refer to $\xi$ as the angle-averaged measurement.

\begin{figure*}
\begin{center}
\includegraphics[width=0.49\textwidth]{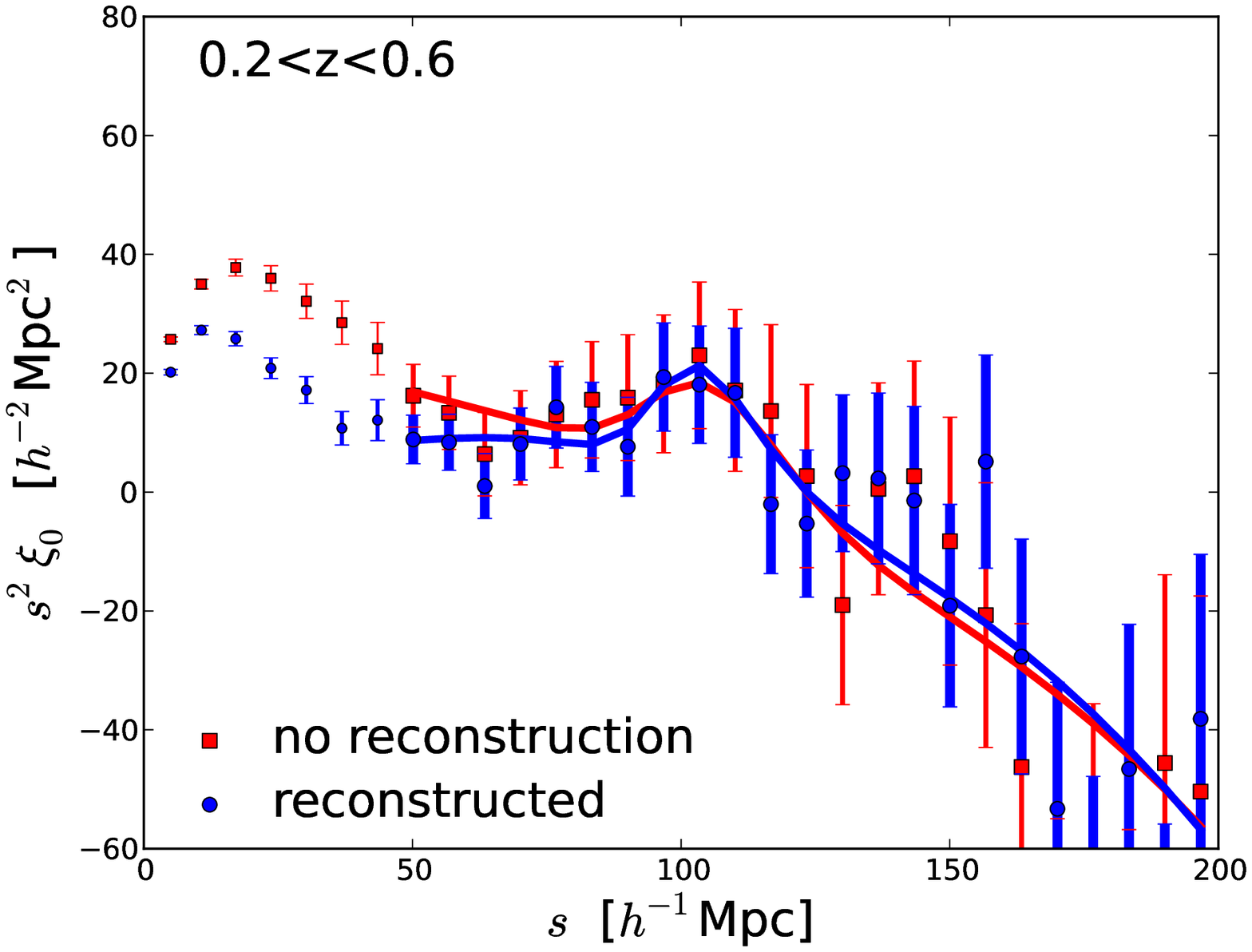}
\includegraphics[width=0.49\textwidth]{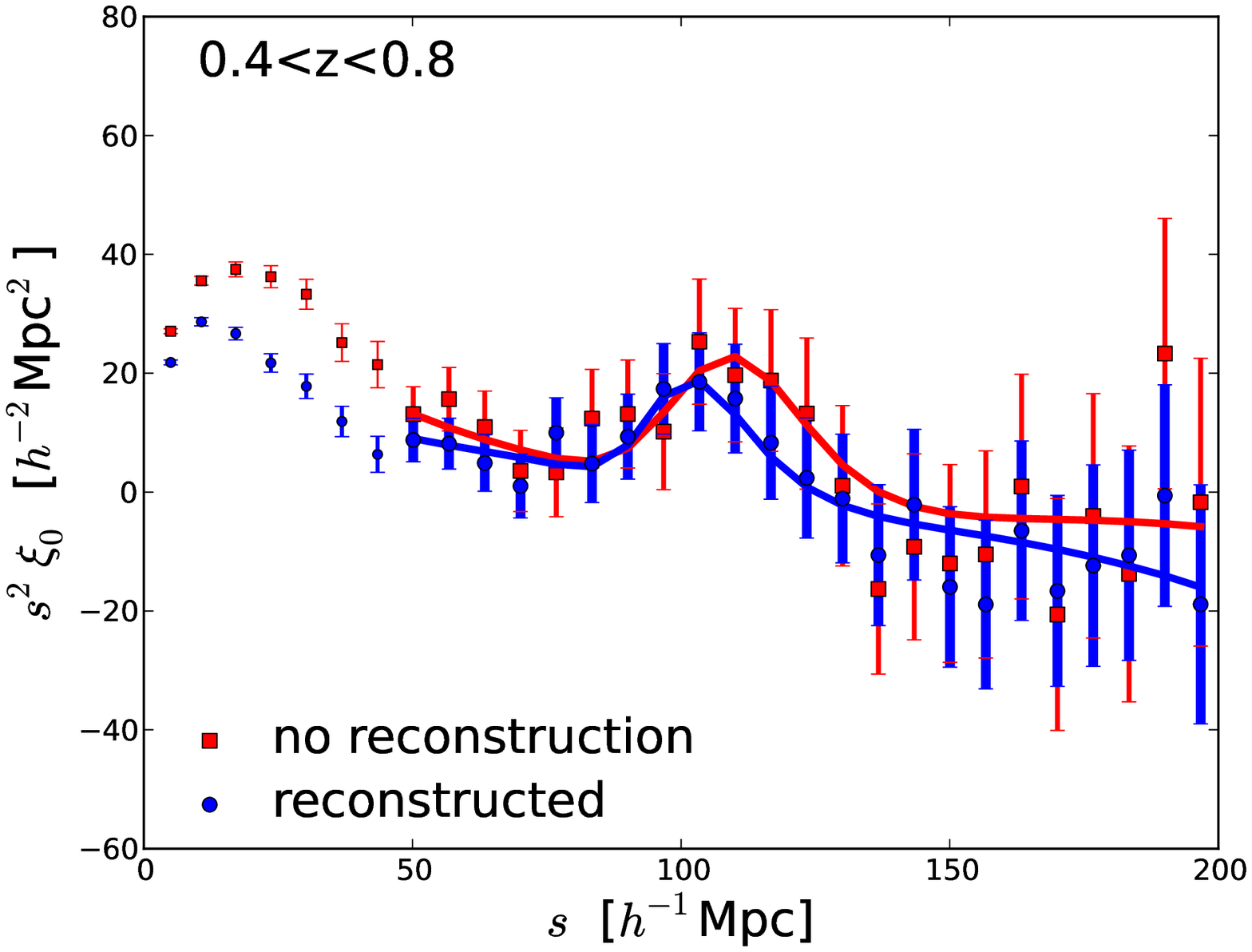} 
\includegraphics[width=0.49\textwidth]{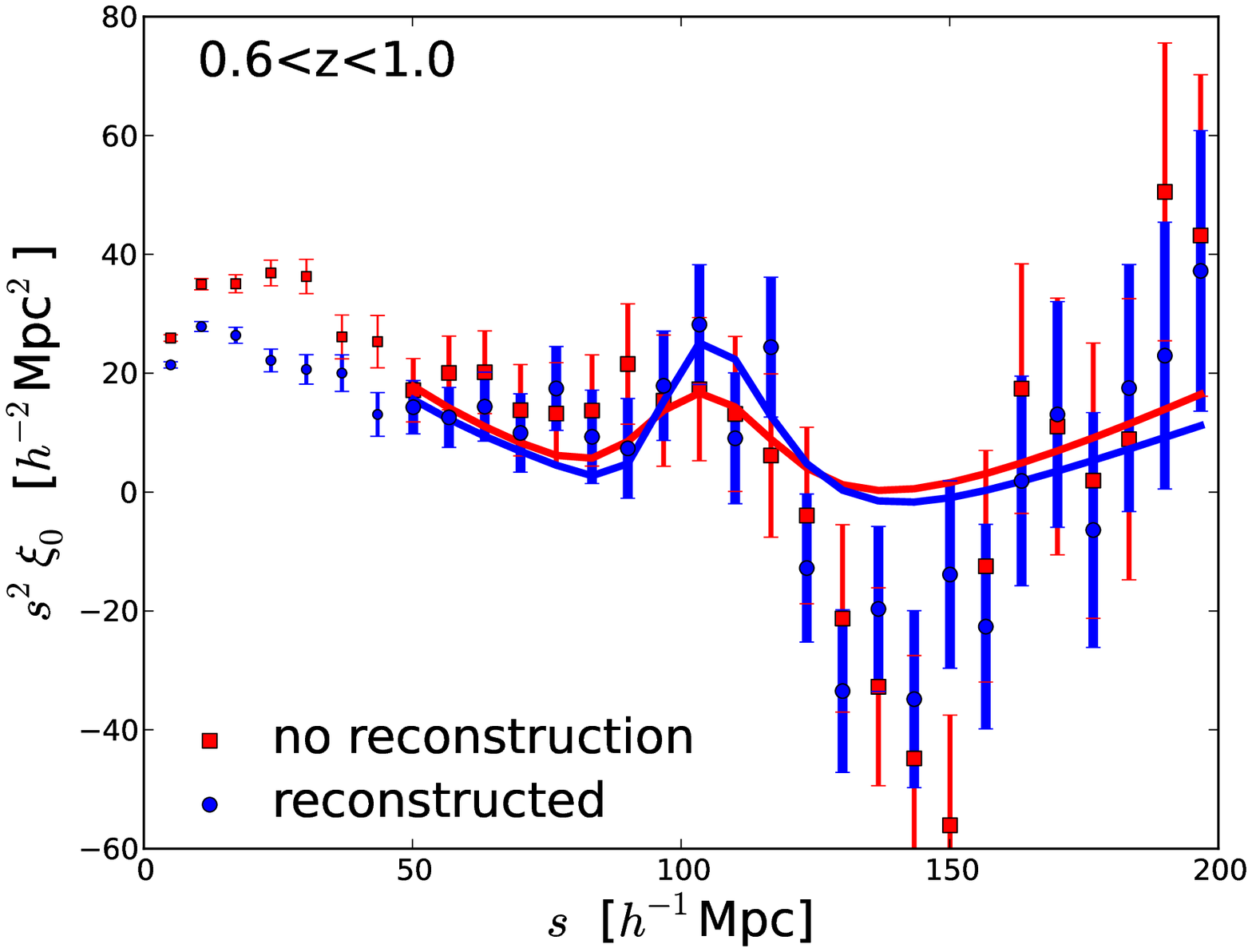}
\includegraphics[width=0.49\textwidth]{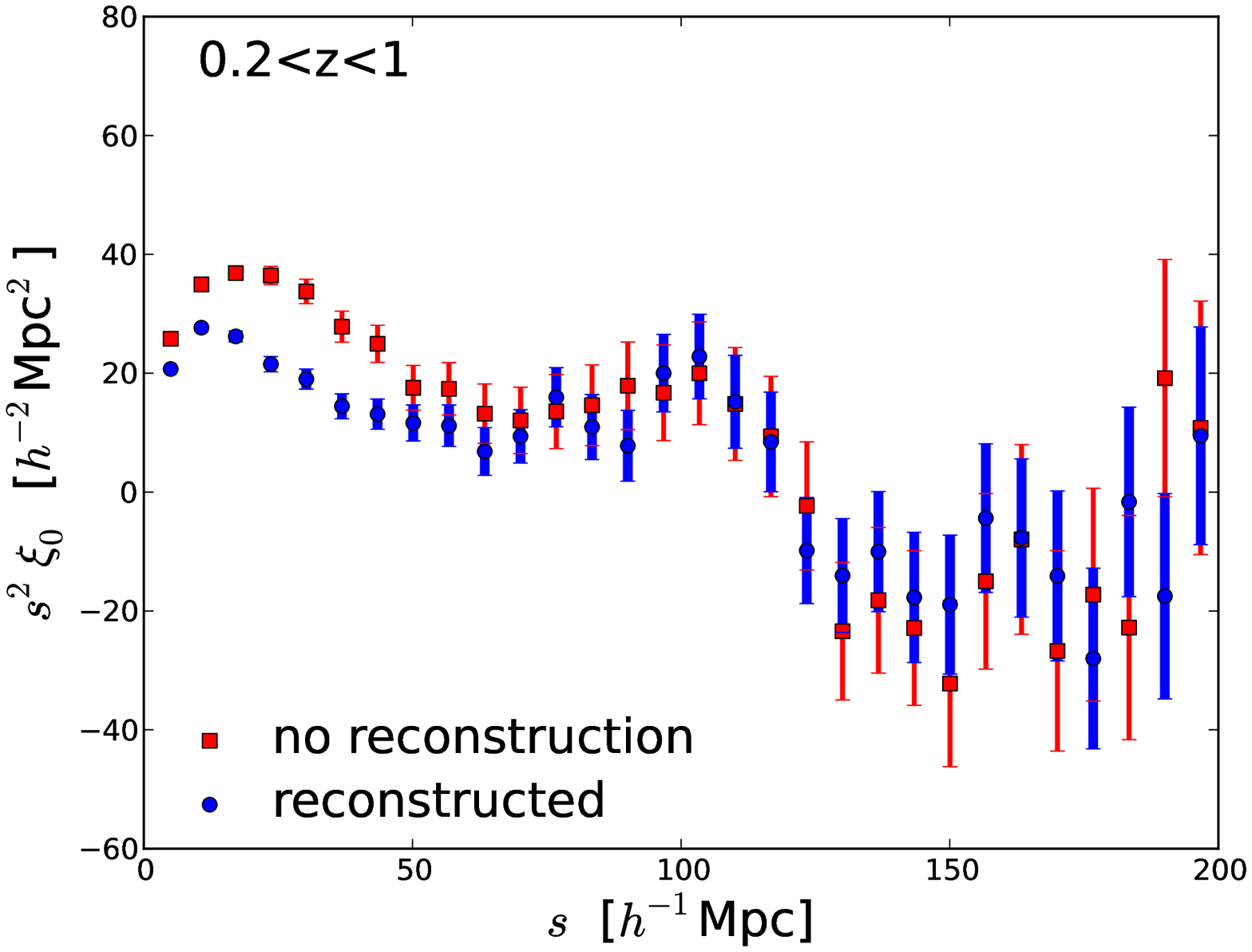} 

\caption{
The WiggleZ two-point correlation functions shown before (red squares) 
and after applying reconstruction (blue circles) for three redshifts bins 
and the full $z$ range, 
as indicated. These are plotted as $\xi s^2$ to emphasize the region of the \bafii. 
The uncertainty bars are the square root of the diagonal elements of the covariance matrix. 
The solid lines are the best fitting models to the range of analysis $50<s<200$\hmpcii. 
We see a clear sharpening of the \baf after reconstruction in all cases.
} 
\label{figure:xi0_pre_post_wigglez} 
\end{center}
\end{figure*}

\begin{figure*}
\begin{center}
\includegraphics[width=0.49\textwidth]{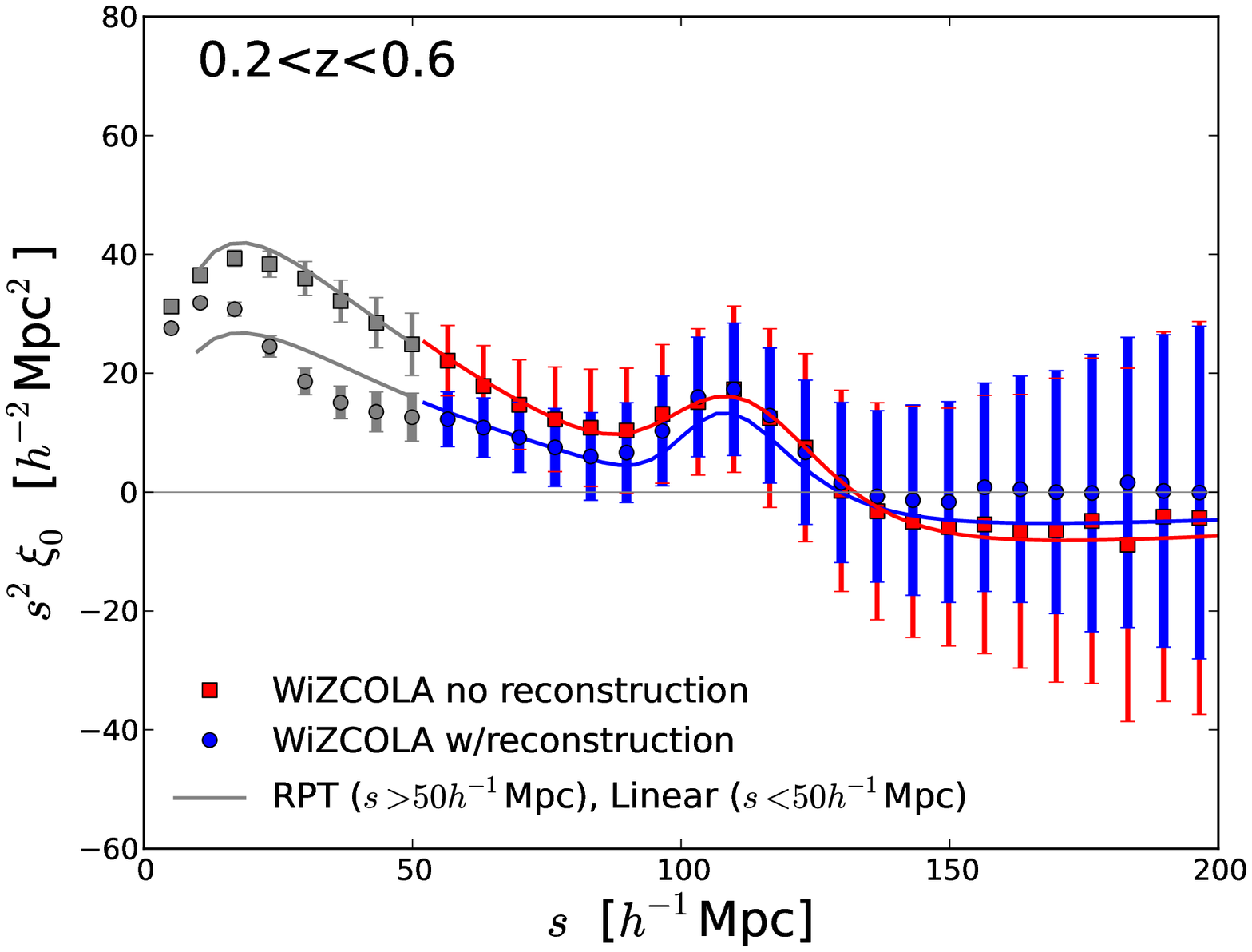} 
\includegraphics[width=0.49\textwidth]{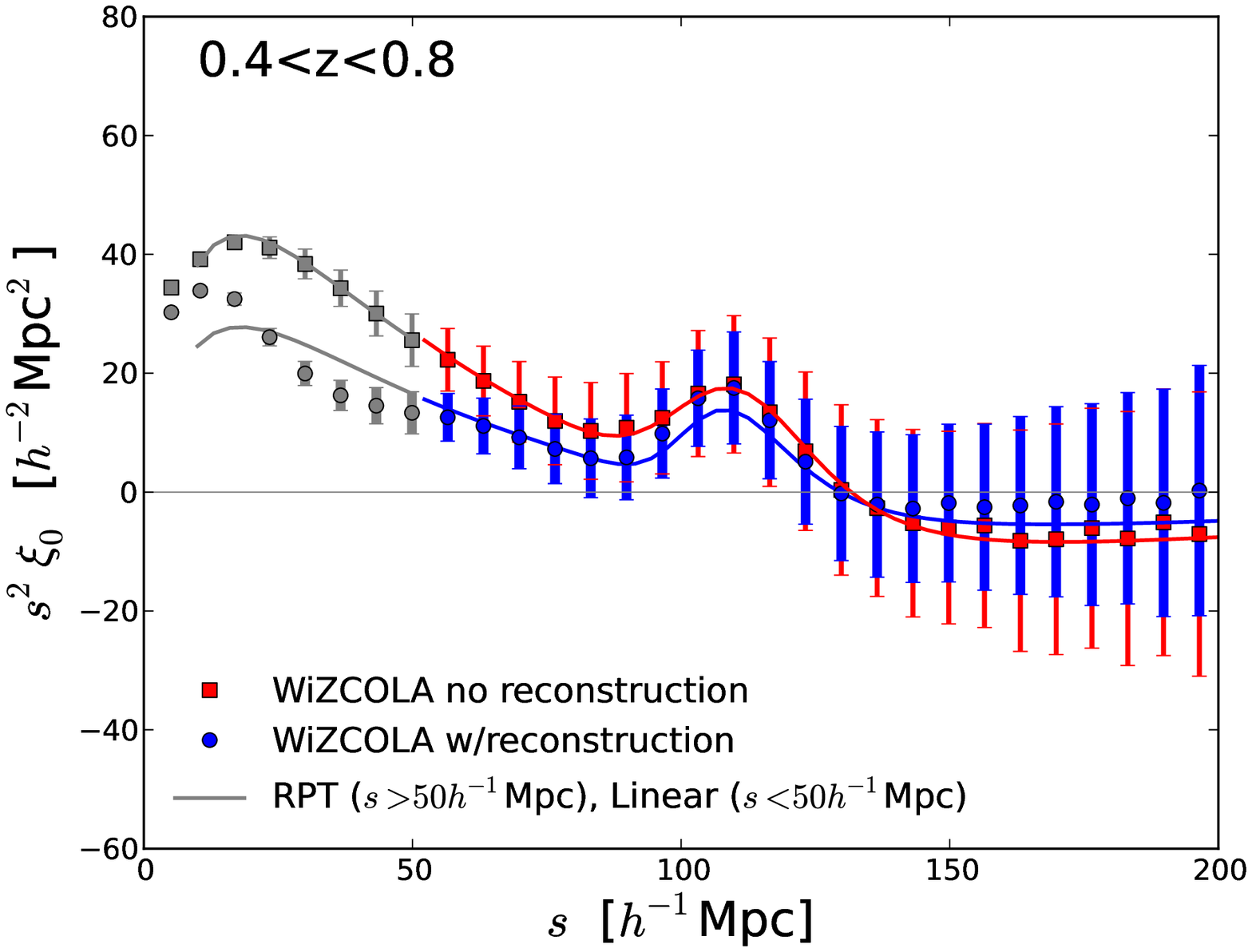}
\includegraphics[width=0.49\textwidth]{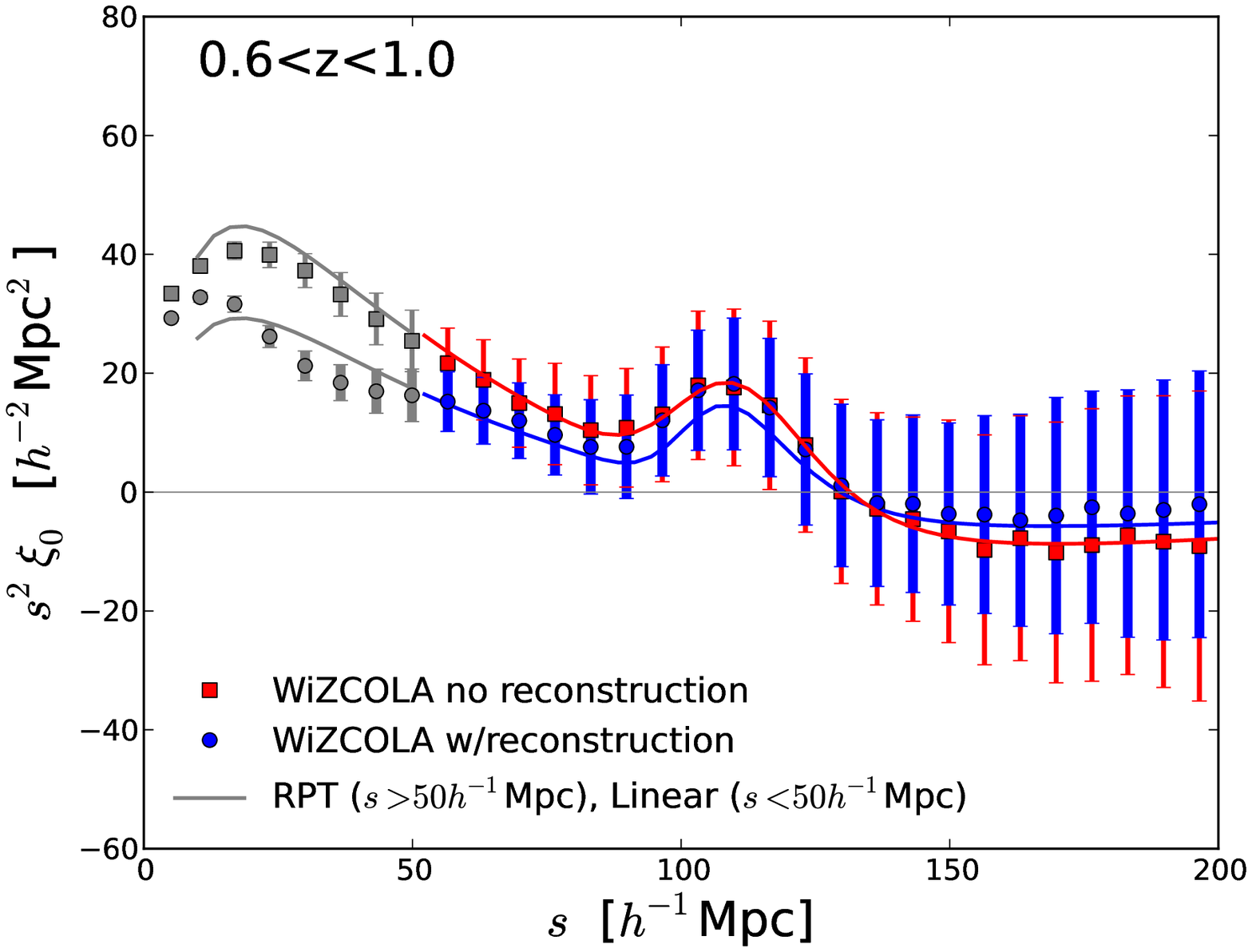}
\includegraphics[width=0.49\textwidth]{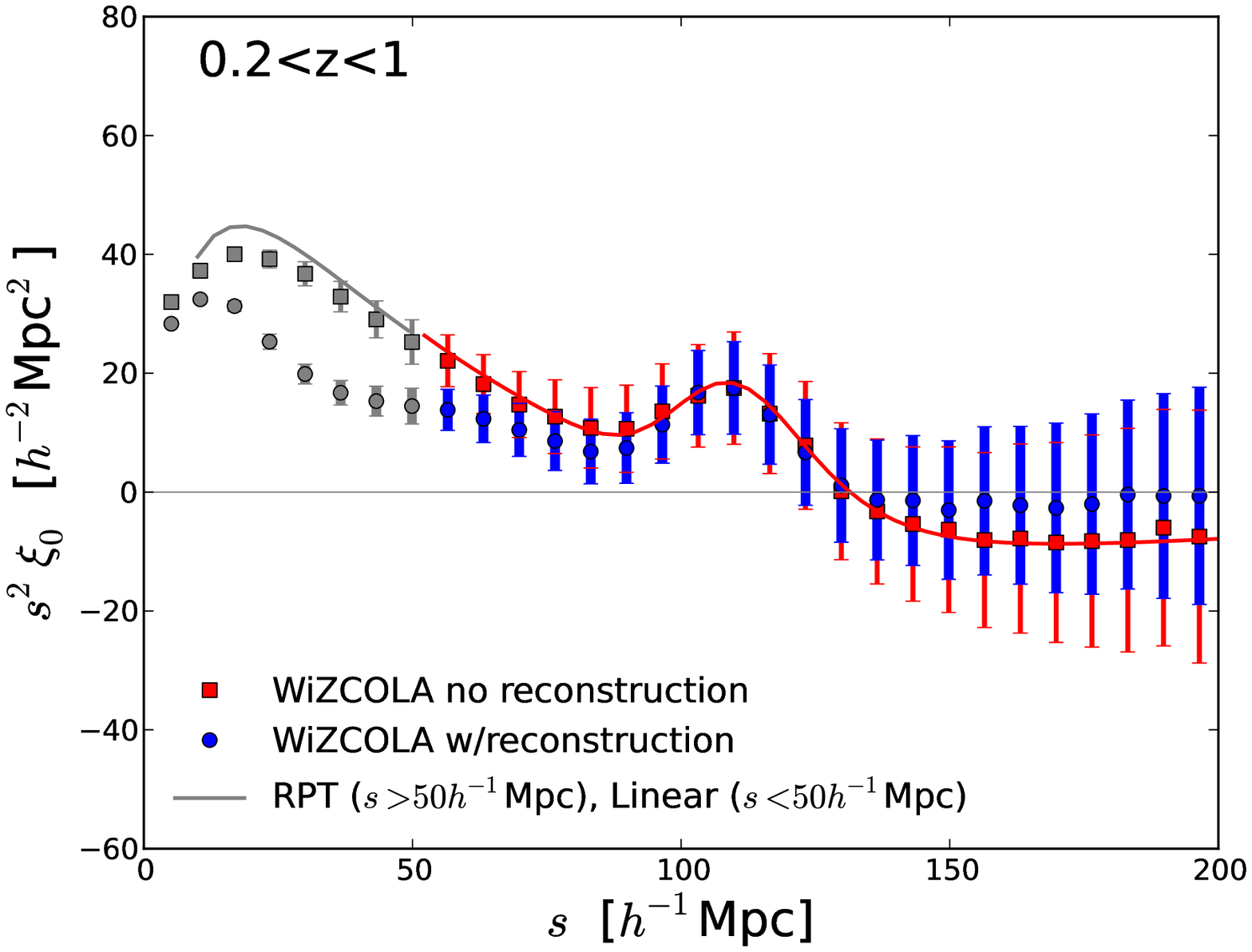}

\caption{
The mean of the simulated two-point correlation functions shown before (red squares) 
and after applying reconstruction (blue circles) for three redshifts bins 
and the full $z$ range, 
as indicated. These are plotted as $\xi s^2$ to emphasize the region of the \bafii. 
The uncertainty bars are the square root of the diagonal elements of the covariance matrix 
(for one \wizcola realization, not the mean). 
The solid lines are the templates $\xi_{\rm T}$ used in the analysis (not the best fit model), 
where we focus on the range of analysis $50<s<200$\hmpcii. 
For the $s<50$\hmpc region we plot a linear model. 
We see a clear sharpening of the \baf after reconstruction in all cases. 
}
\label{figure:xi0_pre_post_wizcola}  
\end{center}
\end{figure*}

In Figure \ref{figure:xi0_pre_post_wigglez} 
we display 
the resulting angle-averaged correlation functions $\xi$ 
from Equation \ref{equation:xi02_combined} for the data 
pre- (red squares) and post-reconstruction (blue circles).  
The corresponding mean signal of the mock simulations $\overline{\xi}$ 
are displayed in Figure \ref{figure:xi0_pre_post_wizcola}.

In each of the three $\Delta z$ bins we see 
a sharpening of the baryonic acoustic peak 
both in the data and in the simulations. 
In \S\ref{section:significance_detection} we quantify this sharpening, 
and in \S\ref{section:distance_constraints} we present the improved 
distance measurements and compare these with expectations 
according to the mocks. 

Comparing results pre- and post-reconstruction 
of the data and mocks, 
we also see a clear reduction post-reconstruction in the amplitude of $\xi$ at  
scales outside the acoustic ring, $s<100$\hmpc  
and $s>140$\hmpcii. This can be explained by the subtraction  
of the linear redshift distortions, when applying reconstruction. 

The negative measurements of 
$\xi$ at large scales for \Dznearii, 
and the positive measurements for 
\Dzfarii, are consistent with the expectations 
of sample variance. 
This is best understood realizing the fact that the 
data points are correlated. 

The various $\xi$ and their covariance matrices can be found 
on the World Wide Web.\footnote{\url{http://www.smp.uq.edu.au/wigglez-data/bao-random-catalogues}}

\section{Methodology}\label{section:method}
\subsection{Modeling $\xi$}\label{section:modeling}

In our previous analysis of this data in \cite{blake11c}, 
we 
treated 
the full shape of $\xi$ as a standard ruler, 
and modelled the whole correlation function. 
In our current analysis
we focus solely 
on the geometrical information contained in the 
\baf \dvrs (defined below) and marginalize over the information encoded in the full shape of $\xi$, 
e.g, $\Omega_{\rm m}h^2$ and the spectral index $n_{\rm s}$. 
This is because the reconstruction procedure  
as described in \S\ref{section:reconstruction}, 
while sharpening the baryonic peak and hence improving distance constraints, 
involves a smoothing process  
which affects the correlation function slope in a manner which is difficult 
to model. 


\scb{
To measure \dvrs 
for each $\Delta z$ bin we compare the data $\xi^{\rm \Delta z}(s_i)$ 
(described in \S\ref{section:xi}) 
to a model $\xi_{\rm m}(s_i)$ defined as: 
}

 \beq\label{equation:model_equation}
  \xi_{\rm m}(s_{\rm f}) =  a_{0}\cdot \xi_{\rm T}(s_{\rm f}/\alpha) + A(s_{\rm f}), 
 \eeq
where $\xi_{\rm T}$ is a template correlation function 
and $A(s)$ is a polynomial, both defined below, 
\scb{and $s_{\rm f}$ is the distance scale in the coordinate system 
of the fiducial cosmology.} 

As we are interested in the geometrical information 
encoded in the \baf position, 
not in the full shape of $\xi$, 
we follow the procedure outlined by \cite{xu12a} in which we 
marginalize over the amplitude and shape parameters $a_i$ ($i=0,1,2,3$)  
as defined by:  
\beq\label{equation:As_equation}
A(s) =a_{1} + \frac{a_2}{s} + \frac{a_3}{s^2}. 
\eeq
All effects on the amplitude, e.g $\sigma_8$, linear bias 
and linear redshift distortions, are contained in 
$a_0$ which we marginalize over.


The $\alpha$ parameter in Equation \ref{equation:model_equation} takes 
into account the 
distortion between distances measured in the fiducial cosmological 
model used to construct the $\xi$ measurement, 
and the trial cosmological model we are testing. 
When applied to the \bafii, \cite{eisenstein05b}  
argued that this distortion may be 
related to the cosmic distance scale as: 
\beq\label{equation:alpha_dvrs}
\alpha=\frac{\left(D_{\rm V}/r_s\right)}{\left(D_{\rm V}/r_s\right)_{\rm fid}},  
\eeq
where the volume-averaged-distance is defined as: 
\beq\label{equation:dv}
D_{\rm V}(z)=\left(\frac{cz(1+z)^2 D_{\rm A}^2}{H}\right)^{1/3},  
\eeq
where $D_{\rm A}(z)$ is the physical angular diameter distance, 
$H(z)$ is the expansion rate and $c$ is the speed of light 
(as defined in \citealt{hogg99cosm}).   
The calculation of the sound-horizon $r_{\rm s}$ is discussed in \S\ref{section:cosmo_implications}. 
Equation \ref{equation:alpha_dvrs} stems from the fact that $\alpha$ 
is the Jacobian of the volume element d$^3s$, 
when transforming between 
the true coordinate system to the fiducial one $s_{\rm f}$.
\cite{anderson13a} showed that this is a fairly good approximation, 
even when there is anisotropic warping.  

The template $\xi_{\rm T}$ we use is based on  
renormalized perturbation theory (RPT), 
as introduced by \cite{crocce08}: 

\beq\label{equation:xi_rpt}
\xi_{\rm T}(s) = \xi_{\rm L} \otimes  e^{-(k_*s)^2} + A_{\rm MC}\xi^{(1)}\frac{{\rm d}\xi_{\rm L}}{{\rm d}s}, 
\eeq
where the $\otimes$ term denotes convolution, 
L means linear, and: 
\beq\label{xi1_equation_fromPk}
\xi^{(1)}(s)= \hat{s}\cdot\nabla^{-1}\xi_{\rm L}=\int_{0}^{\infty} \frac{k}{ 2 \pi^2} P_{\rm L}(k) j_1(ks) {\rm d}k,  
\eeq
where $j_1(y)$ is the spherical Bessel function of order one. 

\scb{This model has been investigated and applied by \cite{sanchez08,sanchez09a,sanchez13a},
who show that it gives an unbiased measurement of $\alpha$, $D_{\rm A}$, $H$,  
and the equation of state of dark energy $w_{\rm DE}$.}

To calculate the linear $P_{\rm L}$ and $\xi_{\rm L}$ 
we use the \texttt{CAMB} package\footnote{\url{http://camb.info}} (\citealt{lewis99a})
using the fiducial cosmology mentioned in \S\ref{section:intro}. 
The input redshifts chosen for each redshift bin 
are the effective values given above.


The first term in Equation \ref{equation:xi_rpt} damps the \baf through the $k_*$ parameter. 
The second term takes into account $k-$mode coupling (MC) via  
the $A_{\rm MC}$ parameter.

In our analysis we fix $k_{*}$ and $A_{\rm MC}$ 
to values corresponding to the best fits to the 
signal of the mock-mean correlation function ($\overline{\xi}$ hereon). 
These fits are performed using the covariance matrix 
of the mock mean,  
and marginalizing over the amplitude.
The value of $A_{\rm MC}$ is set to 0.15,
and the $k_{*}$ values are summarized in Table \ref{table:k_star_values}.

\begin{table} 
\begin{minipage}{172mm}
\caption{$k_*$ values for the RPT $\xi$ templates}
\label{table:k_star_values}
\begin{tabular}{@{}ccc@{}}
 \hline
 Volume                 &     $k_*$ pre-recon    &    $k_*$ post-recon                          \\
 \hline
 \Dznearii: $0.2<z<0.6$ & 0.17 & 0.55 \\
 \Dzmidii: $0.4<z<0.8$  & 0.19 & 0.55 \\
 \Dzfarii: $0.6<z<1$    & 0.20 & 0.55  \\
 \hline
\end{tabular}

\medskip
 $k_*$ in units of $h$Mpc$^{-1}$.
\end{minipage}
\end{table}

In the pre-reconstruction case we notice that $k_*$ 
increases with redshift. 
This is expected because at higher redshift  
galaxies have less time to accumulate a displacement 
from their bulk flows 
and hence the damping scale is smaller. 

The post-reconstruction fits tend to prefer a much 
higher $k_*$ (0.55 $h$Mpc$^{-1}$) due to the sharpening of the peak. 
We test the data and the mock $\overline{\xi}$ and 
verify that the parameter of interest in the analysis, $\alpha$, 
is not correlated with $k_*$ or $A_{\rm MC}$. 
This verifies that our distance constraints do 
not depend on our choice of $k_*$ or $A_{\rm MC}$. 

The resulting templates $\xi_{\rm T}$ are displayed as the solid lines in 
Figure \ref{figure:xi0_pre_post_wizcola}, 
where the upper red is the pre-reconstruction template 
and the bottom blue is post-reconstruction. 
The corresponding data points are the mock $\overline{\xi}$.  
Although the focus of the analysis is the separation range 
$s=50-200$\hmpcii, 
we also extrapolate in gray to the region $s<50$\hmpcii, 
using a linear model $\xi_{\rm L}$ matched in amplitude at $50$\hmpc 
(where RPT is no longer valid; \citealt{sanchez08}).
In an analysis using a similar method \cite{kazin13a}   
demonstrated that the geometric information 
was insensitive to the fitting range as long as the 
lower bound is less 
than $65$\hmpc (see their Figure 13).

In Figure \ref{figure:xi0_pre_post_wizcola} 
the pre-reconstruction templates show excellent agreement   
with the respective $\overline{\xi}$. 
The post-reconstruction template contains 
a slight downward consistent shift in $\xi s^2$ 
compared to the $\overline{\xi}$, 
as the fit tends to be dominated 
by the accurate measurements at lower separations. 
This offset is easily accommodated by the 
$A(s)$ terms, and we verify below 
that any resulting bias in the best-fitting values of 
$\alpha$ is negligible.

\subsection{Statistical methods}\label{section:stat_methods}
Throughout this analysis we define the log-likelihood $\chi^2\equiv -2\log L$, 
calculated by:  
\beq
\chi^2(\Phi)= \sum_{i,j}^{N_{\rm bins}}\left(m_i\left(\Phi\right)-d_i\right)C_{ij}^{-1}(m_j(\Phi)-d_j), 
\eeq
\scb{
where $m$ and $d_i$  are vectors representing the models (Equation \ref{equation:model_equation}) and data 
$d_i=\xi^{\rm \Delta z}(s_i)$ (described in \S\ref{section:xi}), respectively, 
and $\Phi$ is the parameter set which is varied. 
}



The covariance matrix of each redshift bin used 
${\textbfss C^{\Delta z}}$ is the reduced matrix ``0" component   
of ${\textbfss C_{[0,2]}^{\Delta z}}$ 
given in  
Equation \ref{equation:Cij02_combined}.
To correct for 
the bias due to the finite number 
of realizations used to estimate the 
covariance matrix 
and avoid underestimation of 
the parameter confidence limits,  
after inverting the matrix to ${\textbfss C^{-1}_{\rm original}}$  
we multiply 
it by the correction factors (\citealt{hartlap07a, anderson13a}):
\beq
{\textbfss C^{-1}}={\textbfss C^{-1}_{\rm original}}\cdot\frac{(N_{\rm mocks}-N_{\rm bins}-2)}{(N_{\rm mocks}-1)}. 
\eeq
In our analysis we compare separation binning of 
$\Delta s=3.3$\hmpc and $\Delta s=6.7$\hmpcii. 
Using $N_{\rm mocks}=600$ 
and $N_{\rm bins}=23$ and $45$, respectively, between [50,200]\hmpcii, 
we obtain correction factors of 0.96 and 0.92.

\subsection{Parameter space of fitting $\xi$}\label{section:parameter space}
As indicated in Equation \ref{equation:model_equation}, 
the parameter space 
contains five parameters:
\begin{dmath}\label{equation:param_space_xi}
\Phi_{\alpha,a_i}=[\alpha,a_0, a_1, a_2, a_3]. 
\end{dmath}

To sample the probability distributions of the parameter space,  
we use a Markov chain Monte Carlo (MCMC) based on a Metropolis$-$Hastings algorithm. 
We run the MCMC using broad priors in all of these parameters. 
We 
verify that for both the data and mocks that $\alpha$ 
is not correlated with the $a_i$, 
i.e, 
our distance measurements are not affected 
by marginalization of the shape information.

In the analysis of the chains, 
we report results with a prior of $|1-\alpha|\leq 0.2$. 
As shown in \S\ref{section:distance_constraints}, 
this does not have an effect on the posterior of \dvrs  
for well-behaved realizations, i.e, 
realizations with well-defined \baf signatures. 
For lower S/N realizations, 
i.e, for cases of a poor \baf detection, 
this prior helps 
prevent the distance fits from 
wandering to values highly inconsistent 
with other measurements. 
Our choice of $20\%$ is well wider than the 
\cite{planck13xvi} predictions of \dvrs 
at a precision of  $1.1\%-1.5\%$ in our redshift range of interest 
(this is displayed as the yellow band in Figure \ref{plot:dv_comparison},  
which is explained below). 
\section{Results}\label{section:results}



Here we describe results obtained in the 
analysis of $\xi$ for the three redshift bins 
\Dznear ($0.2<z<0.6$), \Dzmid ($0.4<z<0.8$) 
and \Dzfar ($0.6<z<1$). 
All results are compared to those obtained 
when analyzing the $600$ \wizcola mocks. 
Unless otherwise specified, all results described here 
follow the methodology described in \S\ref{section:method}. 

\subsection{Significance of detection of the \baf}\label{section:significance_detection}
To quantify the sharpening of the \baf in the 
data and mock realizations after reconstruction, we analyze 
the significance of its  detection,  
as described below. Although 
we do not use these results for constraining cosmology, 
this analysis yields a first approach to understanding the 
potential improvement due to the reconstruction procedure. 

To quantify the significance of the detection 
of the \baf  
we compare the minimum $\chi^2$ obtained when 
using a physically motivated $\xi$ 
template to that obtained 
when using a featureless template 
not containing baryon acoustic oscillations. 
For the former we use the RPT template described 
in Equation \ref{equation:xi_rpt} and for the latter the 
``no-wiggle" model $\xi_{\rm nw}$ presented in \S 4.2 of \cite{eisenstein98}, 
which captures the broad-band shape information, excluding a \bafii. 

The significance of the detection of the \baf 
is determined by the square root of the 
difference between the minimum $\chi^2$ 
obtained using each template, $\Delta \chi^2$.  
For both calculations we apply the same method, 
i.e, modeling (Equation \ref{equation:model_equation}) 
and parameter space $\Phi_{\alpha,a_i}$ (Equation \ref{equation:param_space_xi}).

\begin{figure*}
\begin{center}
\includegraphics[width=0.32\textwidth]{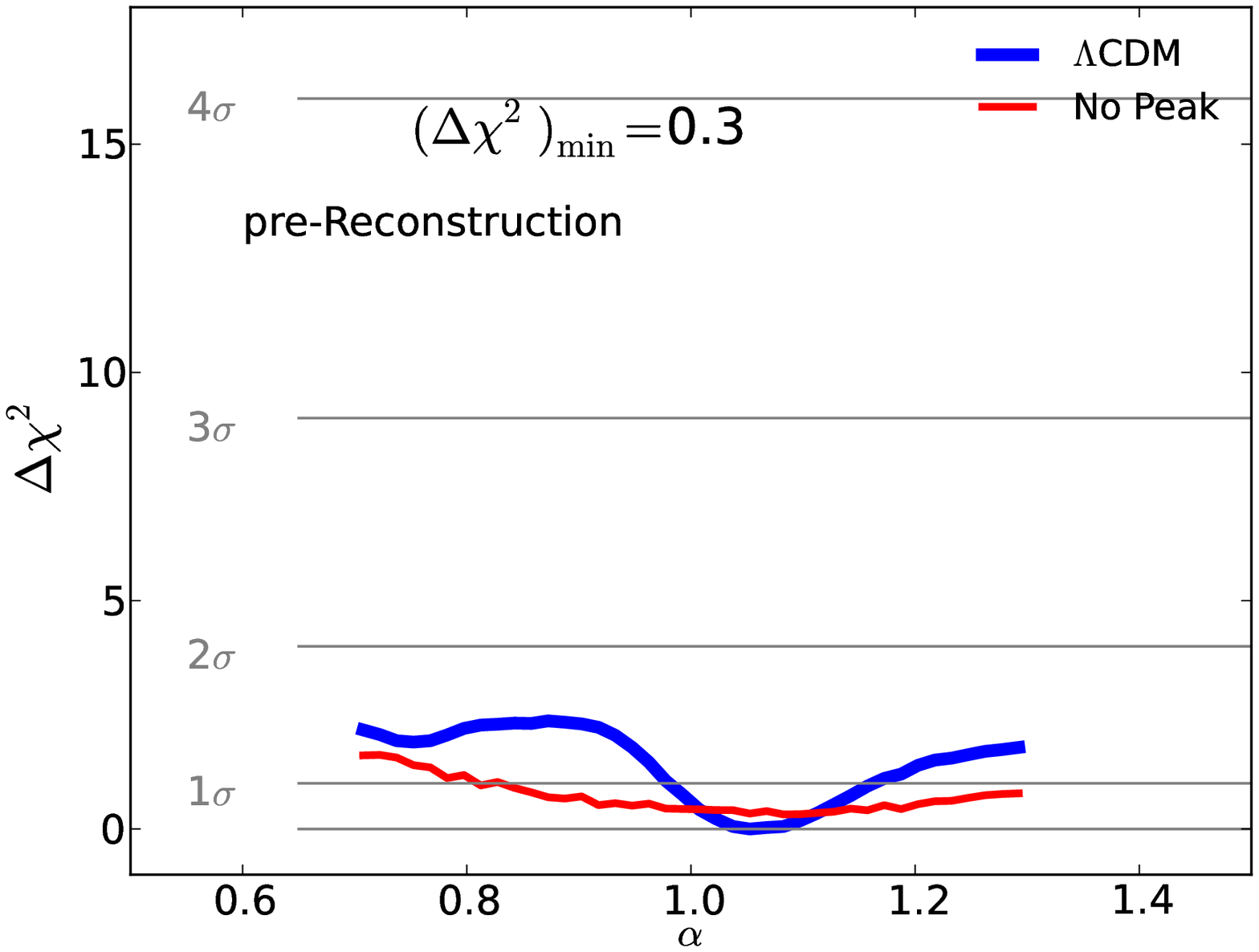}
\includegraphics[width=0.32\textwidth]{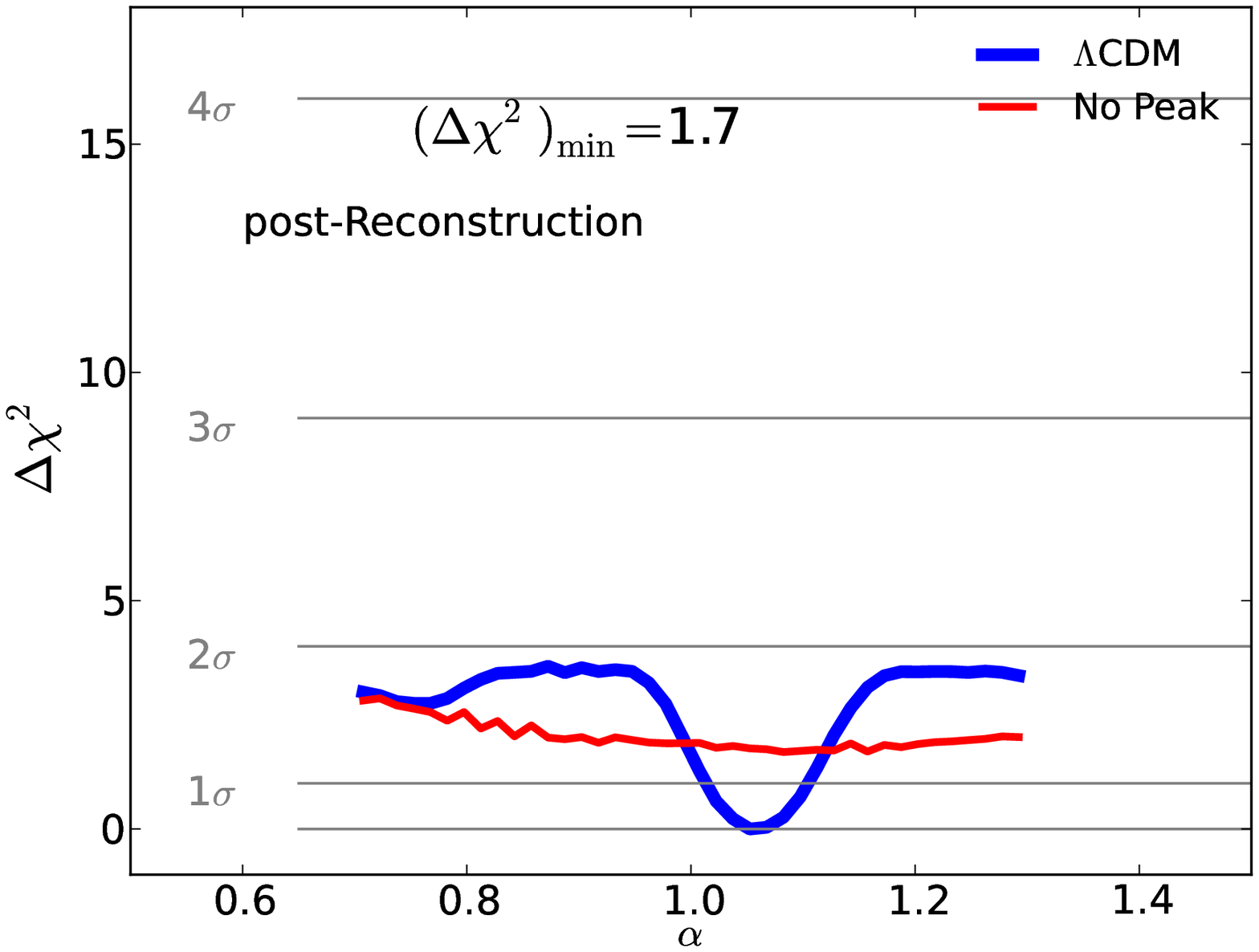}
\includegraphics[width=0.32\textwidth]{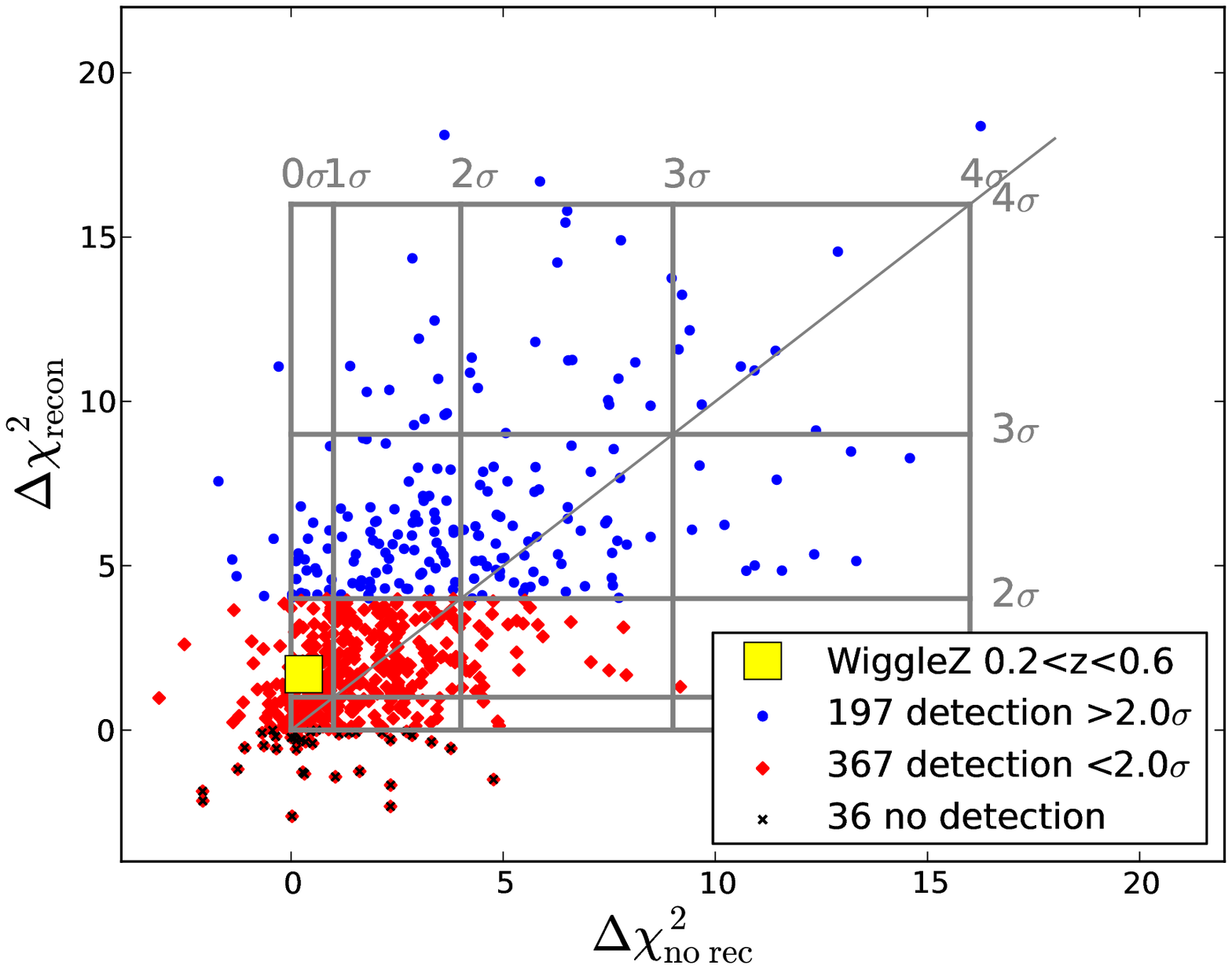}
\includegraphics[width=0.32\textwidth]{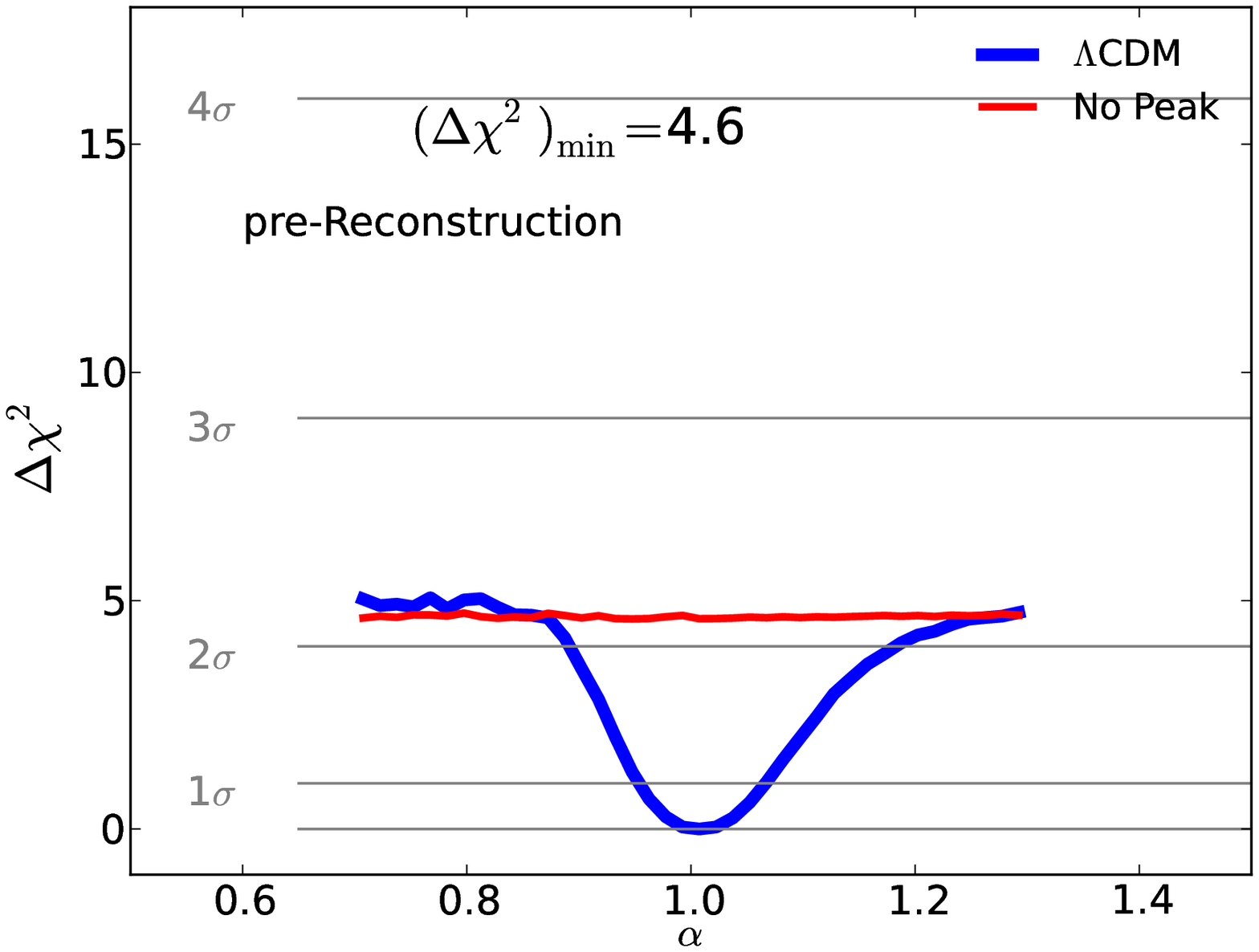} 
\includegraphics[width=0.32\textwidth]{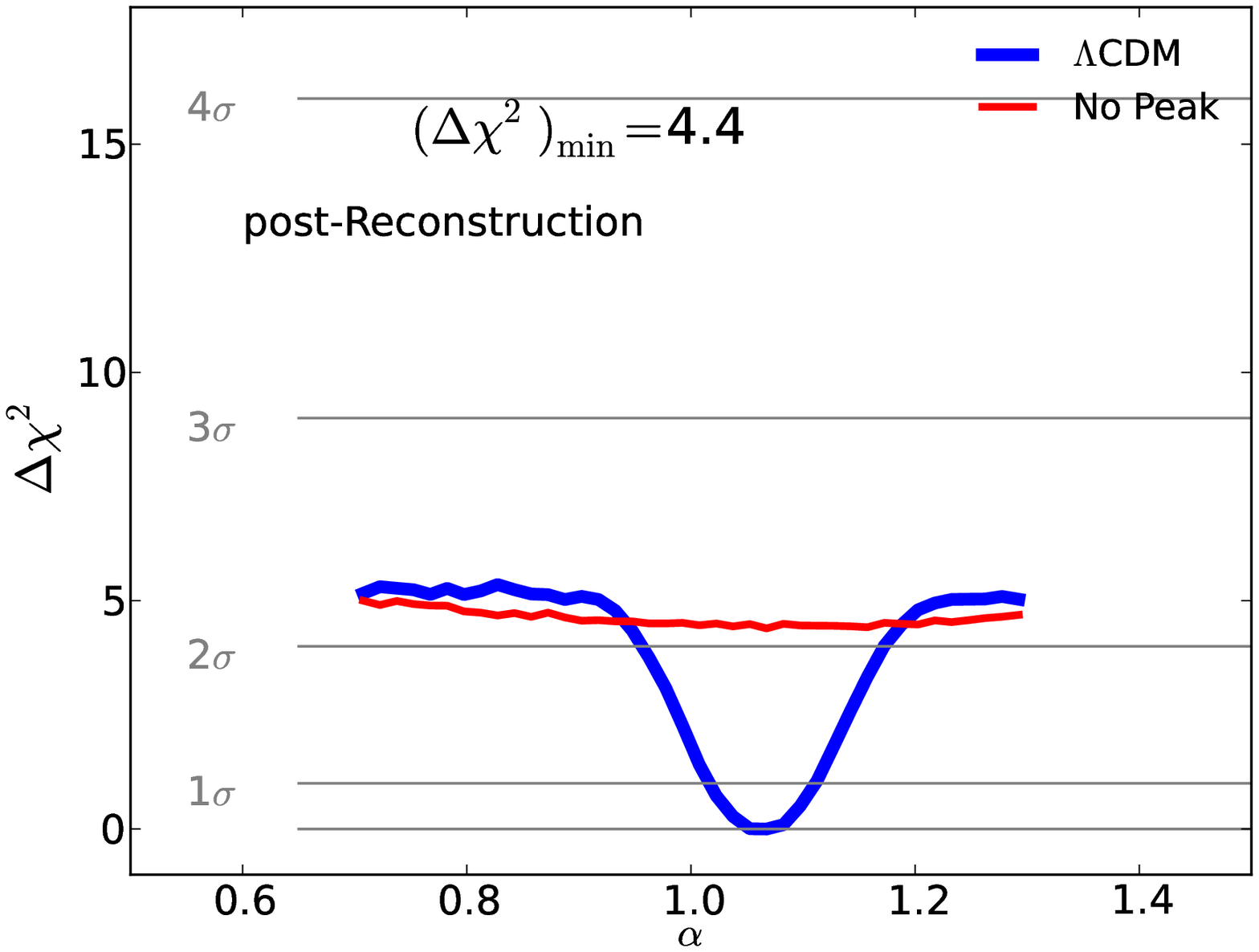}
\includegraphics[width=0.32\textwidth]{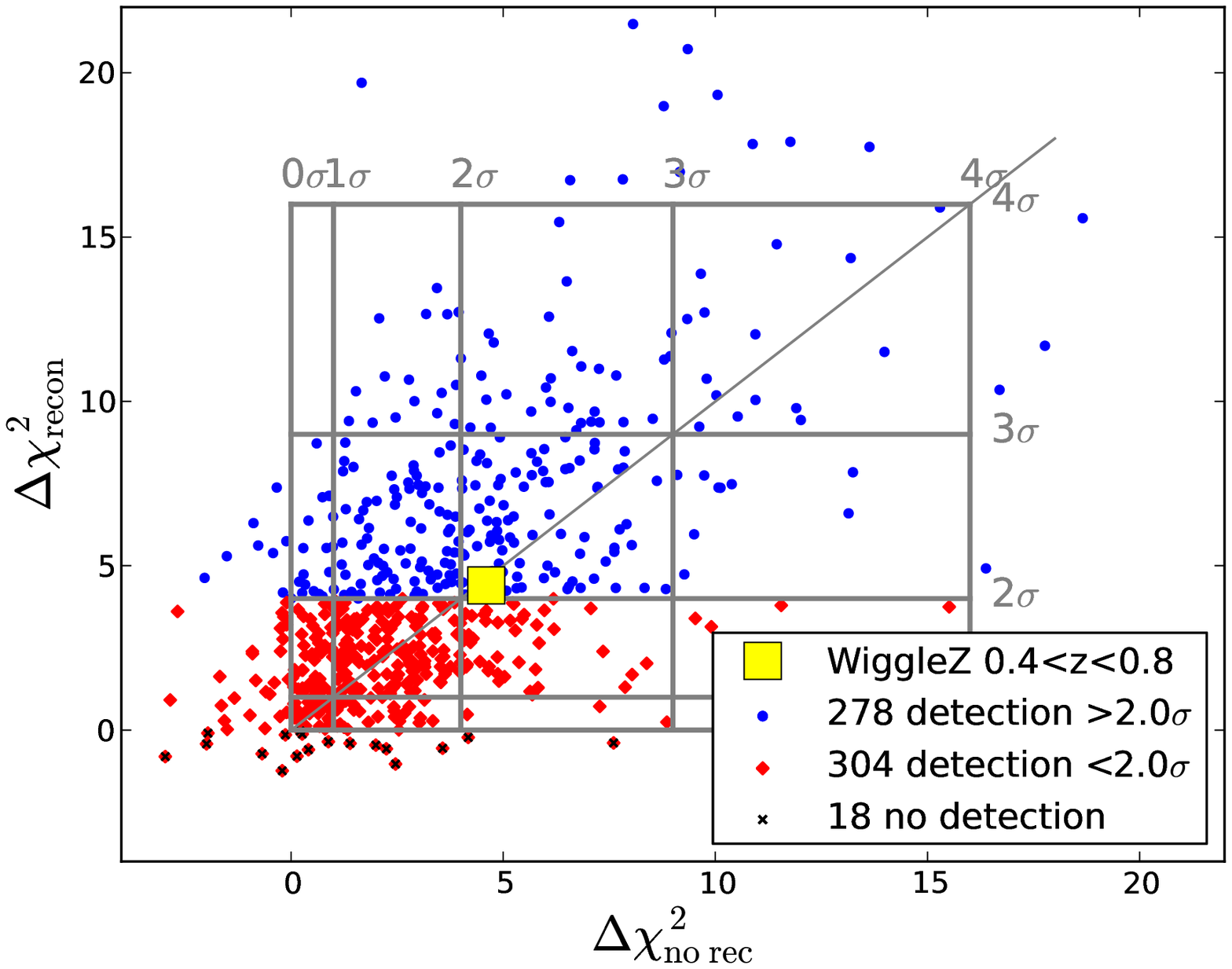} 
\includegraphics[width=0.32\textwidth]{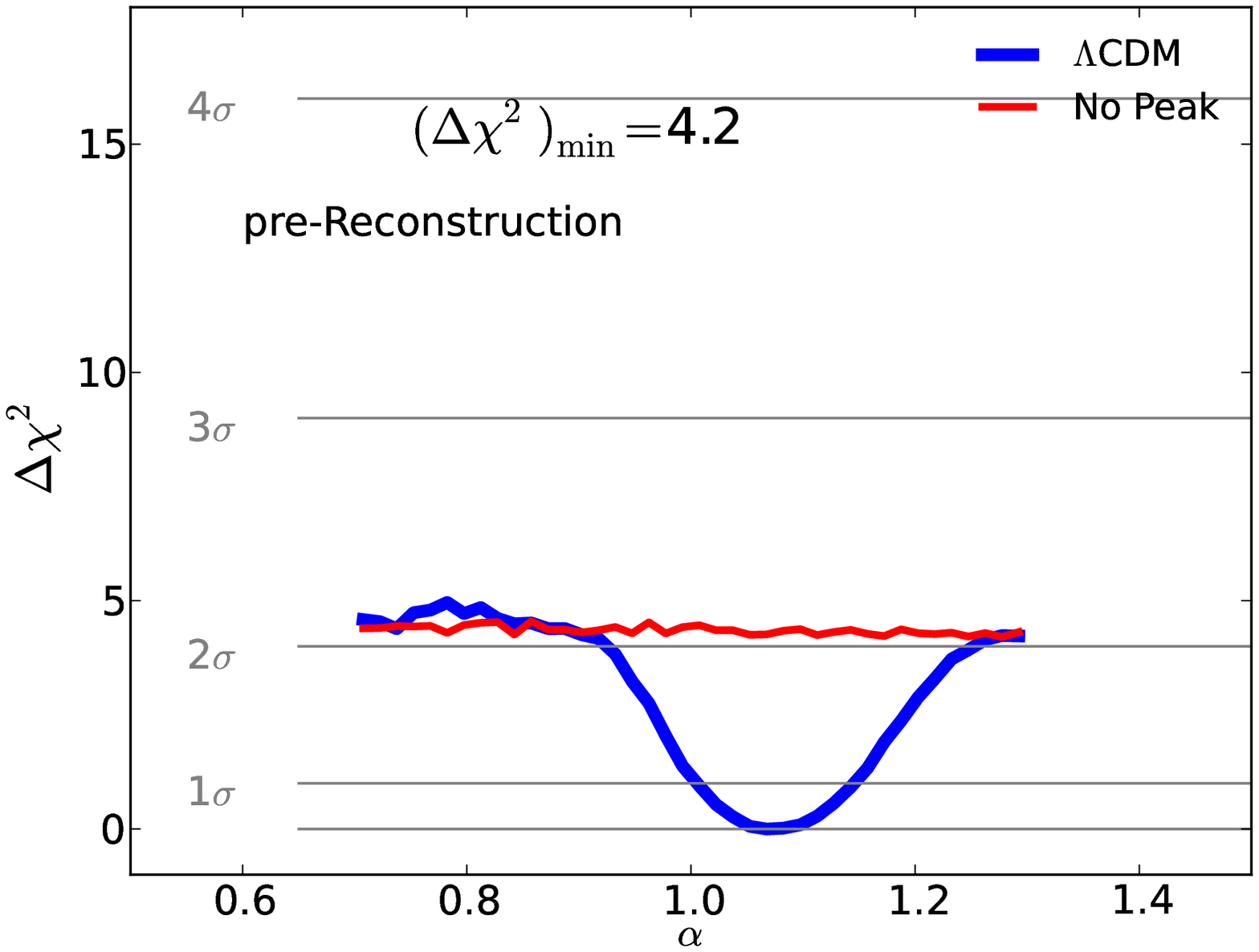}
\includegraphics[width=0.32\textwidth]{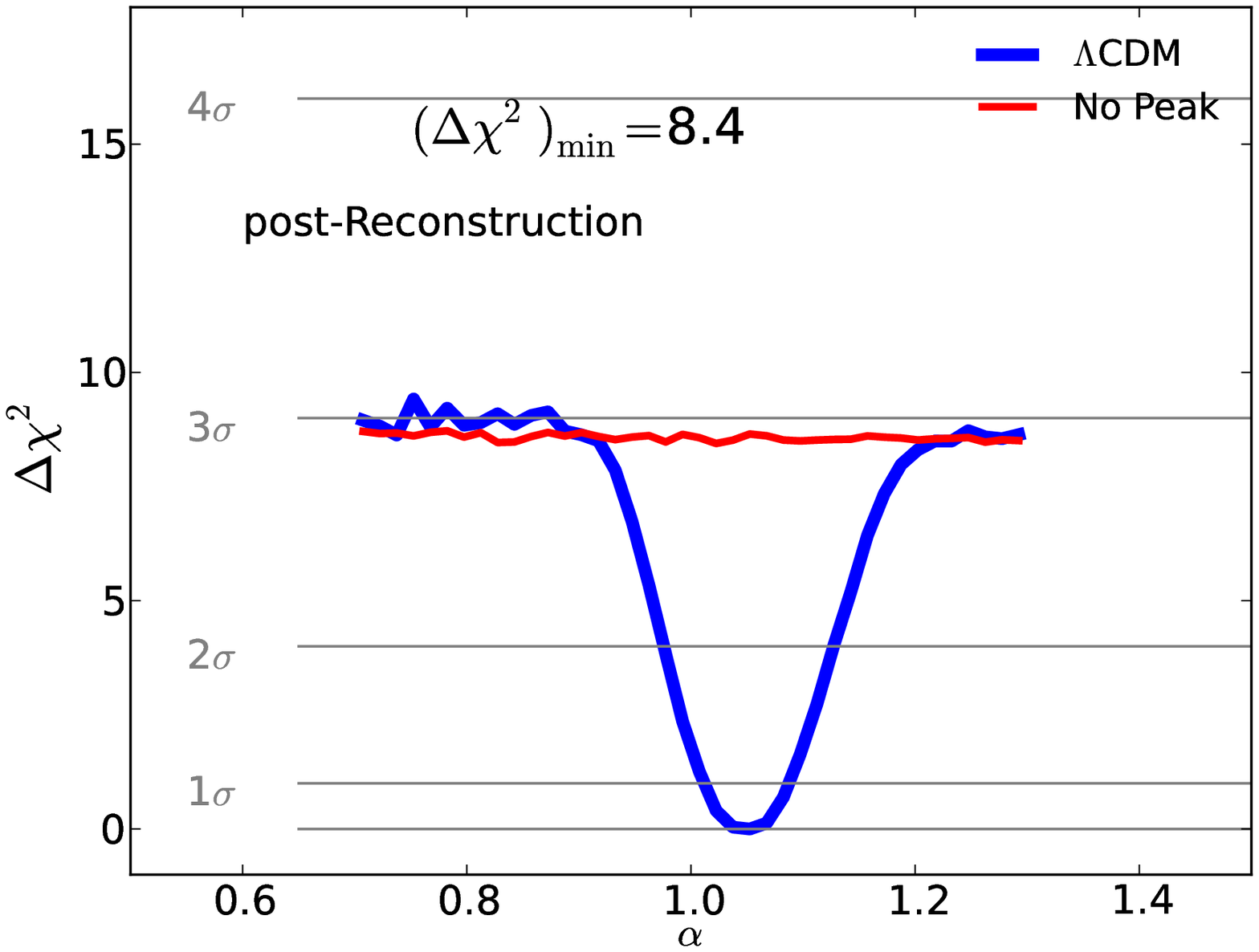}
\includegraphics[width=0.32\textwidth]{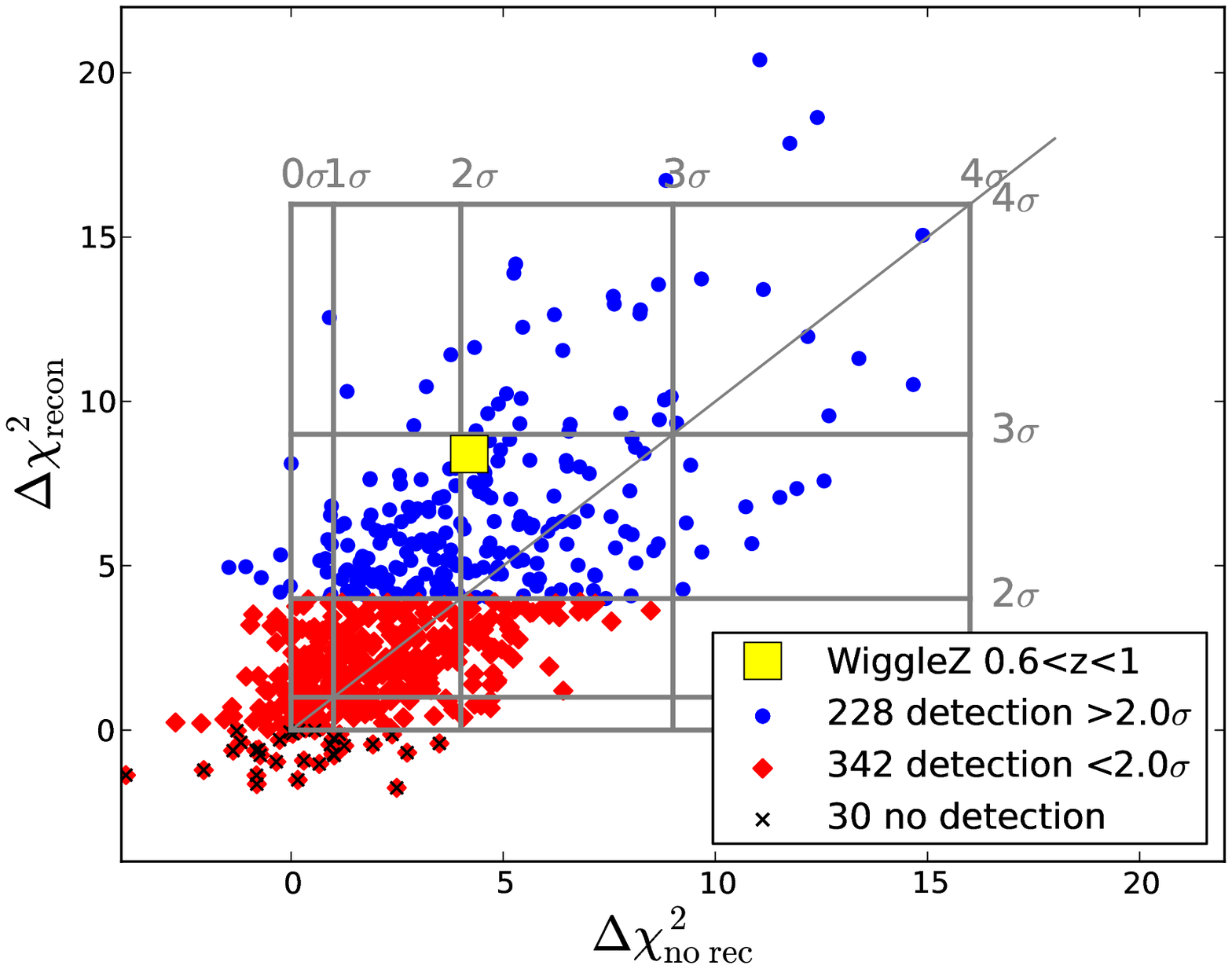}
\caption{
The minimum $\chi^2$ as a function of $\alpha$ 
before (left) and after reconstruction (center) 
for the \Dznear (top), \Dzmid (center), \Dzfar (lower) volumes. 
The thick blue lines are the results when using a physical template, 
and the thin red line when using a no-wiggle template. 
The significance of detection of the \baf is quantified as 
the square root of the 
difference between the minimum values of $\chi^2$ for each template. 
The boundaries are the $|1-\alpha|=0.3$ prior. 
In all cases there is an improvement 
in detection, 
where the most dramatic is in \Dzfar from $2.0\sigma$ to $2.9\sigma$. 
The right panels compare these data results (yellow squares) with 
600 mock $\Delta\chi^2$ results pre- ($x-$axis) and post- ($y-$axis) reconstruction. 
The classification of detection of the significance of the \baf 
is color coded as indicated in the legend and explained in \S\ref{section:significance_detection}.  
A summary of significance of detection values for 
the data and mocks in all redshift bins 
is given in Table \ref{table:significance_detection}.}
\label{plot:significance_of_detection}  
\end{center}
\end{figure*}

Figure \ref{plot:significance_of_detection} 
displays the 
$\Delta \chi^2$ as a function of $\alpha$ 
for the WiggleZ volumes  
before (left panels) and after reconstruction (center panels).  
 
\begin{table*} 
\begin{minipage}{172mm}
\caption{Significance of detection of the \bafii}
\label{table:significance_detection}
\begin{tabular}{@{}ccccc@{}}
 \hline
 Volume                 &     $\sqrt{\Delta\chi^2}$ &  $\chi^2_{\rm phys}$, $\chi^2_{\rm nw}$ &  Expected (All mocks) &  Expected ($>2\sigma$ subsample) \\
 \hline
 \Dznear no recon & 0.5 & 18.0, 18.3 & 1.4$\pm$0.8  (600) &   2.0$\pm$0.8  (197)  \\ 
 \Dznear w/ recon & 1.3 & 24.3, 26.0 & 1.6$\pm$0.9  (600) &   2.4$\pm$0.5  (197)  \\ \\
 \Dzmid no recon  & 2.1 & 20.5, 25.1 & 1.7$\pm$0.9  (600) &   2.1$\pm$0.8  (278)  \\ 
 \Dzmid w/ recon  & 2.1 &  9.1, 13.5 & 1.9$\pm$0.9  (600) &   2.6$\pm$0.6  (278)  \\ \\
 \Dzfar  no recon & 2.0 & 24.3, 28.5 & 1.5$\pm$0.8  (600) &   2.0$\pm$0.7  (228)  \\
 \Dzfar  w/ recon & 2.9 & 24.0, 32.4 & 1.7$\pm$0.8  (600) &   2.5$\pm$0.5  (228)  \\
 \hline
\end{tabular}

\medskip
 All columns, except the second to the left ($\chi^2_{\rm phys},\chi^2_{\rm nw}$), are in terms of $\sigma$ detection.  \\
 The significance of detection in each volume is determined by $\sqrt{\Delta\chi^2}$, where $\Delta\chi^2\equiv\chi^2_{\rm nw}-\chi^2_{\rm phys}$ and $dof=18$. \\
 \Dznearii: $0.2<z<0.6$, \Dzmidii: $0.4<z<0.8$, \Dzfarii: $0.6<z<1$ \\
 The $>2\sigma$ subsample is based on results of the post-reconstruction case. 
\end{minipage}
\end{table*}

Focussing first on the \Dzfar volume, we see 
a significant improvement in 
the detectability of the \baf  
after applying reconstruction.  
The result 
obtained before reconstruction 
shows a low significance of detection 
of $\sqrt{4.2}=2\sigma$ compared 
to that obtained after reconstruction 
$\sqrt{8.4}=2.9\sigma$. 

These results are for a binning of 
$\Delta s=3.3$\hmpcii. 
When using $\Delta s=6.7$\hmpcii, 
both $\Delta\chi^2$ are lower 
(2.3 and 7.2, respectively), 
but the difference between the pre- and post-reconstruction 
values remains similar
$\Delta\left(\Delta\chi^2\right) \sim 4.5$. 

The right panels of 
Figure \ref{plot:significance_of_detection} 
show a comparison of these WiggleZ results 
(yellow square) to that expected from an array 
of 600 \wizcola mocks. 
To facilitate interpretation of 
the results, 
we indicate realizations which contain at least a 
$2\sigma$ detection in the 
post-reconstruction case, 
which is characteristic 
of the data. 
The realizations in which we   
detect a feature better than this threshold are displayed in blue circles 
(\Dznear: 197/600 mocks, \Dzmid: 278/600,  \Dzfar: 228/600)  
compared to those in which we do not in red diamonds (\Dznear: 367/600 mocks, \Dzmid: 304/600,  \Dzfar: 342/600).
The crosses are a subset  
for which the $\Delta\chi^2$ is negative, 
meaning the no-wiggle template fit is better 
than that of the physical template (\Dznear: 36/600 mocks, \Dzmid: 18/304,  \Dzfar: 30/600). 

From these mock results we learn about a few aspects of 
the results in 
the  
\Dzfar volume. 
First, the average mock realization 
yields a fairly low significance of detection, 
where both pre- and post-reconstruction are between $1-2\sigma$, 
and in $5$ per-cent of the mocks the physical $\xi_{\rm T}$ completely 
fails to outperform $\xi_{\rm nw}$. 

Second, we see that after applying reconstruction,  
there is a 
moderate improvement 
in the detectability of the \bafii. 
This can be quantified by a change in the median 
detectability of $1.4\sigma$  to $1.7\sigma$, both with an r.m.s of $0.8\sigma$ 
(the negative $\Delta\chi^2$ values are set to zero in this calculation). 
When focusing on the $>2\sigma$ detection subsample 
(where the threshold is applied to the post-reconstruction results), 
the improvement is slightly better, from a median of 
$2.0\sigma$ (r.m.s of $0.7\sigma$) to a median $2.6\sigma$ with an r.m.s of $0.5\sigma$. 
In \S\ref{section:distance_constraints} we find,  
that, on average, 
this translates into 
an improvement in accuracy of the \dvrs measurement. 

Third, whereas the pre-reconstruction  
detection significance in the data appears similar to an average realization, 
the post-reconstruction detection is on the   
fortunate side (top 8 percentile of all 600 mocks). 
We show the corresponding improvement in the measurement of \dvrs 
in \S\ref{section:distance_constraints}. 
These data and mock results, as well as those for \Dznear and \Dzmidii,  
are summarized in Table \ref{table:significance_detection}. 

We turn now to examine the other two redshift bins. 
In the top panels of Figure \ref{plot:significance_of_detection}  
and in Table \ref{table:significance_detection} 
we see that the detection in the WiggleZ  
$\Delta z^{\rm Near}$ volume 
improves from no clear preference of $\xi_{\rm T}$ over $\xi_{\rm nw}$  
before reconstruction, to a weak detection of $1.3\sigma$ 
after. 
In the pre-reconstruction case  
this volume appears to be under-performing compared to 
the mock results.  
In the post-reconstruction case 
its performance appears to be within 
expectations of the mocks. 

According to the mock catalogues, 
the performance of the \Dzmid volume 
should be the best 
amongst the three $\Delta z$ volumes. 
This is evidenced by the fact that 
the $>2\sigma$ subset is larger (278/600) than the others 
(197 and 224). 
This reflects the fact 
that this redshift range contains the highest effective volume, 
i.e, the 
best combination of shot-noise and sample variance 
of the three. The effective volume numbers are evaluated at $k=0.1 \ h/$Mpc in 
units of \hgpcii: 
0.096  (\Dznearii), 0.130  (\Dzmidii), 0.089 (\Dzfarii). 
This does not, however, translate into notable improvements in the 
average significance of detection or constraints on \dvrs 
in the mocks or in the data. 
In the data, as we shall continue to see, the redshift bin that  
benefits the most from the reconstruction procedure is \Dzfarii. 
The mock results suggest that this is due to sample variance reasons.  

From Table \ref{table:significance_detection} we also learn 
that reconstruction improves the significance of 
detection 
of the \baf for the average mock 
by $\sim 0.2-0.3\sigma$, 
whereas the $>2\sigma$ subsample improves by $0.4-0.5\sigma$. 
We also note that the scatter of the significance of detection 
in the generic case does not vary, 
but in the  $>2\sigma$ subsample  improves from $0.7-0.8\sigma$ 
pre-reconstruction to $0.5-0.6\sigma$ post-reconstruction. 

\cite{blake11c} reported pre-reconstruction 
significance of detections 
1.9$\sigma$, 2.2$\sigma$ 2.4$\sigma$,  
which are slightly higher than those reported here. 
Their results are expected to yield 
a higher detection significance 
through using a fixed shape of $\xi$, 
whereas we vary the shape in the fit 
(as described in \S\ref{section:method}). 
\scb{
This could be understood, e.g, by the fact 
that the full shape of $\xi$ analysis assumes a cosmology, 
and hence explores a smaller parameter space, 
leading to a higher significance of detection. 
In our analysis we make no assumption of a prior cosmology, 
effectively marginalizing over a much larger parameter space, 
and hence we report a more model-independent significance of detection. 
}

To summarize, 
we find that reconstruction improves the detectability 
of the \baf in the majority of the \wizcola volumes. 
For the \Dznear volume we find improvement of detectability 
for 373/600 of the mocks, in \Dznear for 389/600 and in \Dzfar 
for 378/600. 
Hence we learn that there is 
a $\sim 65\%$ probability of improvement 
of detection 
of the \baf in WiggleZ volume due to reconstruction. 
In the case of the data, 
we find moderate improvement for \Dznearii, 
no improvement for \Dzmid and significant improvement 
for \Dzfarii. 

\subsection{Distance constraints}\label{section:distance_constraints}
\begin{figure*}
\begin{center}
\includegraphics[width=0.32\textwidth]{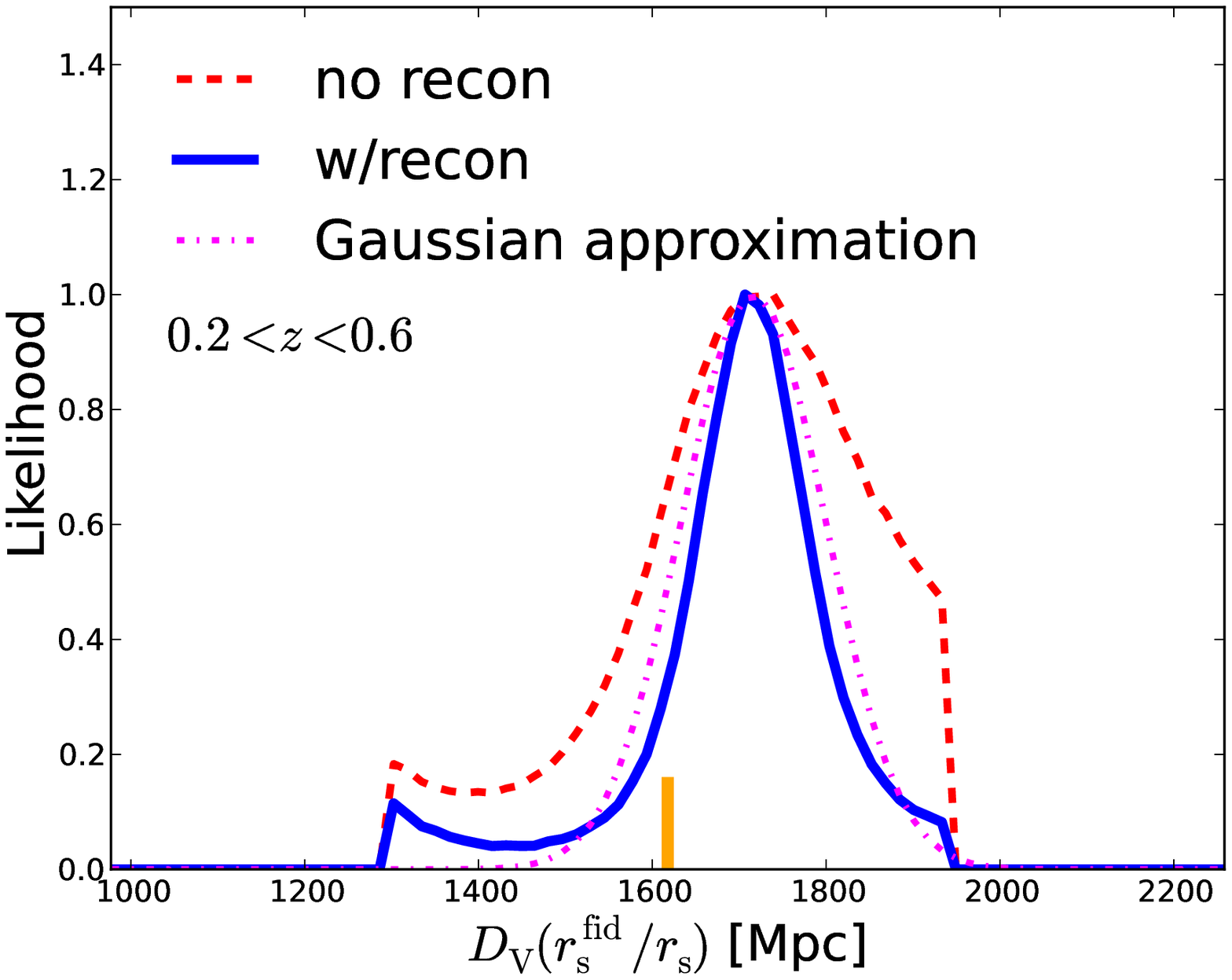}
\includegraphics[width=0.32\textwidth]{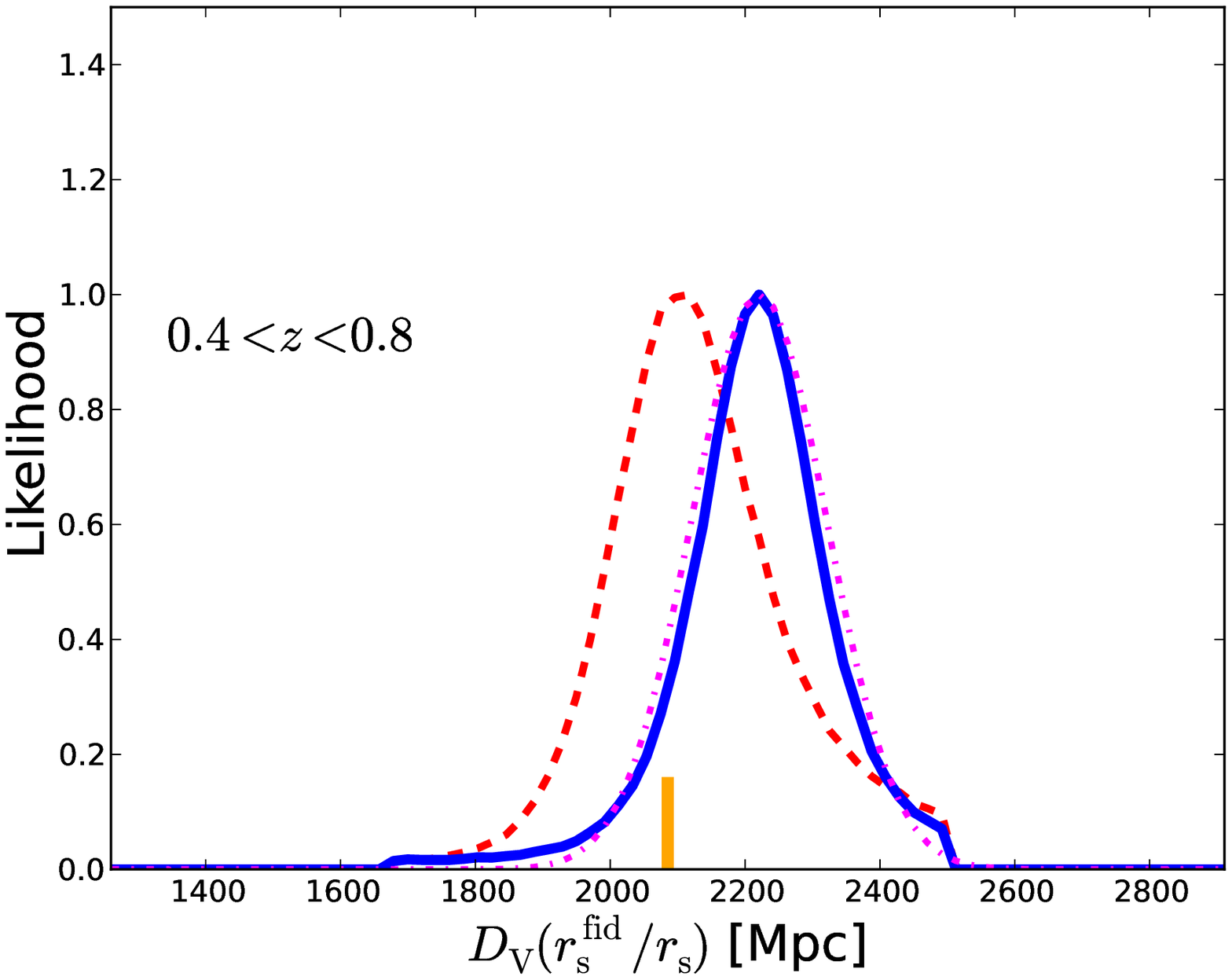}
\includegraphics[width=0.32\textwidth]{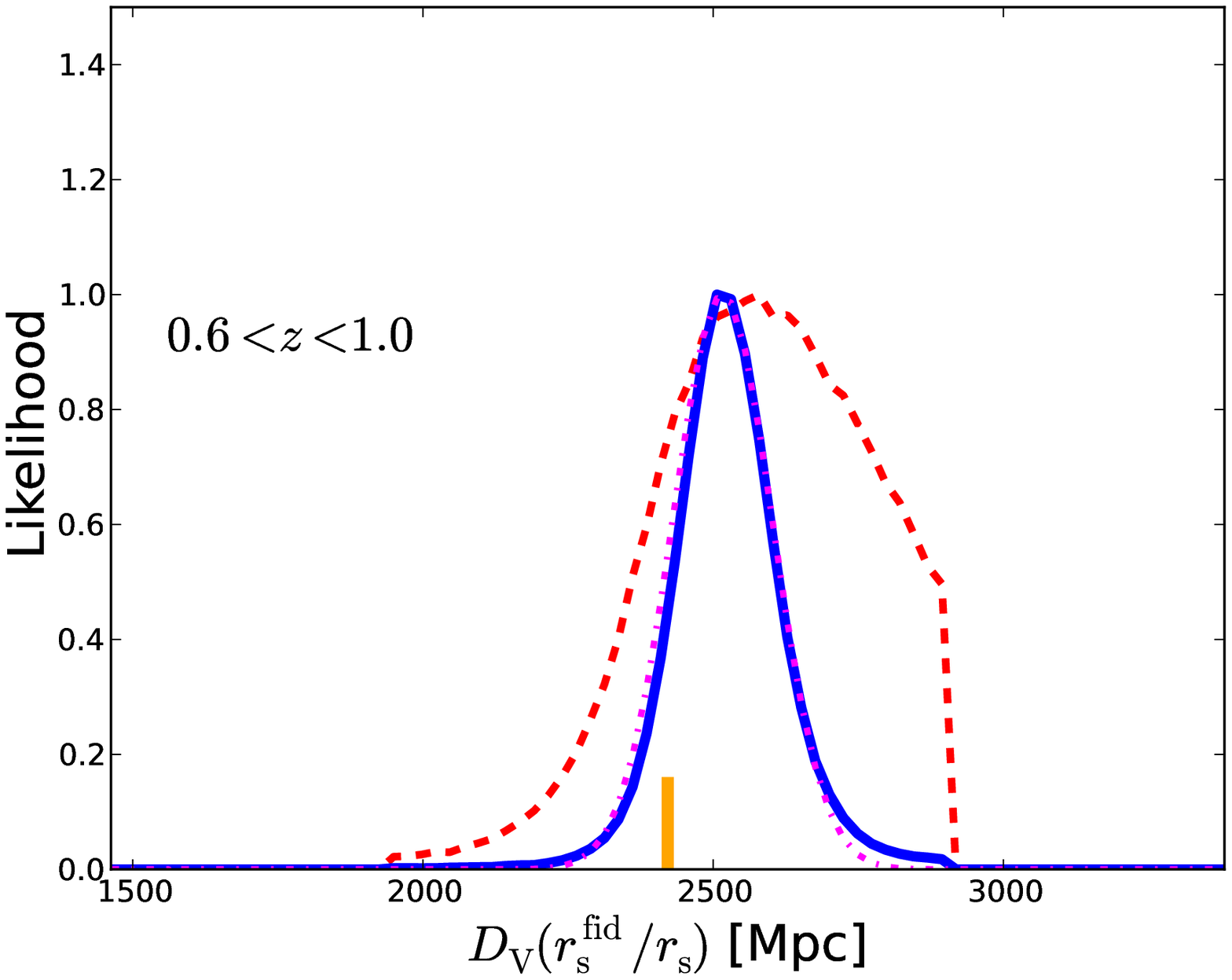} 
\caption{
The \dvrsrsfid posterior probability distributions of the 
three WiggleZ $\Delta z$ volumes (as indicated), for both pre- (dashed red) 
and post-reconstruction (solid blue). 
Gaussian approximations based on the mode and 
standard deviation values of the post-reconstruction cases  
are shown in dot-dashed magenta. In each panel we 
quote \dvrsrsfid and its $68\%$ confidence region, 
the $\alpha\equiv ($\iidvrsii$)/($\iidvrsii$)_{\rm fid}$ value, 
and plot the orange vertical line at 
the fiducial value $\alpha=1$ for comparison.  
The sharp cut-off in some of the results is  
due to the $|1-\alpha|<0.2$ prior. 
The improvement due to reconstruction is apparent in all $\Delta z$ bins. 
These results are summarized in Table \ref{table:dvrs_results}.
}
\label{plot:dvrs_wigglez}  
\end{center}
\end{figure*}

We now turn to using the \baf 
to constrain \dvrsii. 
We quote the final results in terms of 
\dvrsrsfid 
in order not to assume the 
sound horizon obtained with the fiducial cosmology $r_{\rm s}^{\rm fid}$. 
This is further discussed in \S\ref{section:dist z summary}. 
Figure \ref{plot:dvrs_wigglez} displays the 
posterior probability distributions of \dvrsrsfid 
for all three WiggleZ $\Delta z$ bins, 
both pre- (dashed red) and post-reconstruction (solid blue). 
The dotted magenta lines are Gaussian distributions based on 
the mode values and the half width of the $68\%$ confidence region 
of the post-reconstruction case (not the best fit Gaussian to the posterior). 
A summary of the statistics may be found in Table \ref{table:dvrs_results}, 
as well as in the panels of Figure \ref{plot:dvrs_wigglez}. 

\begin{table*} 
\begin{minipage}{172mm}
\caption{Distance measurment summary}
\label{table:dvrs_results}
\begin{tabular}{@{}ccccc@{}}
 \hline
 effective $z$ &  data $\alpha$  ($\%$) & data \dvrsrsfid [Mpc]              & mock $\alpha$ results  & mock $\sigma_\alpha$ results (\# mocks) \\
 \hline
 0.44 no recon & 1.065 (7.9\%)& 1723$^{+122}_{-151}$               & 1.005$\pm$0.067  &  0.051$\pm$0.027 (197) \\ 
 0.44 w/ recon & 1.061 (4.8\%)& {\bf 1716}$^{\bf +73}_{\bf -93}$   & 1.005$\pm$0.048  &  0.034$\pm$0.010 (197) \\
 0.60 no recon & 1.001 (6.0\%)& 2087$^{+156}_{-95}$                & 1.002$\pm$0.051  &  0.049$\pm$0.023 (278) \\ 
 0.60 w/ recon & 1.065 (4.5\%)& {\bf 2221}$^{\bf +97}_{\bf -104}$  & 1.003$\pm$0.037  &  0.032$\pm$0.010 (278) \\
 0.73 no recon & 1.057 (7.2\%)& 2560$^{+215}_{-157}$               & 1.0004$\pm$0.059 &  0.050$\pm$0.022 (228) \\ 
 0.73 w/ recon & 1.039 (3.4\%)& {\bf 2516}$^{\bf +94}_{\bf -78}$   & 1.003$\pm$0.050  &  0.037$\pm$0.013 (228) \\
 \hline
\end{tabular}

\medskip
 The columns marked by `data' are the WiggleZ results, and those by `mock' are simulated. \\
 The effective $z$ are for volumes \Dznearii: $0.2<z<0.6$, \Dzmidii: $0.4<z<0.8$, \Dzfarii: $0.6<z<1$  \\
 $\alpha\equiv (D_{\rm V}/r_{\rm s})/(D_{\rm V}/r_{\rm s})_{\rm fid}$ \\ 
 The figures in brackets in the `data $\alpha$' column is the half-width of the 68\% confidence region. \\
 \scb{To convert $\alpha$ to \dvrsrsfid  we use fiducial values of $D_{\rm V}^{\rm fid}$ for the three $\Delta z$ (in Mpc): 1617.7, 2085.2, 2421.7, respectively.}  \\
 The $^+_-$ values for the \dvrsrsfid column are the 68\% confidence region, as calculated from the edges inwards. \\
 The cross-correlation of the \dvrsrsfid results is indicated in Table \ref{table:dv_rs_iCij}. \\
 The mock median and std results for $\alpha$ and $\sigma_\alpha$ are from the $>2\sigma$ detection subsamples, as indicated. These are not Gaussian. 
\end{minipage}
\end{table*}

We find that in all three 
redshift bins, 
the \dvrsrsfid constraints 
improve with the application of reconstruction. 
As noted above, the most dramatic improvement is  
for \Dzfar ($0.6<z<1$) which is shown 
in the right panel of Figure \ref{plot:dvrs_wigglez} 
(as well as the left and center of the bottom panels of Figure \ref{plot:significance_of_detection}). 
As indicated in Table \ref{table:dvrs_results} 
the width of the $68\%$ confidence region improves from $7.2$ to $3.4$ per-cent accuracy. 
This improvement can be attributed to the clear sharpening of the \baf as seen in Figure \ref{figure:xi0_pre_post_wigglez}, 
which makes the peak-finding algorithm much more efficient. 
Here we fix the damping parameter $k_{*}$ and $A_{\rm MC}$. When relaxing this assumption 
we obtain similar results. 
Here we use a binning of $\Delta s=6.7$\hmpcii, 
but find consistent results for $\Delta s=3.3$\hmpcii. 


The clear cutoff that is seen in some of the posteriors 
(mostly the pre-reconstruction) is due to the $|1-\alpha|<0.2$ 
flat prior described in \S\ref{section:parameter space}.  
This prior does not appear to have an effect on the post-reconstructed posteriors. 
We attribute the elongated wings of the posteriors seen in some cases 
to the low significance of detection of the \baf 
in the pre-reconstruction cases for \Dznear and \Dzfarii. 

\begin{figure*}
\begin{center}
\includegraphics[width=0.32\textwidth]{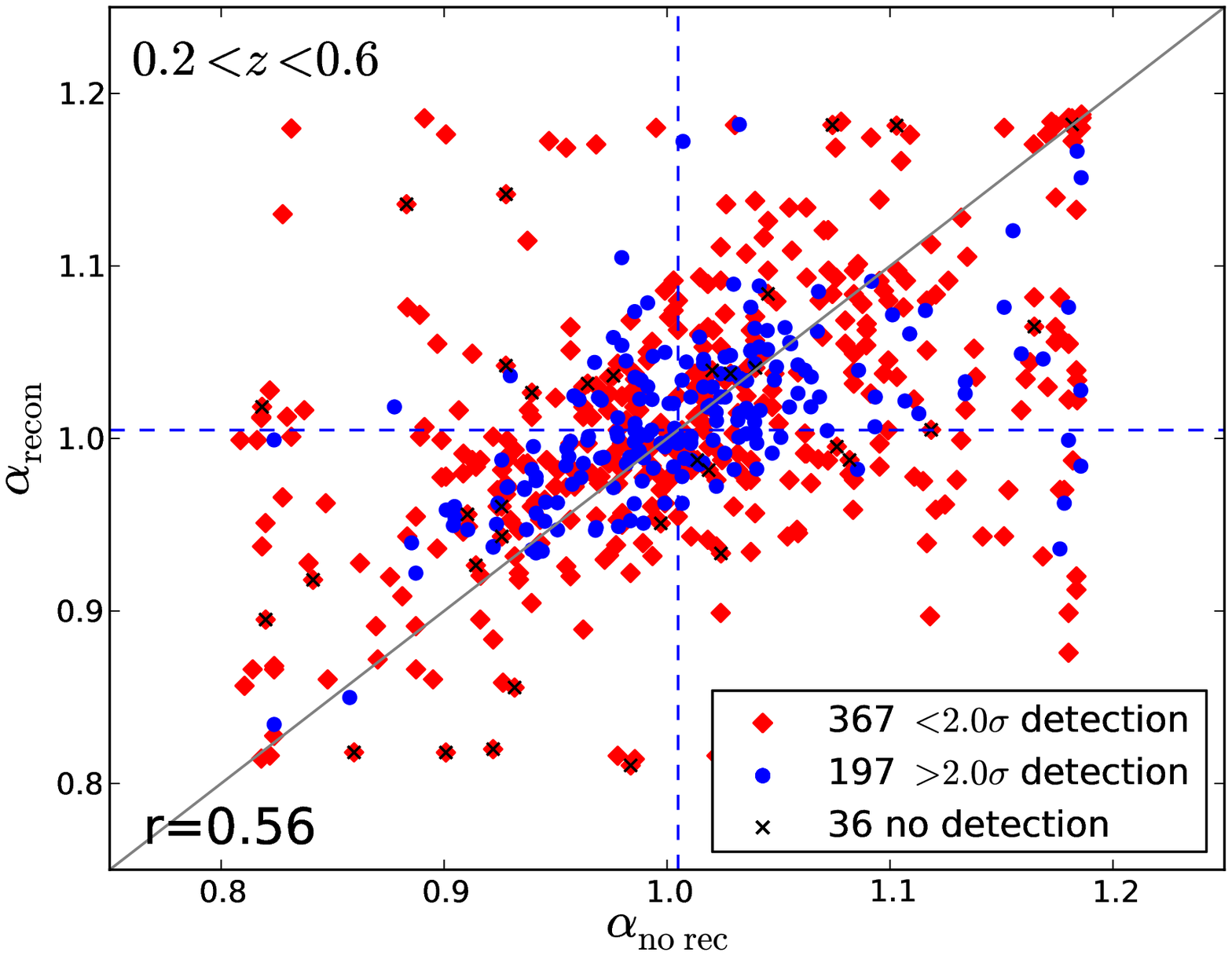}
\includegraphics[width=0.32\textwidth]{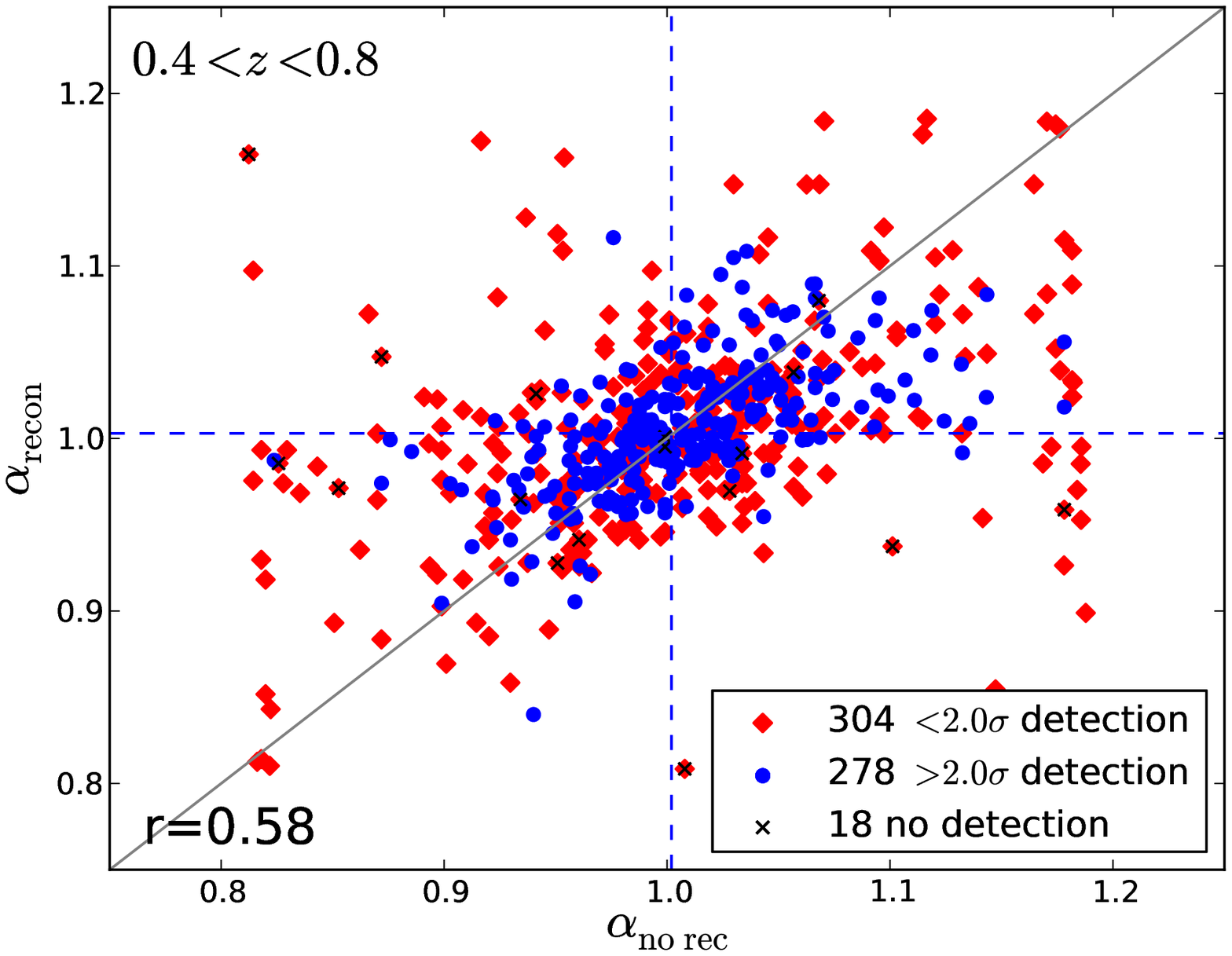}
\includegraphics[width=0.32\textwidth]{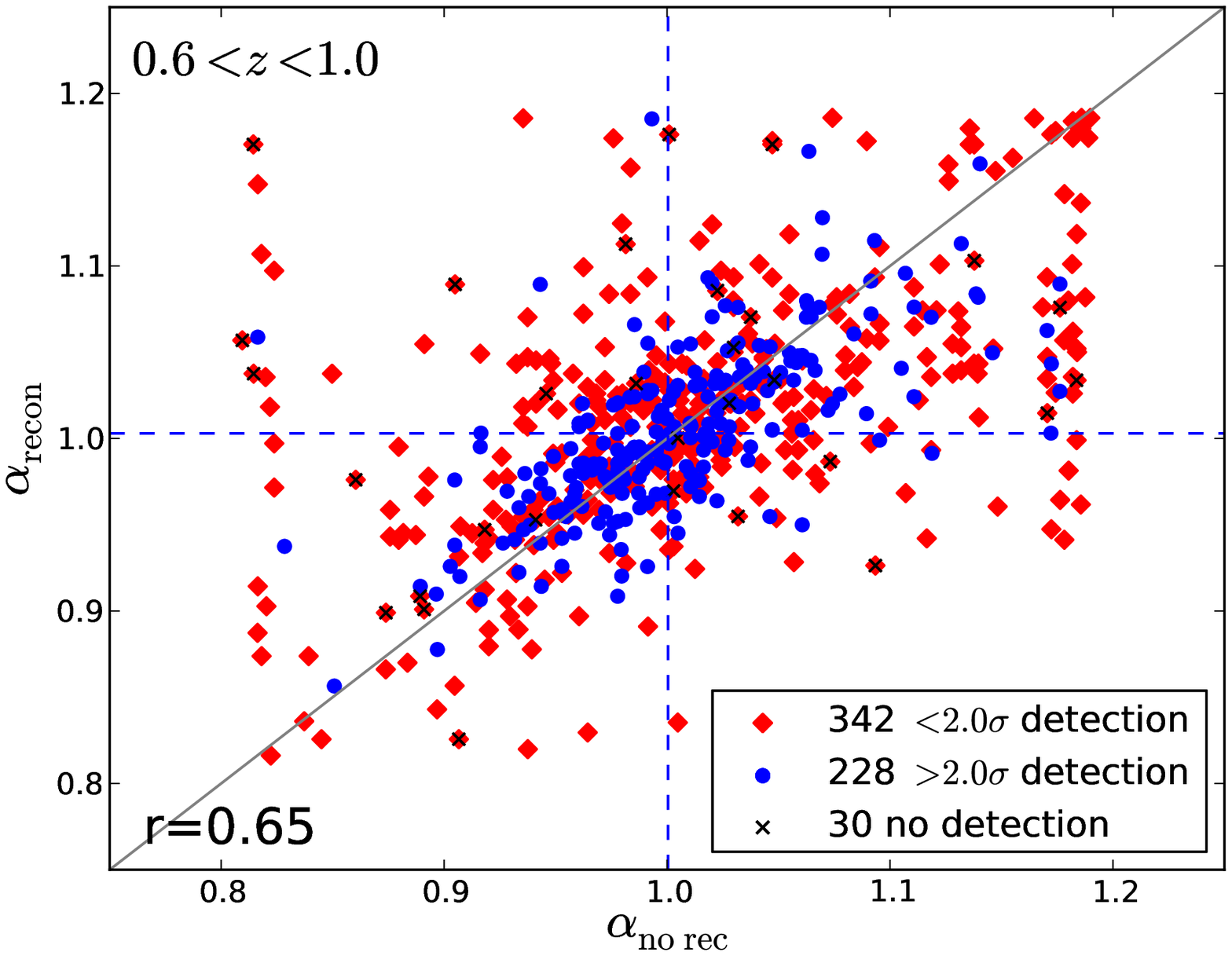} 
\includegraphics[width=0.32\textwidth]{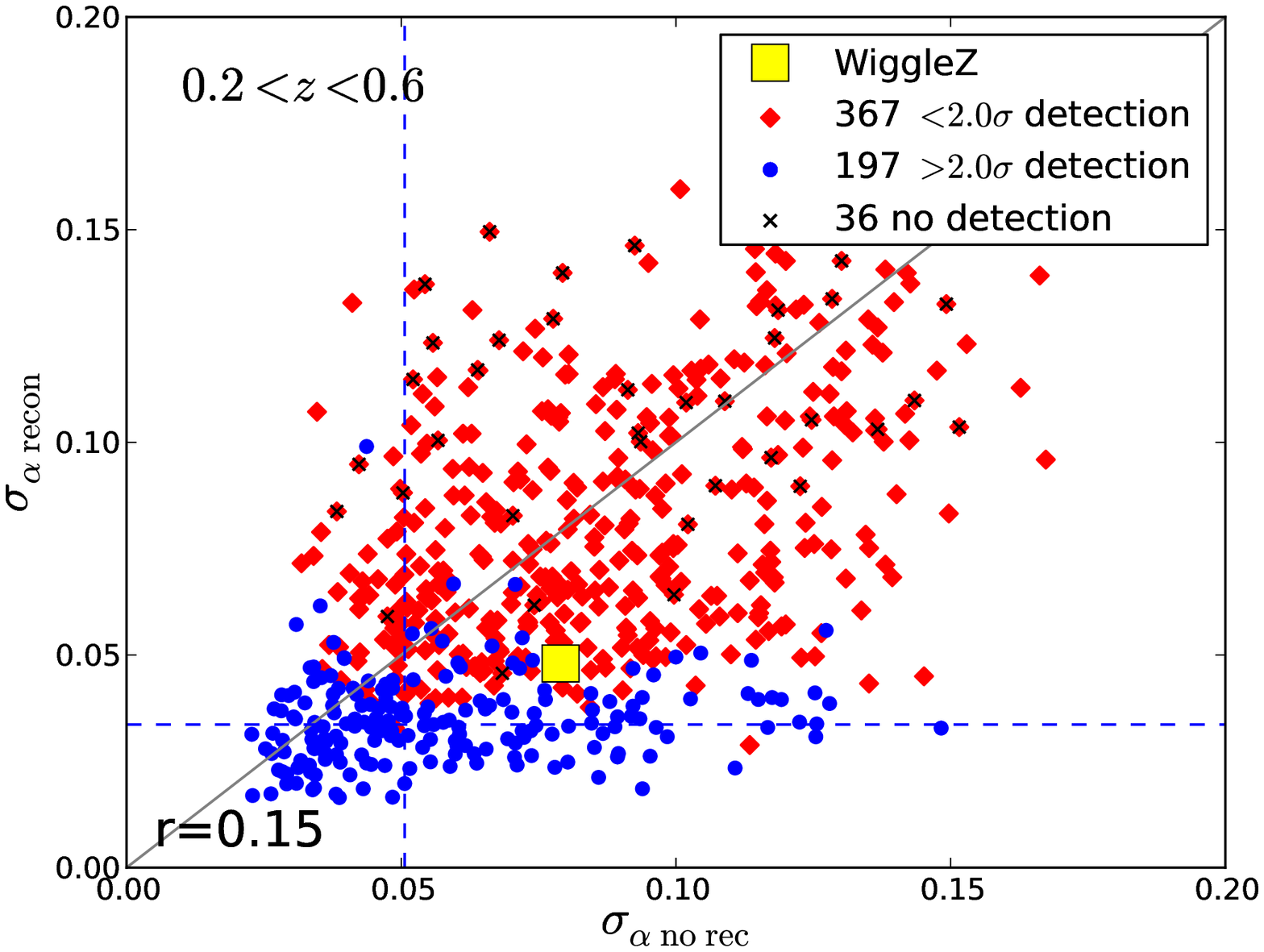}
\includegraphics[width=0.32\textwidth]{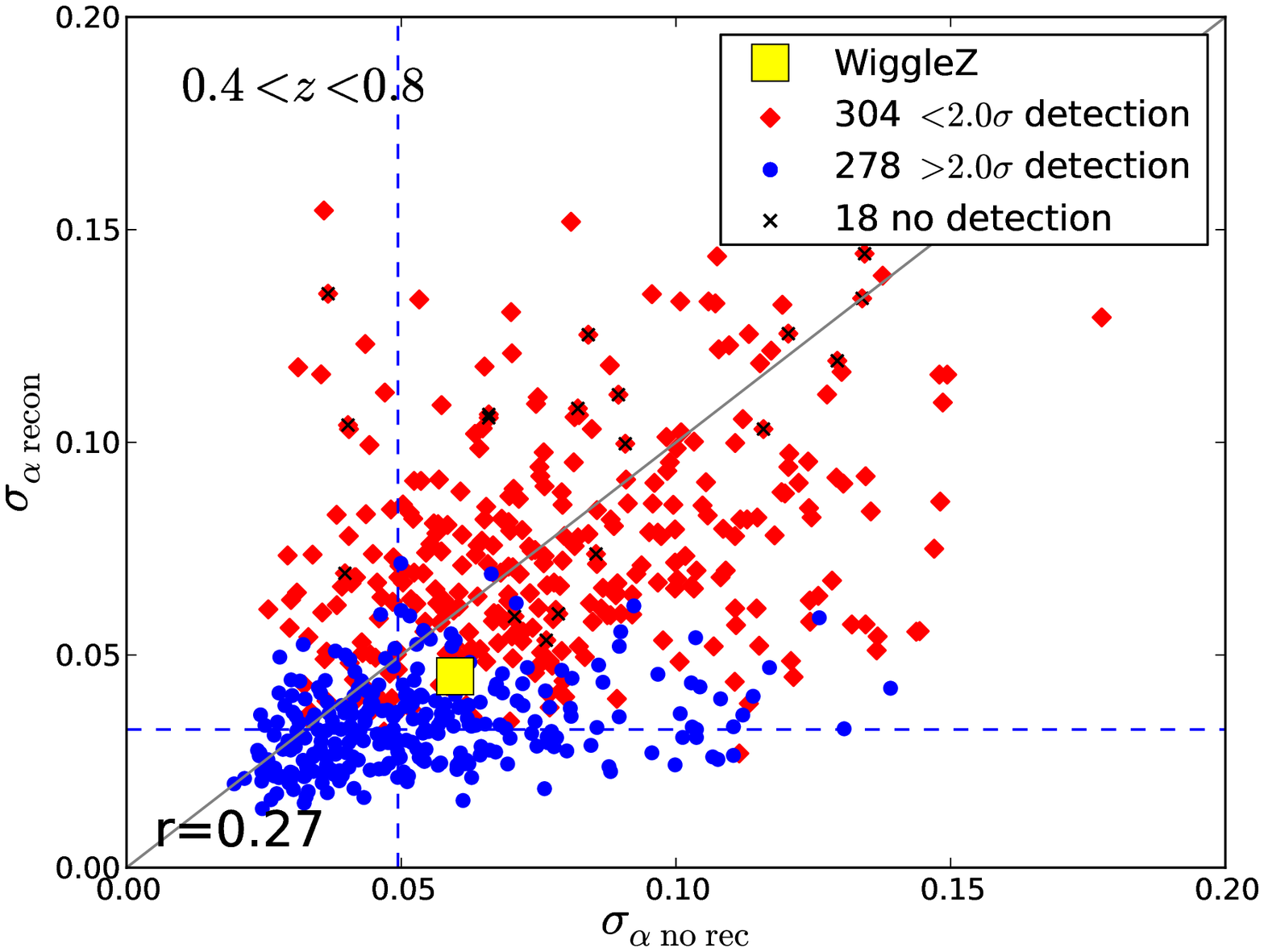}
\includegraphics[width=0.32\textwidth]{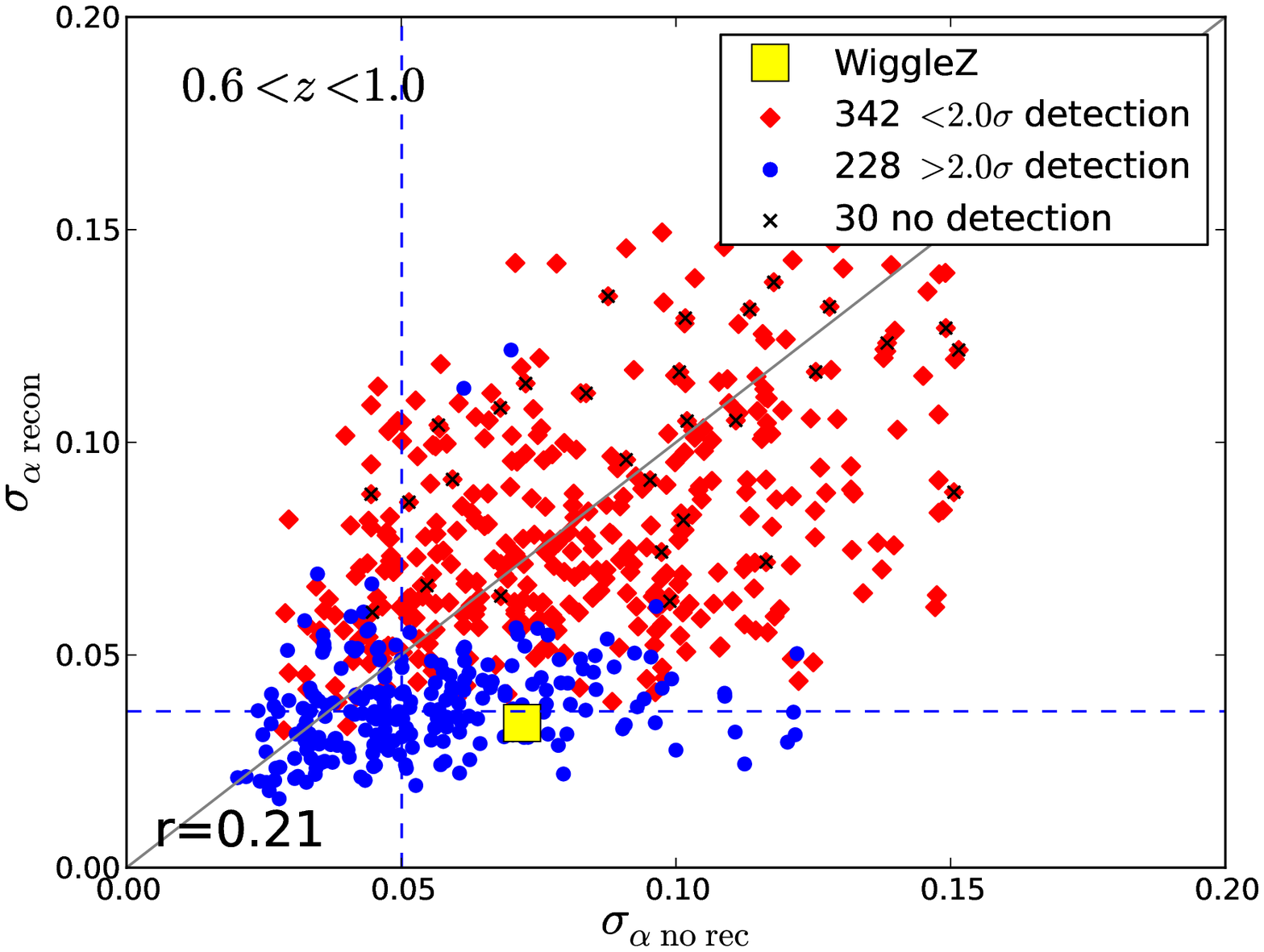} 
\caption{
The top row shows the 
distribution of best-fitting $\alpha$ for 
the 600 mocks for the three 
redshift bins as indicated before ($x-$axis) and after ($y-$axis) reconstruction. 
The bottom row is the same for the uncertainties $\sigma_\alpha$ of the mocks, as 
well as the WiggleZ data (yellow squares). 
The blue circles are results of realizations in which the significance of detection 
of the \baf after reconstruction is better than $2\sigma$, 
and the red diamonds are for mocks below this threshold. 
The marked Xs are realizations in which the $\xi_{\rm nw}$ template 
outperforms the physical one. 
The dashed lines indicate the median of each statistic for the $>2\sigma$ detection sub-samples, 
and $r$ is the correlation coefficient of this sub-sample.  
There is a clear trend of the $>2\sigma$ detection realizations yielding 
tighter $\sigma_\alpha$ constraints.  
WiggleZ results and summaries of the mocks are in Table \ref{table:dvrs_results}. 
}
\label{plot:alpha_distributions}  
\end{center}
\end{figure*}

We find that the maximum likelihood values of \dvrsrsfid  
at all redshifts are consistent 
before and after reconstruction, within the $68\%$ confidence regions, 
and see a clear overlap of the posteriors. 
This is in agreement with predictions from mock catalogs,
which indicate that we would expect a cross correlation of $0.55-0.65$ between the 
\dvrsrsfid measurements before and after reconstruction 
(see top panels of Figure \ref{plot:alpha_distributions}, which is described below).

To better understand expectations 
of results in the three WiggleZ volumes, we apply our analysis pipeline 
to 600 \wizcola mocks in each $\Delta z$ volume.  
Results are displayed in Figure \ref{plot:alpha_distributions}. 
Each column represents results of a different $\Delta z$ 
bin, as indicated. In the top row are the $\alpha$ 
distributions pre- and post-reconstruction, 
and the panels in the bottom row are the  
distribution of the uncertainty in the fit to each realization $\sigma_{\alpha}$. 
Similar to the right panel of Figure \ref{plot:significance_of_detection}, 
the color coding is such that realizations 
with a detection of the \baf above the threshold of $2\sigma$ 
in the reconstruction case
are in blue circles, below are in red diamonds, and no detection are marked by X. 
Also displayed are dashed lines which indicate the median values of the 
$>2\sigma$ subset, as well as the cross-correlation 
values $r$ of this subset. 
In the bottom row we also indicate the WiggleZ  
$\sigma_{\alpha}$ results for comparison in the yellow boxes. 
In Table \ref{table:dvrs_results} we summarize statistics for these distributions 
for the $>2\sigma$ subset,  
which can be compared to the data. 

In the top row of Figure \ref{plot:alpha_distributions} we notice in all $\Delta z$ bins 
groupings along the boundaries of boxes with sides at $|1-\alpha|=0.2$ 
from the center, the hard prior we set in the analysis. 
These indicate failures of determining $\alpha$ 
in these realizations, which is dominantly from the $<2\sigma$ subsets, 
i.e, when the S/N ratio is low. 

Compared to the fiducial cosmology of the mocks $\alpha=1$ 
the distribution of fitted $\alpha$ 
yields a median bias 
between $0.04-0.5\%$, 
which is much smaller than the statistical uncertainties. 
We also test the peak finding algorithm on the 
mock $\overline{\xi}$ 
and find fairly good agreement with the median $\alpha$ 
results of the $>2\sigma$ subsample reported in Table \ref{table:dvrs_results}. 

The reconstruction cases demonstrate a clear improvement in the scatter of $\alpha$, 
as seen in Table \ref{table:dvrs_results}. 
For the $>2\sigma$ subset, the scatter is reduced  
from $5-6.5\%$ to $3.5-5\%$.  A similar improvement in the scatter is obtained   
when examining the full sample.

In the bottom row of Figure \ref{plot:alpha_distributions} 
we see that reconstruction results in 
moderate to dramatic improvements 
in most of the $\sigma_\alpha$ results. 
The $2\sigma$ threshold of detection of the \baf 
also shows clear trends that the $<2\sigma$ subsample (red diamonds)
does not constrain $\alpha$ as well as the $>2\sigma$ subsample (blue circles). 
This dramatic improvement is also shown in the right column in Table \ref{table:dvrs_results}, 
where the median $\sigma_\alpha$ improves in all $z$ bins from $5\%$ with a scatter of $\sim 2.2-2.7\%$ 
to  $3.2-3.7\%$ with a scatter of $1\%$. 
Examining the full $600$ mocks in each $\Delta z$, 
there is a similar improvement in the median, but not in the scatter.

\begin{figure*}
\begin{center}
\includegraphics[width=0.32\textwidth]{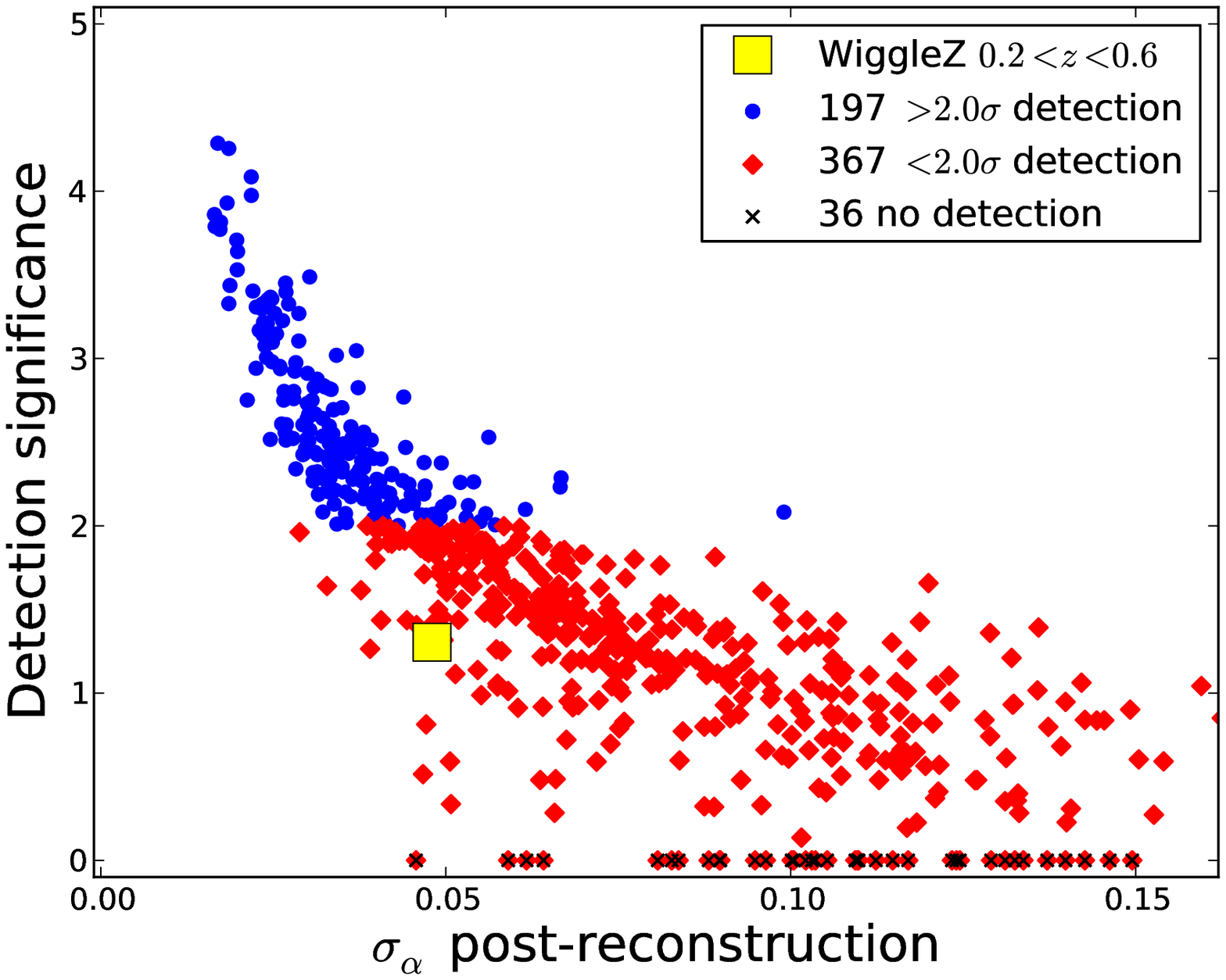}
\includegraphics[width=0.32\textwidth]{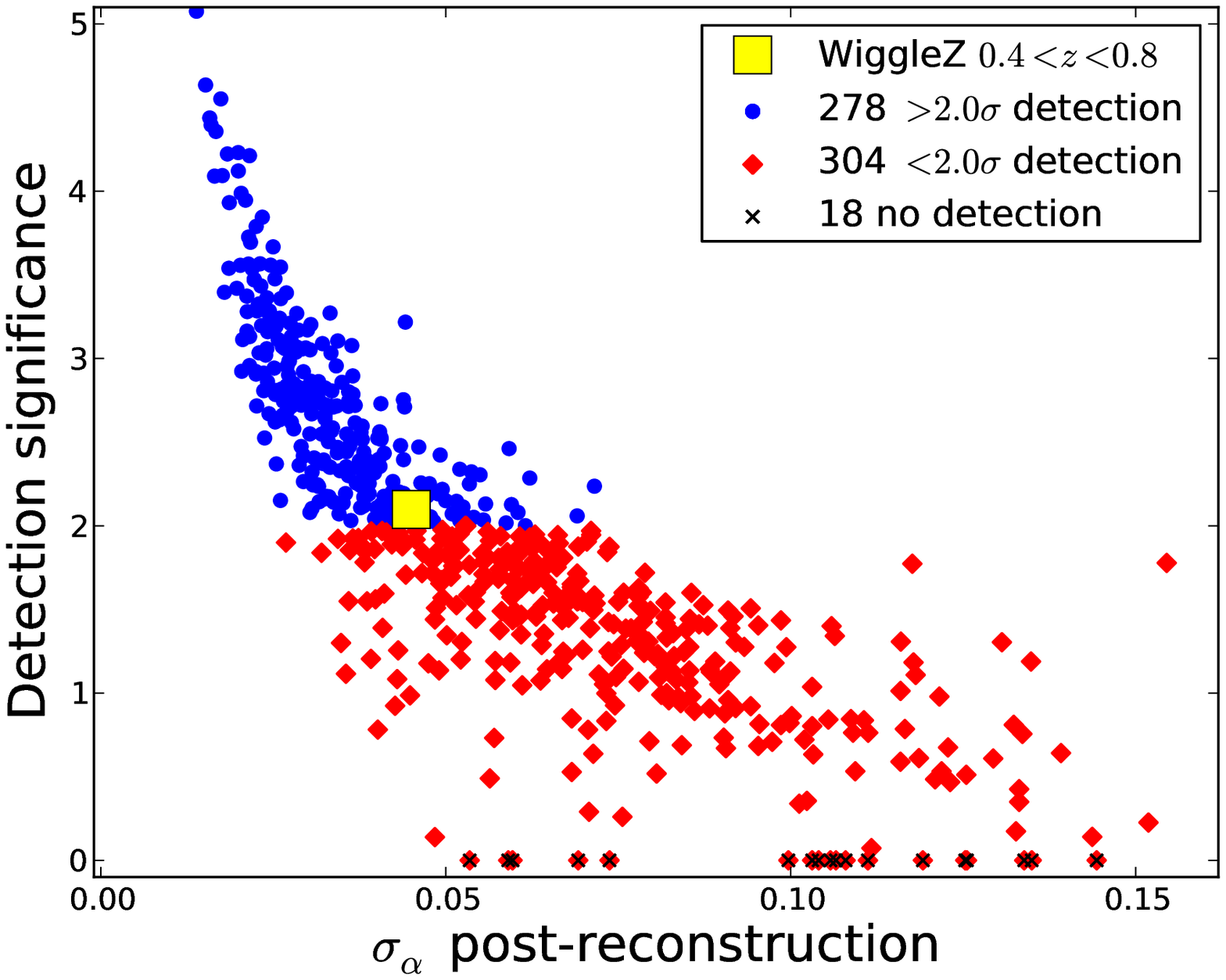}
\includegraphics[width=0.32\textwidth]{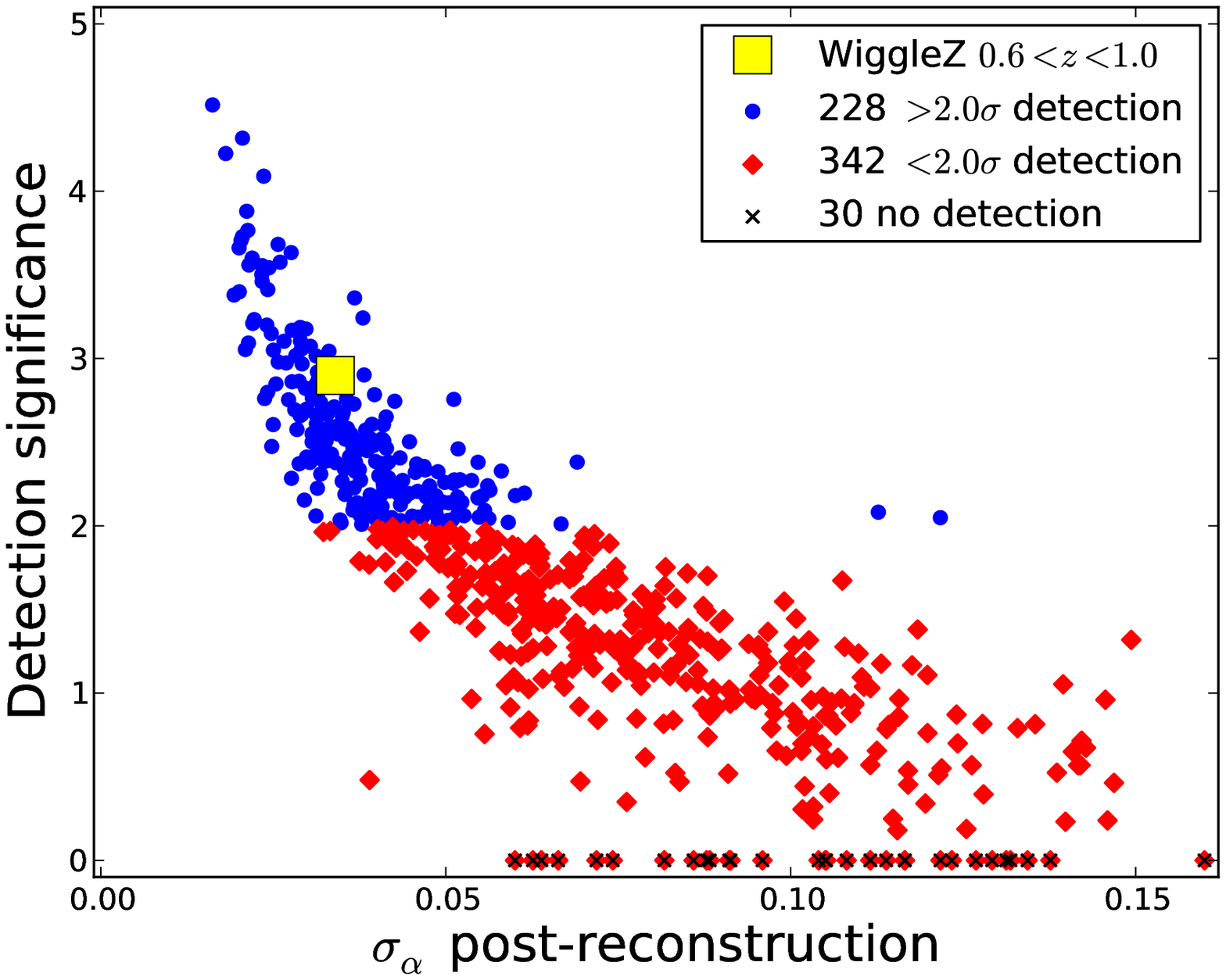}
\caption{
For each volume we plot the post-reconstruction 
significance of detection of the \baf against the resulting $\sigma_\alpha$ 
for each mock realization. 
As in previous figures, the color coding is such that 
blue circles are realizations in the $>2\sigma$ subsample, 
red diamonds are from the $<2\sigma$ subsample, and Xs do not yield a detection. 
The WiggleZ data points are indicated by the yellow squares.
}
\label{plot:sigmaalpha_significance}
\end{center}
\end{figure*}

Distributions of $\alpha$ and $\sigma_\alpha$ 
across the mocks show significant non-Gaussian tails. 
We attribute this to the effect of 
low-significance detection of the \bafii. 
We perform Kolmogorov$-$Smirnov tests for Gaussianity 
of $\alpha$ and $\sigma_\alpha$ and find the $p-$values to be 
negligible. 
In the regime where the \baf is being 
just resolved, 
there is a steep non-linear relation between the significance of detection of the \baf 
and the uncertainty in $\alpha$, 
which is demonstrated in Figure \ref{plot:sigmaalpha_significance}.
Here 
we display the significance of detection 
of the \baf and the resulting $\sigma_\alpha$ 
of all realizations 
for the post-reconstruction case in all three 
$\Delta z$ volumes. 
We see a transition from a somewhat linear relationship 
for the $>2.5\sigma$ significance of detection realizations 
to a more non-linear relationship below this threshold. 

%

The values of the uncertainties of \dvrsrsfid obtained for the 
WiggleZ data in each redshift slices lie within 
the range covered by the mocks in both pre- and post-reconstruction cases. 


We next briefly discuss cosmological implications of these 
improved measurements. 

\subsection{Distance-redshift relation summary}\label{section:dist z summary}

\begin{figure*}
\begin{center}
\includegraphics[width=0.49\textwidth]{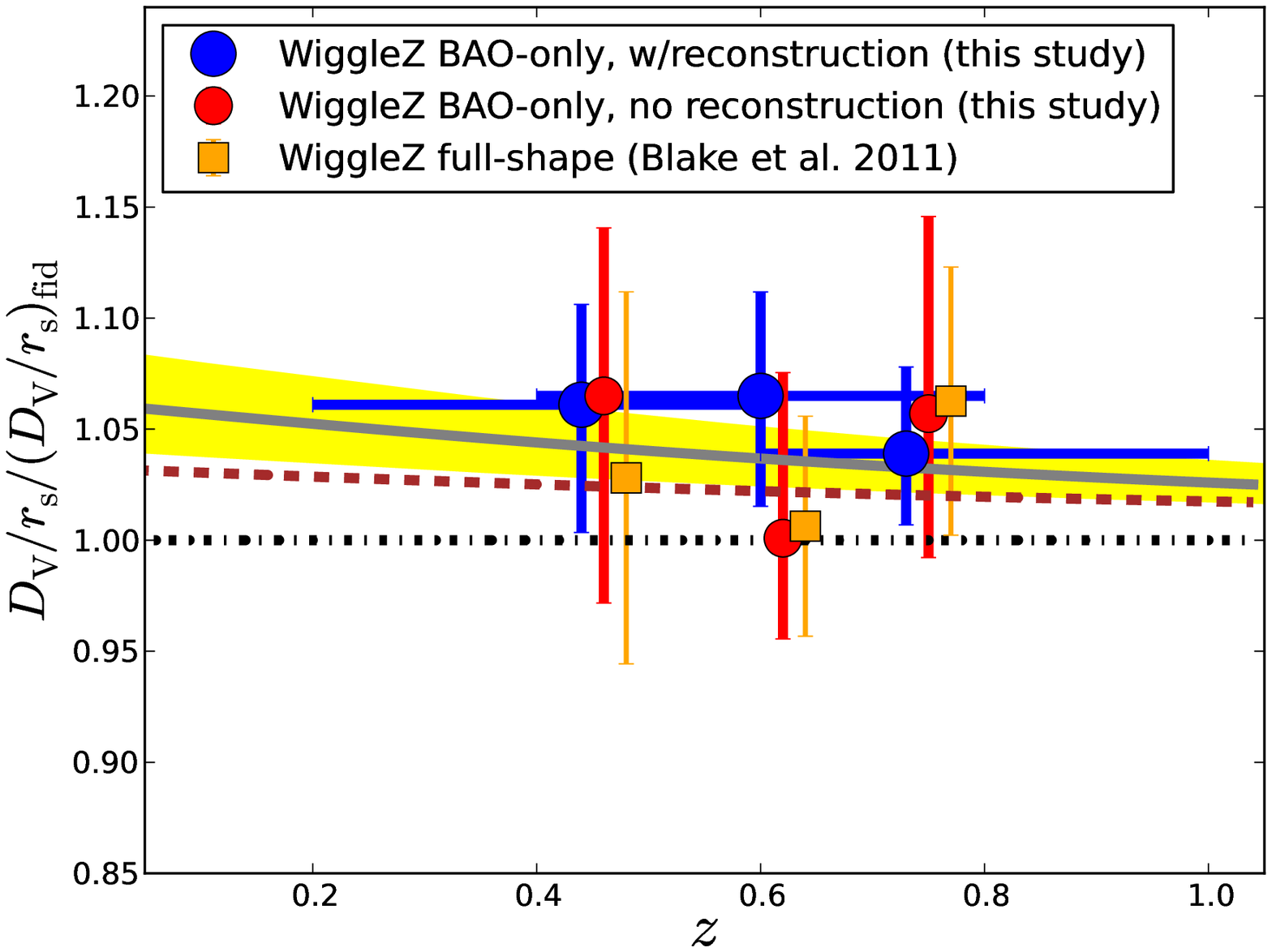}
\includegraphics[width=0.49\textwidth]{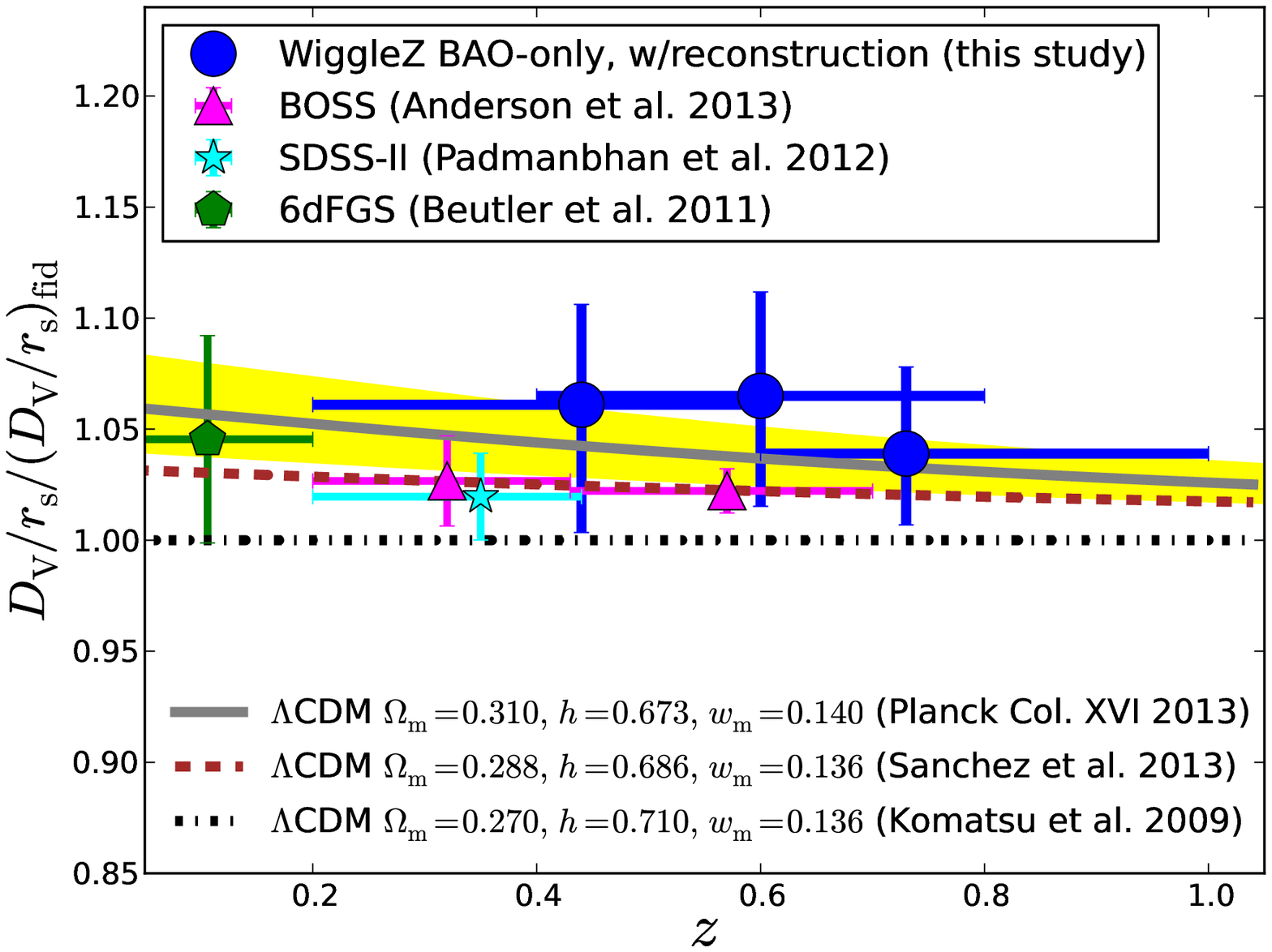}
\caption{
Both panels display the volume-average-distance to sound-horizon ratios $D_{\rm V}/r_{\rm s}$ 
normalized by the fiducial value, 
where the post-reconstruction results are indicated by the large blue circles. 
In the left panel, the no reconstruction BAO-only results (red circles) 
and the $\xi$ shape analysis results (\citealt{blake11c}, orange squares) are slightly shifted for clarity. 
In the right panel we compare with two results from the SDSS-II (cyan star, $0.2<z<0.44$; \citealt{padmanabhan12a})   
and SDSS-III (magenta triangles $0.2<z<0.43$ $0.43<z<0.7$; \citealt{anderson13b}), 
as well as the result obtained by the 6dFGS ($z\sim 0.1$; \citealt{beutler11a}). 
In both panels the cosmology prediction lines are best-fit flat $\Lambda$CDM results ($\Lambda$CDM)
obtained by:   
Planck (\citealt{planck13xvi}; solid) where the yellow band is the 
$68\%$ confidence region, 
SDSS-BOSS (\citealt{sanchez13a}; dashed), 
WMAP (\citealt{komatsu09a}; dot-dashed). 
The $y-$axis uncertainty bars are the $68\%$ confidence region, 
and those on the $x-$axis indicate the redshift range of analysis.}
\label{plot:dv_comparison}  
\end{center}
\end{figure*}

Figure \ref{plot:dv_comparison} summarizes the model-independent \dvrs results 
obtained here pre- (red; left panel) and post-reconstruction (blue; both panels). 
All results are divided by the 
distance-redshift relation for the fiducial cosmology 
used for analysis. 
These new WiggleZ measurements (blue and red) are also indicated in Table \ref{table:dvrs_results}. 

Also plotted in the left panel of Figure \ref{plot:dv_comparison} 
are the WiggleZ $d_z\equiv r_{\rm s}/D_{\rm V}$ results from the \cite{blake11c} analysis: 
$(0.0916 \pm 0.0071, 0.0726 \pm 0.0034, 0.0592 \pm 0.0032)$ for $z_{\rm eff}=0.44,0.6,0.73$, respectively.
There are a few differences in methodology between our pre-reconstruction analysis  and theirs. 
The most important difference is that they focus on the information 
in the full shape of $\xi$, where we marginalize over shape  
and focus only on the peak position, 
making our results model-independent.   
However, despite these differences, the results of the two analyses are consistent. 

For comparison in the right panel of Figure \ref{plot:dv_comparison} we 
plot \dvrs measurements by 
\cite{padmanabhan12a} ($8.88 \pm 0.17$; $z=0.35$), 
\cite{anderson13b} (\dvrsrsfid=1264$\pm$25 Mpc, 2056$\pm$20 Mpc at  $z=0.32, \ 0.57$, respectively)
and $d_z(z=0.106)=0.336\pm0.015$ from \cite{beutler11a}. 
As pointed out by \cite{mehta12a}, there are discrepancies in the literature 
regarding the calculation of $r_{\rm s}$. A common approximation is  
using Equations 4-6 in \cite{eisenstein98}. A more generic treatment is obtained by 
using the full Boltzmann equations as used in the \texttt{camb} package (\citealt{lewis99a}) 
(e.g, this takes into account the effect of neutrinos). Calculations show that these 
differ by over $2\%$, which is now worse than the current $0.4\%$ accuracy measurements of \cite{planck13xvi}. 
Although \cite{mehta12a} show that differences in methods do not yield significant variations of 
$r_{\rm s}/r_{\rm s}^{\rm fid}$ when varying a cosmology from a fiducial,  direct comparisons of results 
require a uniform method. For this reason, because our choice 
of preference is using \texttt{camb}, we re-scale the \dvrs results of 
\cite{padmanabhan12a} and \cite{beutler11a} 
by $r_{\rm s \ {\rm EH98}}^{\rm fid-study}/r_{\rm s \ {\rm CAMB}}^{\rm fid-study}$, 
according to the fiducial cosmologies reported in the each study, fid-study (1.025 and 1.027, respectively). 
For the \cite{anderson13b} results we use their calculation of $r_{\rm s \ {\rm CAMB}}^{\rm fid-study}=149.28$ Mpc. 

In Figure \ref{plot:dv_comparison} 
we also plot predictions for models 
based on flat $\Lambda$CDM,  
according to best-fit parameters obtained by 
\cite{komatsu09a} (dot-dashed line; this is our fiducial cosmology), \cite{sanchez13a} (short dashed line) 
\cite{planck13xvi} (solid line), where the wide yellow band  
shows the $68\%$ confidence region using \texttt{cosmomc}. 
The $\Omega_{\rm m}, \ h$, and $w_{m}\equiv\Omega_{\rm m}h^2$ of each model are indicated 
in the legend. 

%


Of the three predictions, 
our results appear to agree best with those reported 
by \cite{planck13xvi} (solid line), 
which obtain $h\sim 0.67$, 
where the local expansion rate is defined by $H_0\equiv 100h$(km)(Mpc)$^{-1}$(s)$^{-1}$. 
When analyzing various data-sets below in \S\ref{section:cosmo_implications} we show 
that our $h$ results are similar to those of \cite{planck13xvi} rather than the 
lower value obtained by WMAP (dot-dashed) of $h\sim 0.71$.  

\subsection{Covariance matrix of \dvrs }
\begin{figure*}
\begin{center}
\includegraphics[width=0.49\textwidth]{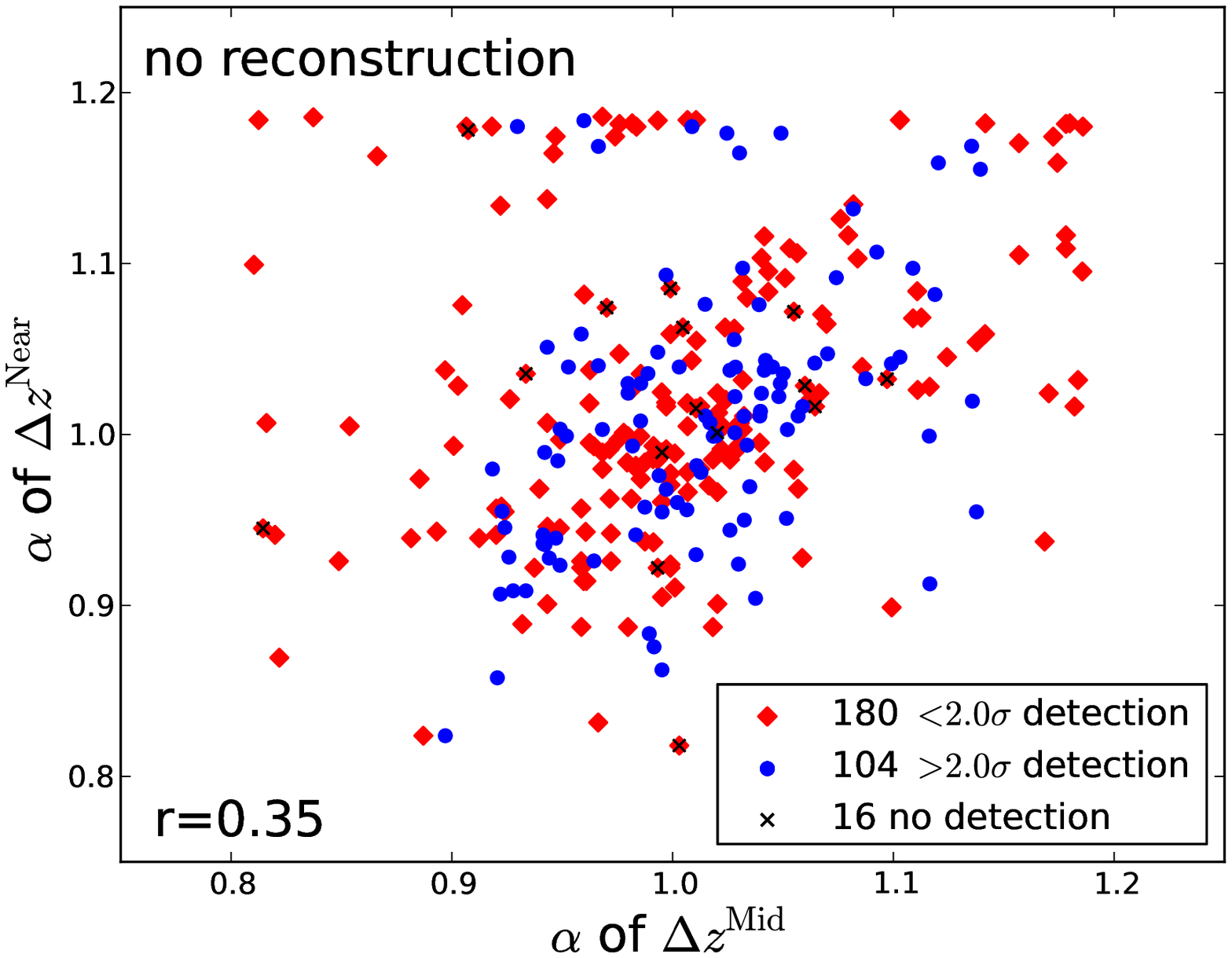}
\includegraphics[width=0.49\textwidth]{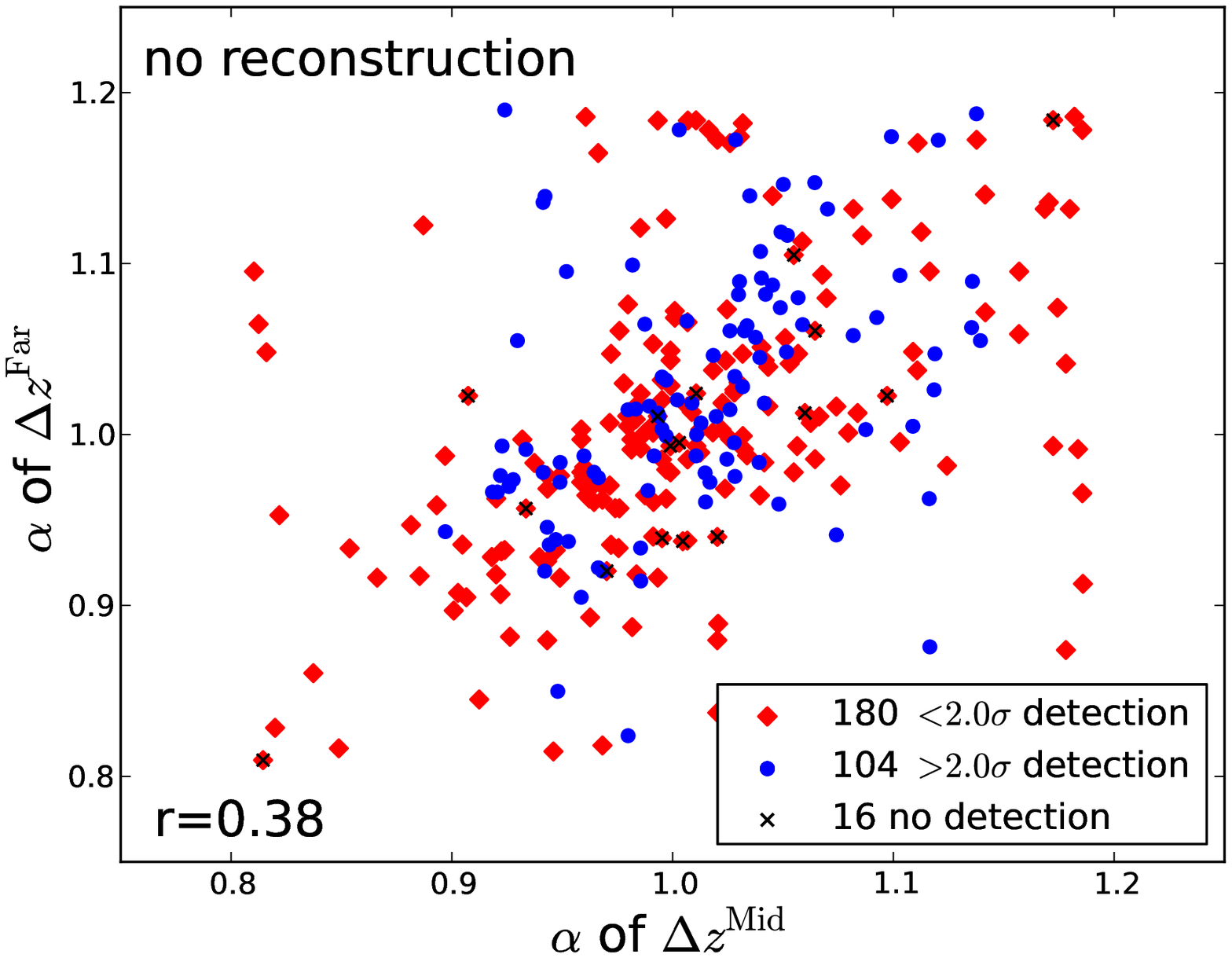}
\includegraphics[width=0.49\textwidth]{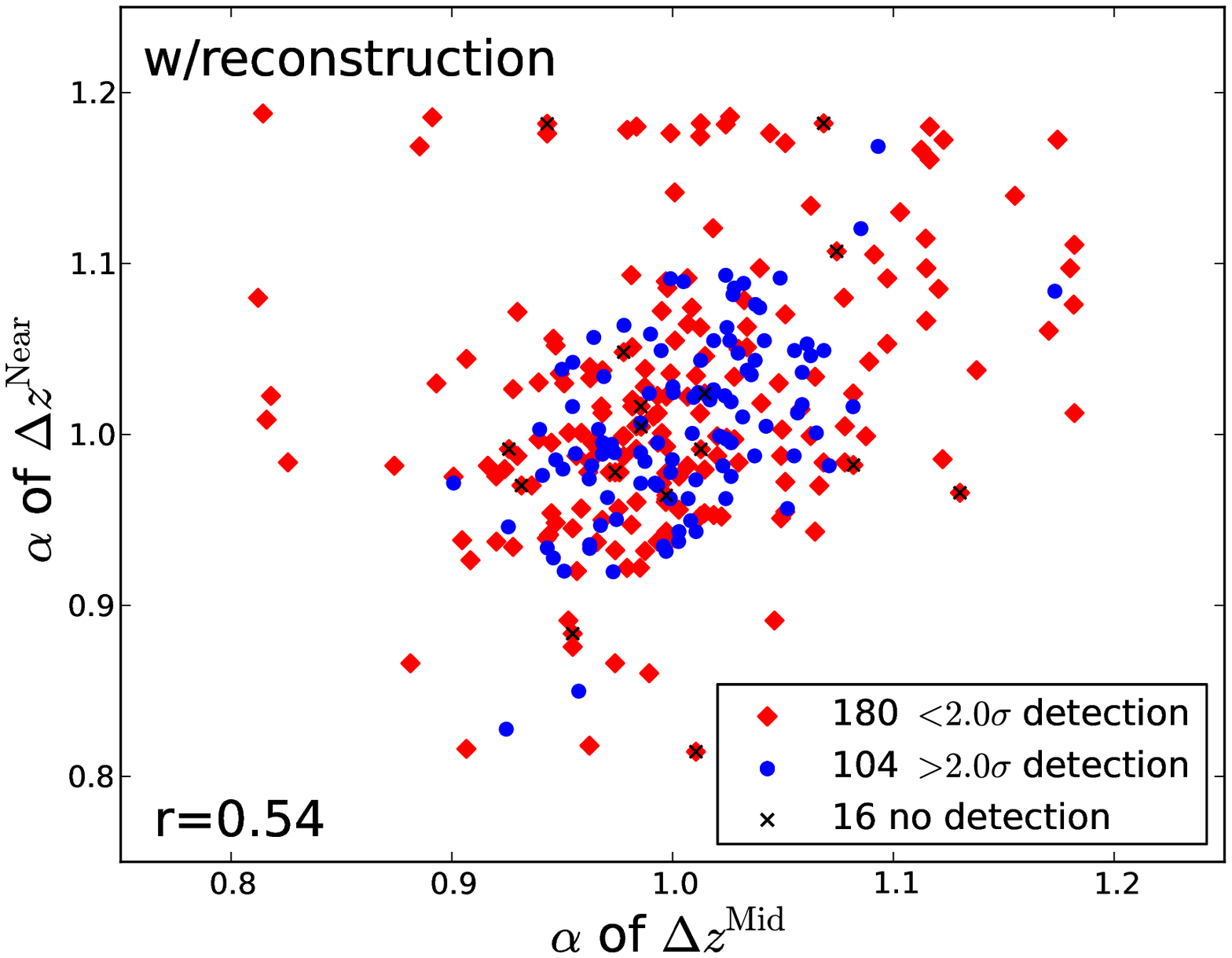} 
\includegraphics[width=0.49\textwidth]{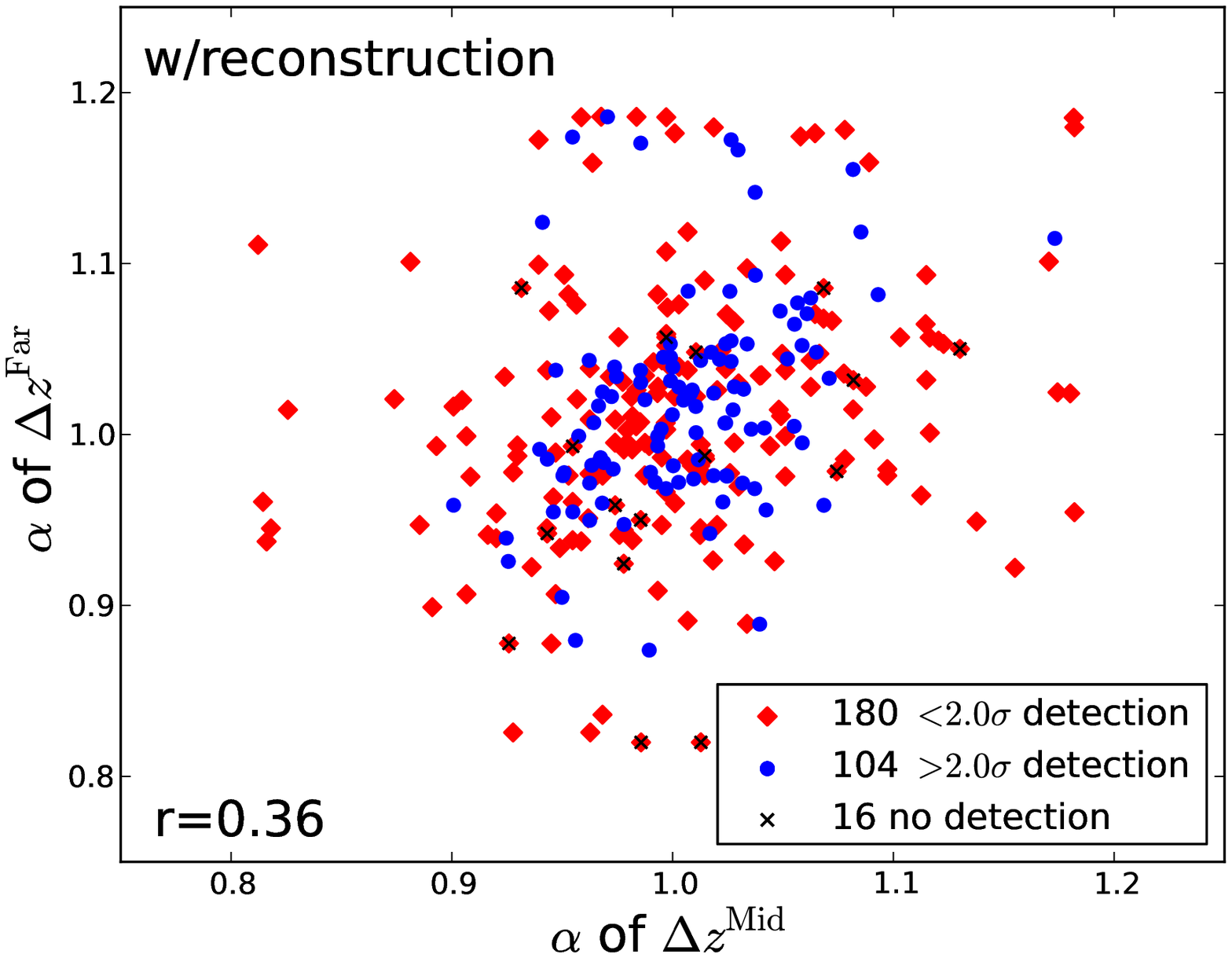}
\caption{
The top row shows the $\alpha$ distribution of the 300 mocks for the no reconstruction 
case and the bottom  for  post-reconstruction.  
In each, the $x-$axes values are those obtained with the \Dzmid ($0.4<z<0.8$) realizations, 
and the $y-$axes values are for 
\Dznear ($0.2<z<0.6$; left panels) and \Dzfar ($0.6<z<1$; right panels), accordingly. 
The blue circles are results of realizations in which the significance of detection 
of the \baf after reconstruction is better than $2\sigma$, 
and the red diamonds are for mocks below this threshold, 
Xs indicate realizations with no detection. 
The correlation coefficient $r$ for the  $>2\sigma$ subsample 
is indicated in the bottom left of each panel. 
}
\label{plot:alpha_distributions_stitched}  
\end{center}
\end{figure*}

Before presenting cosmological implications, 
we first discuss the calculation 
of the covariance between measurements in different redshift slices. 
Due to the overlap between \Dzmid ($0.4<z<0.8$) and the 
other redshift bins, we calculate the correlation coefficients 
between the $\alpha$ results obtained using the 300
{\it stitched} \Dzmid mock catalogs (see \S\ref{section:wizcola}) 
and the corresponding \Dznear and \Dzfar catalogs. 
We apply the same $\alpha$ fitting algorithm as before 
and present comparisons of the results in Figure \ref{plot:alpha_distributions_stitched}. 
The top two panels are before reconstruction and the bottom panels  are after reconstruction. 
For all panels, the $x-$axis values are the $\alpha$ results when using the stitched \Dzmid volume, 
and the $y-$axis values are for the corresponding \Dznear (left column) and \Dzfar (right) volumes. 
As before, we color-code the results according to the significance of detection of the \bafii, 
where the reference subsample for this classification is the stitched \Dzmid case.  

Focusing on the $>2\sigma$ subsample in each case we find that the 
correlation coefficient between the stitched \Dzmid and its 
overlapping neighbors is $r\sim 0.35-0.45$. 
We verify that between \Dznear and \Dzfar $r\sim 0$. 
We use these 
and the uncertainties in Table \ref{table:dvrs_results} to 
construct the covariance matrix of the WiggleZ post-reconstruction 
\dvrsii. The inverse covariance matrix is presented in Table \ref{table:dv_rs_iCij}.

%
\begin{table} 
\caption{The inverse covariance matrix of the \dvrsrsfid measurements 
from the reconstructed WiggleZ survey data. 
The volume-average distance is defined in Equation \ref{equation:dv} and 
$r_{\rm s}$ is the sound horizon at $z_{\rm drag}$, 
and the fiducial cosmology assumed is given in \S\ref{section:intro}.  
These measurements are performed in three overlapping redshift slices 
$0.2<z<0.6$, $0.4<z<0.8$, $0.6<z<1$ with effective redshifts of $0.44, \ 0.6, \ 0.73$ respectively. 
The data vector is \dvrsrsfid $=[ 1716.4, \ 2220.8, \ 2516.1]$ Mpc as listed in Table \ref{table:dvrs_results}. 
As the matrix is symmetric we quote the upper diagonal, 
and for brevity multiply by a factor of $10^4$Mpc$^{2}$. I.e, the user should multiply 
each element by this factor, e.g, the first element would be 2.17898878 10$^{-4}$ Mpc$^{-2}$.} 
 
\label{table:dv_rs_iCij}
\begin{tabular}{@{}cccc@{}}
 \hline
 Redshift Slice                 &     $0.2<z<0.6$    &    $0.4<z<0.8$  &  $0.6<z<1$                        \\
 \hline
$0.2<z<0.6$ & 2.17898878 & -1.11633321 &  0.46982851 \\
$0.4<z<0.8$ &            &  1.70712004 & -0.71847155 \\
$0.6<z<1.0$ &            &             &  1.65283175 \\
 \hline
\end{tabular}

\medskip
\end{table}

\subsection{Cosmological Implications}\label{section:cosmo_implications}
We next examine cosmological implications of the 
new distance-redshift measurements. 
In this analysis we use the reconstructed WiggleZ 
\dvrsrsfid results listed in Table \ref{table:dvrs_results}, 
and their inverse covariance matrix (Table \ref{table:dv_rs_iCij}). 

Our base model 
corresponds to an energy 
budget consisting of baryons (b), radiation (r), 
cold dark matter (CDM), 
and the so-called dark energy.  
The primordial density fluctuations are adiabatic 
and Gaussian with a power law-spectrum of Fourier 
amplitudes. 

We investigate four models. 
The first is the 
flat cosmological constant cold dark matter paradigm, 
where the equation of state of dark energy is set to $w=-1$ ($\Lambda$CDM). 
We then relax the assumption of flatness ($o\Lambda$CDM). 
We also investigate the variation of 
$w$ both when assuming flatness ($w$CDM), as well as without ($ow$CDM)

The main advantage of using information from 
low redshift surveys $z<1$ 
is their ability to constrain the equation of state of 
dark energy $w$ and the curvature $\Omega_{\rm K}$, 
which are otherwise degenerate when analyzing the CMB on its own. 
This is understood through the relationship between the expansion rate $H(z)$ 
and the cosmic composition:  
\begin{dmath} 
H(z)^2=H_0^2\left(  \Omega_{\rm M} \left( 1+z \right)^3+\Omega_{\rm K} \left( 1+z \right)^2+\Omega_{r} \left( 1+z \right)^4+\Omega_{\rm {\rm DE}}  e^{3\int_0^z{\frac{1+w\left( z' \right)}{1+z'}dz'}} \right),
\end{dmath}
where $\sum_i \Omega_i=1$ for $i=$m, K, r, DE. 
According to the definition of $D_{\rm V}$  (Equation \ref{equation:dv}), 
our \dvrsrsfid measurements yield degeneracies between $H$, 
$D_{\rm A}$, and the sound horizon at the end of the 
drag epoch $r_{\rm s}$. 

The physical angular diameter distance\footnote{Note that this is generic because $i\sin(ix)=-\sinh(x)$.} 
\beq\label{da_equation}
D_{\rm A}=\frac{1}{1+z}\frac{c}{H_0} \frac{1}{\sqrt{-\Omega_{\rm K}}} \sin\left(   \sqrt{-\Omega_{\rm K}} \frac{\chi}{c/H_0}  \right)  
\eeq
integrates over 
$H$ through the definition of the comoving distance:
\beq\label{comoving_equation}
\chi(z)=c\int_0^{z}{\frac{dz'}{H(z')}}.  
\eeq 
We calculate the sound-horizon $r_{\rm s}$ 
and the end-of-drag redshift $z_{\rm d}$ 
by using \texttt{camb} (\citealt{lewis99a}). 
For our fiducial cosmology we obtain $r_{\rm s}^{\rm fid}=$148.6 Mpc. 
We point out that another popular choice of calculating 
$r_{\rm s}$ is by using 
Equation 6 
of \cite{komatsu09a} and  $z_{\rm d}$ 
with their Equations 3-5. 
With this we obtain $r_{\rm s}^{\rm fid}=$152.3 Mpc. 
We do not use this last calculation in our analysis. 
See \S\ref{section:dist z summary} 
for a discussion regarding these differences across other survey results.  

Information from the CMB is required to break the degeneracy 
with the sound horizon scale $r_{\rm s}$. For this purpose 
we use the 
Planck CMB temperature anisotropies (\citealt{planck13xv}), 
and the CMB polarization measurements from WMAP9 (\citealt{bennett12a}). 
When analyzing the CMB information we vary 
the physical baryon density $w_{\rm b}\equiv \Omega_{\rm b}h^2$, 
the physical cold dark matter density $w_{\rm c}\equiv \Omega_{\rm c}h^2$, 
the ratio of the sound horizon 
to the angular diameter distance at the last-scattering surface $\Theta$, 
the Thomson scattering optical depth due to reionization $\tau$, 
the scalar power-law spectral index $n_{\rm s}$ and  
the log power of the primordial curvature perturbation $\ln (10^{10} A_{\rm s})$ (at $k=0.05$Mpc$^{-1}$). 

The CMB anisotropies also depend on the following parameters, 
which we fix:  
the sum of neutrino masses $\sum m_{\nu}=$0.06eV, 
the effective number of neutrino-like relativistic degrees of freedom 
$N_{\rm eff}$=3.046, the fraction of baryonic mass in helium $Y_{\rm P}=0.24$, 
the amplitude of the lensing power relative to the fiducial value $A_{\rm L}=1$. 
We also set to zero 
the effective mass of sterile neutrinos $m^{\rm eff}_{\nu, \ \rm sterile}$, 
the tensor spectrum power-law index $n_{\rm t}$, 
the running of the spectral index $dn_{\rm s}/d\ln k$ and 
the ratio of tensor primordial power to curvature power $r_{0.05}$. 
\citealt{planck13xvi} describe the nuisance parameters 
that are marginalized when fitting 
the CMB data. 

In addition we use the 6dFGS BAO 
measurement $r_{\rm s}/D_{\rm V}=0.336\pm0.015$ obtained by \cite{beutler11a}. 
Lastly, 
to quantify the improvements due to using 
the reconstructed WiggleZ \dvrsrsfidii, 
we compare all results to those obtained when using 
the $A(z)\propto D_{\rm V}\sqrt{w_{\rm M}}$ measurements of  
\cite{blake11c}. 
They conclude that,  
when using the full shape of $\xi$ as 
a standard ruler, 
the $A(z)$ parameter, 
as introduced by \cite{eisenstein05b}, 
is a more appropriate representation of the BAO 
information. 
The values used here at $z=0.44,\ 0.6, \ 0.73$ 
are listed in their Table 5, 
and their inverse covariance matrix in their Table 2. 

We use the \texttt{cosmomc} package (October 2013 version; \citealt{lewis02a}) 
to calculate the posteriors. 
The algorithm explores cosmological parameter space by Monte-Carlo sampling 
data sets where it does accurate 
calculations of theoretical matter power spectrum 
and temperature anisotropy $C_{\ell}$ calculations using \texttt{camb} (\citealt{lewis99a}). 

In our MCMC runs we test the following combinations of data: 
\begin{enumerate}
\item CMB: Planck temperature fluctuations (\citealt{planck13xv}) and WMAP9 polarization (\citealt{bennett12a}). 
\item CMB+(WiggleZ pre-recon): CMB with the $A(z)$ pre-reconstruction constraints from \cite{blake11c}.
\item CMB+(WiggleZ post-recon): CMB with post-reconstruction  \dvrsrsfid results investigated here.
\item CMB+(WiggleZ post-recon)+6dFGS: Same as CMB+(WiggleZ post-recon) with the addition of 
the \baf results from the 6dF Galaxy Survey. 
\end{enumerate}
For comparison we also test CMB with the 6dF Galaxy Survey results without information from WiggleZ. 

Here we report results for   
the local expansion rate $H_0$, 
the density of matter $\Omega_{\rm m}$, 
the equation of state of dark energy $w$ 
and the curvature parameter $\Omega_{\rm K}$, 
as relevant in the tested models.  

\begin{figure*}
\begin{center}
\includegraphics[width=0.49\textwidth]{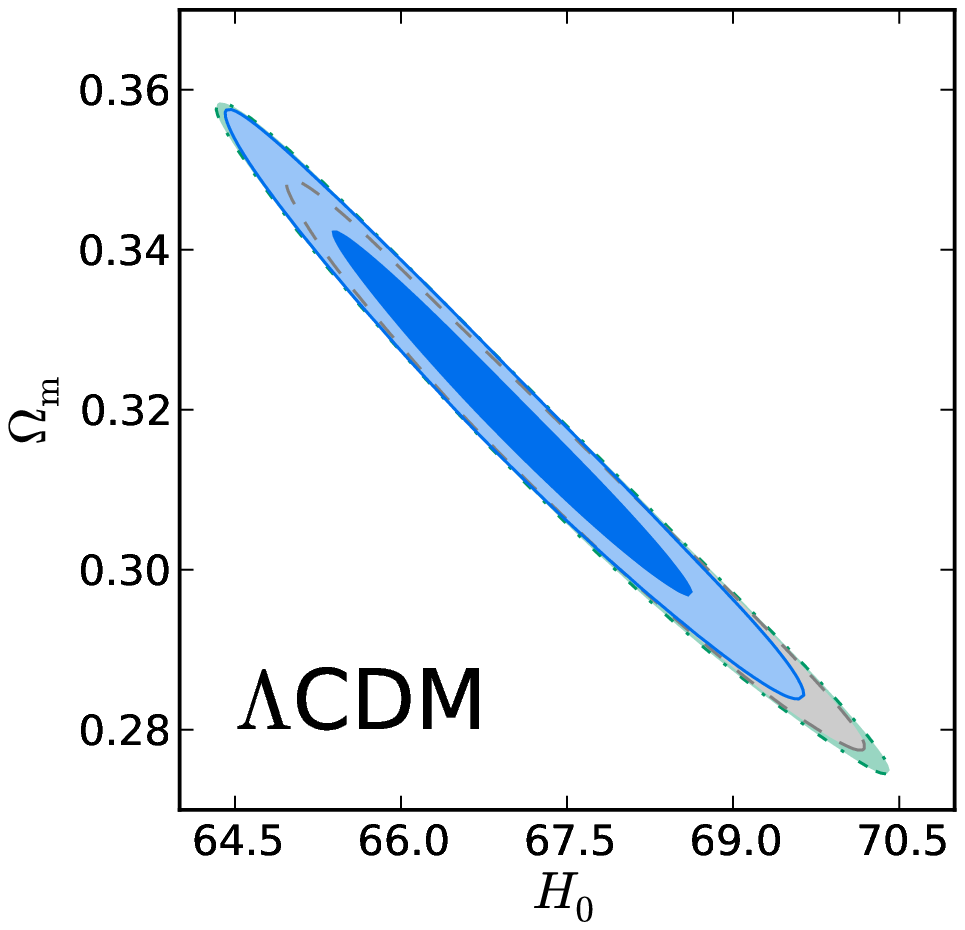}
\includegraphics[width=0.49\textwidth]{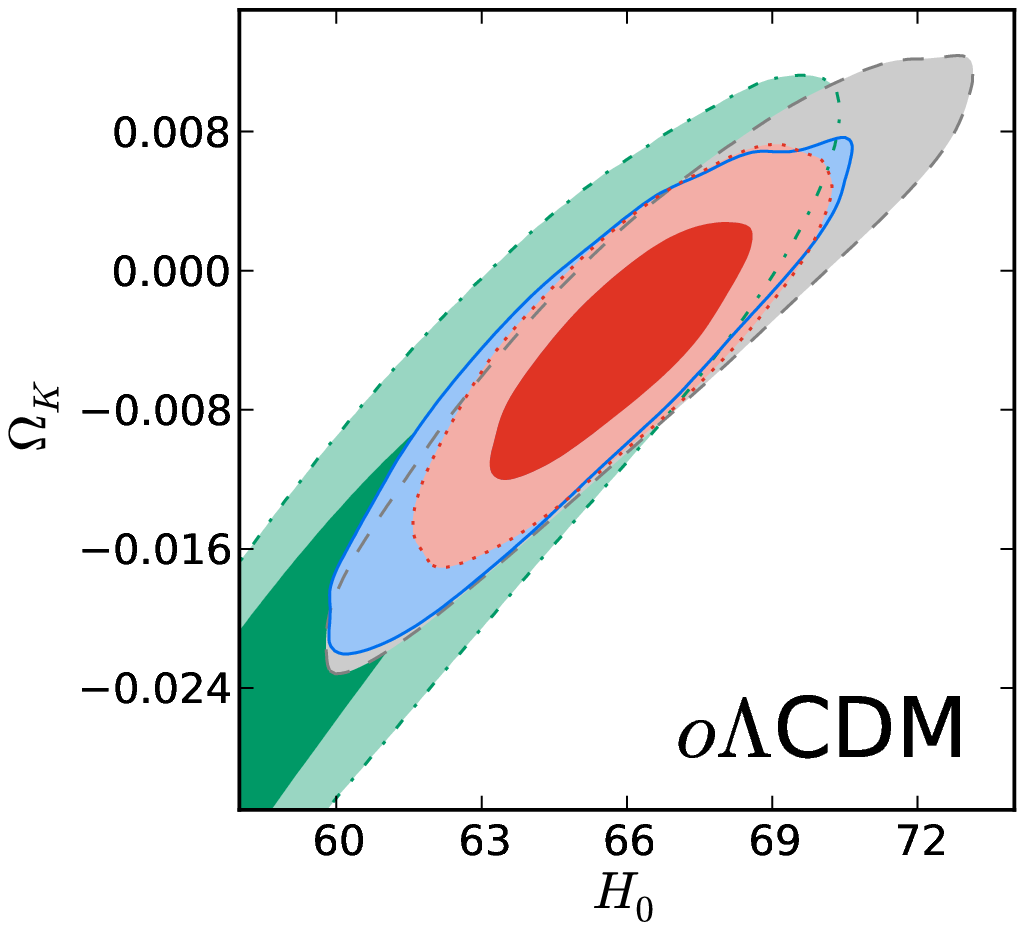}
\includegraphics[width=0.49\textwidth]{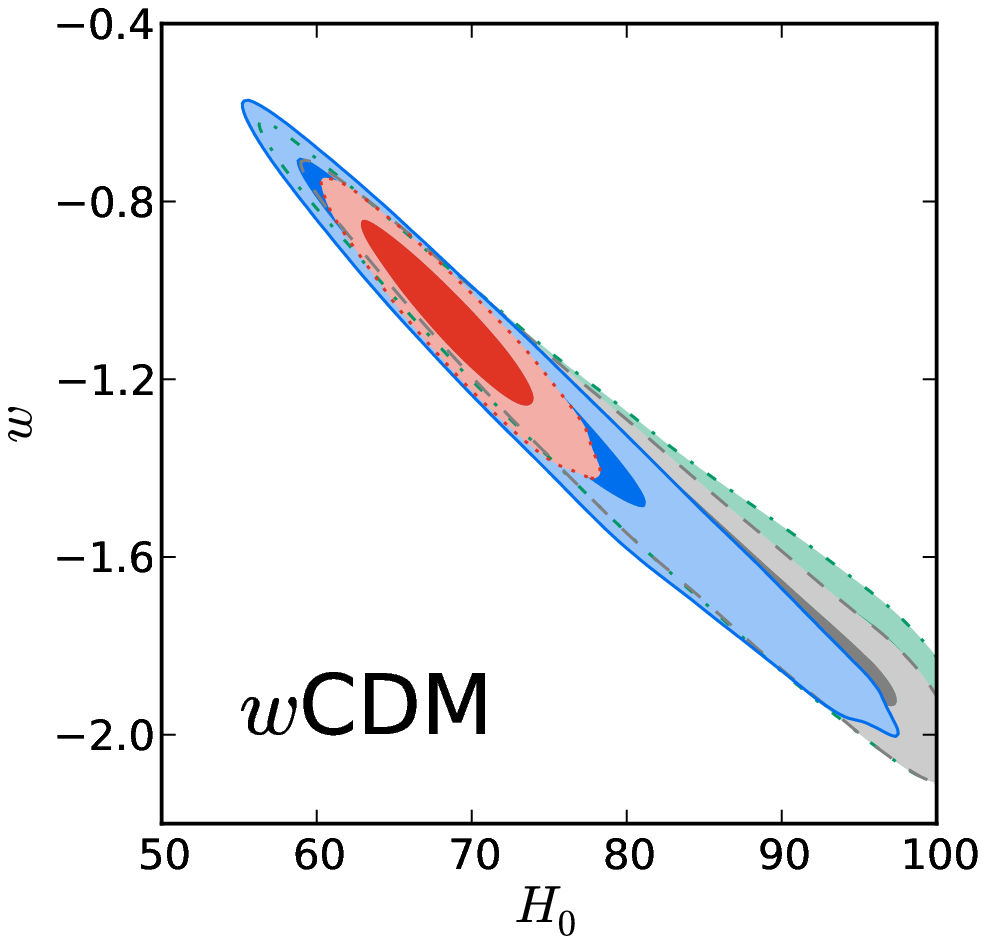}
\includegraphics[width=0.49\textwidth]{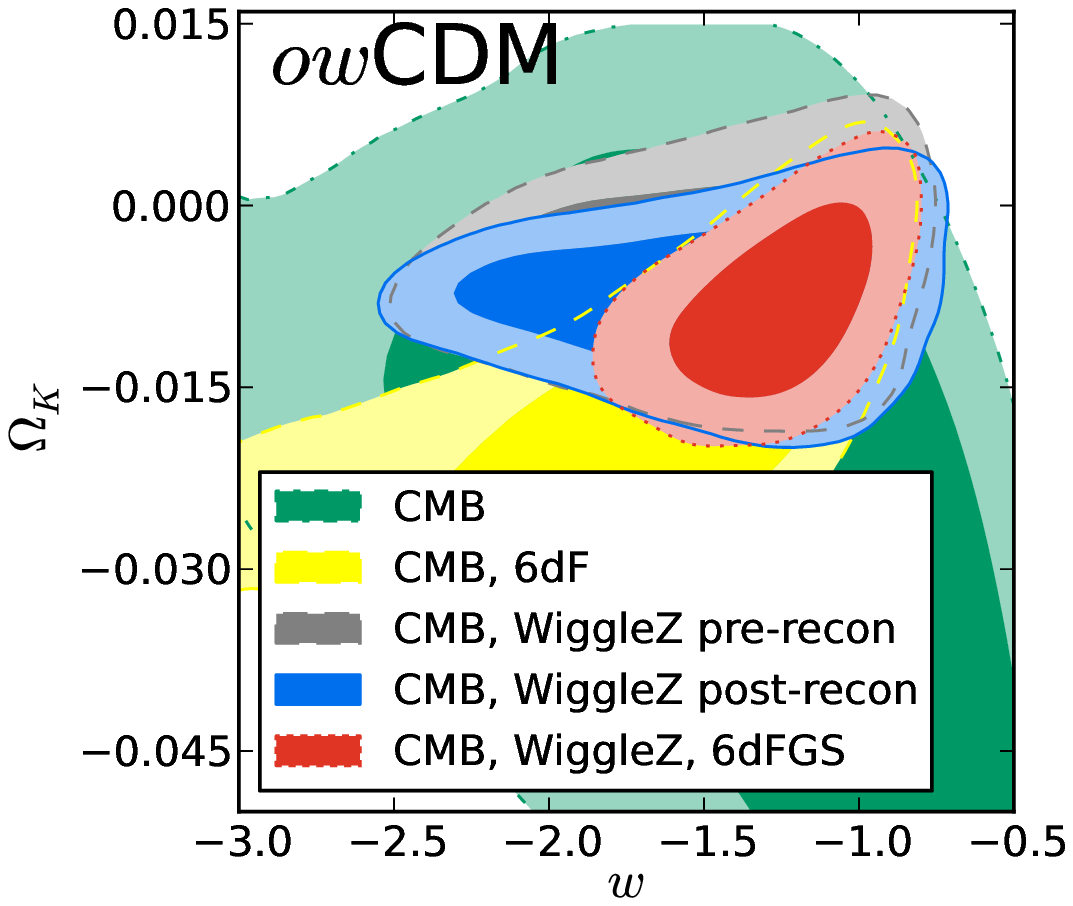}
\caption{
Marginalized 68\% and 95\% joint confidence regions of cosmological parameter pairs, as indicated.  
In the left panels we assume flatness, 
where the top left panel is $\Lambda$CDM and the bottom left  
is $w$CDM, where $w$ is the equation of state of dark energy.   
In the right panels we let the curvature $\Omega_{\rm K}$ vary, where  
the top right panel is $o\Lambda$CDM and the bottom right is $ow$CDM.   
The expansion rate $H_0$ is in units of kms$^{-1}$Mpc$^{-1}$ and  
$\Omega_{\rm m}$ is the matter density. 
In all panels the dot-dashed green contours 
are when using information only from the CMB: Planck temperature fluctuations (\citealt{planck13xv}) and WMAP9 polarization (\citealt{bennett12a}). 
CMB, WiggleZ pre-recon (dashed gray) is when adding $A(z)$ information from the WiggleZ $\xi$ full-shape analysis (\citealt{blake11c}). 
CMB, WiggleZ post-recon (solid blue) is when adding to CMB our post-reconstruction  \dvrsrsfid results. 
The CMB, WiggleZ, 6dFGS results (dotted red) is when we add to CMB, WiggleZ post-recon 
BAO results from the 6dF Galaxy Survey (\citealt{beutler11a}). 
For comparison, in the bottom right we also 
show results of CMB+6dFGS without WiggleZ information. 
}
\label{plot:cosmo_constraints_i}  
\end{center}
\end{figure*}

Our results are summarized in 
Table \ref {table:cosmo_constraints} and in Figure \ref{plot:cosmo_constraints_i}. 
All the results show consistency with the flat ($\Omega_{\rm K}=0$) 
cosmological constant ($w=-1$) cold dark matter paradigm. 
In the following subsections we describe the main results of the four models tested.

\begin{table*} 
\begin{minipage}{172mm}
\caption{Constraints assuming  flat $\Lambda$CDM}
\label{table:cosmo_constraints}
\begin{tabular}{@{}ccccc@{}}
 \hline
 Parameter/Data-set(s)                 &     CMB    &    CMB+(WiggleZ no-recon)  &  CMB+(WiggleZ w/recon)  &  CMB+(WiggleZ w/recon)+6dFGS  \\
\hline
{\bf $\Lambda$CDM} &&&&\\ 
 \hline
$H_0$&67.26$^{+1.19}_{-1.20}$&67.52$^{+1.05}_{-1.03}$&67.00$^{+1.02}_{-1.03}$&67.15$^{+0.99}_{-0.97}$\\
$\Omega_{\rm m}$&0.316$^{+0.016}_{-0.018}$&0.312$^{+0.014}_{-0.014}$&0.319$^{+0.014}_{-0.016}$&0.317$^{+0.013}_{-0.015}$\\
 \hline
$-2\ln(L)$ &9805.3&9805.2&9805.4&9804.9\\
 \hline
\hline
{\bf $w$CDM}  &&&&\\
 \hline 
$H_0$&83.36$^{+14.70}_{-7.29}$&81.15$^{+9.67}_{-11.60}$&72.33$^{+5.09}_{-10.48}$&69.04$^{+3.26}_{-4.01}$\\
$\Omega_{\rm m}$&0.217$^{+0.023}_{-0.078}$&0.227$^{+0.035}_{-0.074}$&0.285$^{+0.067}_{-0.059}$&0.304$^{+0.030}_{-0.033}$\\
$w$&-1.49$^{+0.25}_{-0.42}$&-1.44$^{+0.33}_{-0.34}$&-1.18$^{+0.36}_{-0.19}$&-1.08$^{+0.15}_{-0.12}$\\
\hline
$\Delta AIC$ & 0.3&-0.1&-2.8&-2.6\\
 \hline
 \hline
{\bf $o\Lambda$CDM} &&&&\\
 \hline 
$H_0$&56.13$^{+5.26}_{-6.05}$&66.24$^{+2.61}_{-2.60}$&64.92$^{+2.03}_{-2.05}$&65.84$^{+1.69}_{-1.70}$\\
$\Omega_{\rm m}$&0.462$^{+0.072}_{-0.107}$&0.324$^{+0.024}_{-0.028}$&0.337$^{+0.022}_{-0.024}$&0.327$^{+0.017}_{-0.019}$\\
$100\Omega_{\rm K}$&-3.83$^{+2.91}_{-1.78}$&-0.39$^{+0.74}_{-0.66}$&-0.64$^{+0.62}_{-0.55}$&-0.43$^{+0.47}_{-0.47}$\\
 \hline
$\Delta AIC$ &0.5&-2.5&0.0&-2.7\\
 \hline
 \hline
 {\bf $ow$CDM} &&&&\\  
 \hline 
$H_0$&61.24$^{+9.28}_{-21.01}$&80.26$^{+9.41}_{-12.47}$&76.40$^{+7.29}_{-13.06}$&70.38$^{+3.43}_{-4.55}$\\
$\Omega_{\rm m}$&0.451$^{+0.119}_{-0.289}$&0.230$^{+0.037}_{-0.079}$&0.255$^{+0.057}_{-0.080}$&0.289$^{+0.032}_{-0.032}$\\
$w$&-1.23$^{+0.84}_{-0.47}$&-1.55$^{+0.44}_{-0.37}$&-1.50$^{+0.51}_{-0.33}$&-1.27$^{+0.24}_{-0.18}$\\
$100\Omega_{\rm K}$&-4.18$^{+4.23}_{-1.55}$&-0.54$^{+0.46}_{-0.47}$&-0.78$^{+0.42}_{-0.43}$&-0.83$^{+0.44}_{-0.55}$\\
 \hline
$\Delta AIC$ &-0.8&-1.5&-1.4&-2.4\\
 \hline
  
\end{tabular}

\medskip
CMB refers to temperature fluctuations of \cite{planck13xv} and WMAP9 polarization (\citealt{bennett12a}). \\
WiggleZ no-recon refers to the pre-reconstruction $A(z)$ obtained by using the full shape of $\xi$ (\citealt{blake11c}). \\
WiggleZ w/recon refers to the post-reconstruction \dvrsrsfid measurements presented here.\\
6dFGS refers to the \baf measurements of that survey (\citealt{beutler11a}). \\
In the $\Lambda$CDM section we quote the maximum likelihood as -$2\ln L$.\\
In the $w$CDM, $o\Lambda$CDM, $ow$CDM sections we quote the $\Delta AIC \equiv AIC_{\Lambda {\rm CDM}}-AIC_{{\rm M}}$ of each model $M$, as explained in the text. 
A positive $\Delta AIC$ indicates a preference for the model M over $\Lambda$CDM and vice versa. 
The relative likelihood of the model $M$ can be quantified as 
$\exp (\Delta AIC/2)$.\\
\end{minipage}
\end{table*}

\subsubsection{$\Lambda$CDM results}
The top left panel of Figure \ref{plot:cosmo_constraints_i} 
presents the joint posterior probability distribution of $H_0$ 
and $\Omega_{\rm m}$, 
and the marginalized results are summarized in 
Table \ref {table:cosmo_constraints}. 
These measurements follow the degeneracy line of constant    
$\Omega_{\rm m}h^3$ (e.g, \citealt{percival02a,sanchez13b}). 
All combinations of data sets tested yield consistent results.
There is a moderate improvement when adding 
the reconstructed WiggleZ \dvrsrsfid information to that of the CMB. 
This can be quantified by the marginalized measurement of $H_{0}$ 
improving from 1.8\% accuracy to 1.5\% accuracy, 
and $\Omega_{\rm m}$ from 5.4\% accuracy to 4.7\% accuracy.  
Comparing CMB+(WiggleZ no recon) to the other combinations, 
we conclude that 
the reconstruction of WiggleZ and 
the additional information from 6dFGS does little to improve the 
$H_{0}$ and $\Omega_{\rm m}$ measurements. 


\subsubsection{$w$CDM results}
We now allow $w$ to vary as a constant 
(i.e, no dependence on $z$). 
The bottom left panel of Figure \ref{plot:cosmo_constraints_i} 
presents the joint posterior probability of $H_0$ 
and $w$. 
Here we see that 
the CMB alone does not constrain this combination well,  
showing a large allowed range towards the lower 
region of $w$.  
Adding 
the pre-reconstruction WiggleZ information 
does little to improve these measurements. 
Replacing with the post-reconstruction 
WiggleZ \dvrsrsfidii, 
we see a slight improvement 
of the $w$ measurement 
on its low side of the 68$\%$ confidence region 
(but there is no improvement on the high side). 
A further substantial improvement is achieved when adding information from the 
6dFGS \baf resulting in $w=-1.08^{+0.15}_{-0.12}$, 
a $\sim 13\%$ accuracy measurement. 
This can be explained by 
the fact that the low redshift \dvrs 
is particularly sensitive to $H_0$, 
helping to break the degeneracy. 

\subsubsection{o$\Lambda$CDM results}
When allowing for variation of $\Omega_{\rm K}$ and 
assuming $w=-1$, 
we notice 
some improvement in constraints when adding the 
WiggleZ pre-reconstruction to that of the CMB.  
When replacing the WiggleZ pre-reconstruction $A(z)$ 
by the post-reconstruction \dvrsrsfidii, however, we see substantial improvement 
in the measurements on the high side of $\Omega_{\rm K}$. 
Further improvement to measurements on the low side of  $\Omega_{\rm K}$ 
are obtained when adding information from the 
6dFGS \bafii. 
These are shown in the top right panel of Figure \ref{plot:cosmo_constraints_i} 
which displays the joint posterior probability of $H_0$ 
and $\Omega_{\rm K}$.  

\subsubsection{$ow$CDM results}
Lastly, we 
allow both $w$ and $\Omega_{\rm K}$  to vary 
and find results to be consistent with 
the flat cosmological constant paradigm. 
This is shown in the bottom right panel 
of Figure \ref{plot:cosmo_constraints_i} 
which displays the joint posterior probability  
of these parameters. 
As expected, the CMB-only results do not 
constrain these parameters well, 
and the addition of the WiggleZ information 
yields substantial improvement. 
As noticed in the case of 
$o\Lambda$CDM case, we obtain a clear improvement 
in the higher end of the confidence region of $\Omega_{\rm K}$  
when adding to the CMB the reconstructed WiggleZ \dvrsrsfidii, 
compared to adding the pre-reconstruction $A(z)$. 
The marginalized $68\%$ confidence region of $\Omega_{\rm K}$ is limited to 
[-0.0121,-0.0036].  

Adding the \baf from the 6dFGS does not improve constraints 
on the curvature but does substantially reduce the allowed 
space for $w$, as seen in the $w$CDM case. 
In the case of CMB+(WiggleZ post-recon)+6dFGS we obtain 
a marginalized result of $w=-1.27^{+0.24}_{-0.18}$, 
a $17\%$ accuracy measurement. 

To better understand contributions from 
WiggleZ compared to those from 6dF, 
when added to the CMB information, 
in the bottom right panel 
of Figure \ref{plot:cosmo_constraints_i}
we plot in yellow dashed constraints obtained with 
CMB+6dF without 
WiggleZ data. This result shows that 
CMB+6dF alone is not enough to simultaneously 
constrain 
$w$ and $\Omega_{\rm K}$. 
We do find in the $o\Lambda$CDM case, however, that  
CMB+6dF constrains $H_{0}$ and $\Omega_{\rm K}$ 
in a similar manner to results obtained using 
CMB+(WiggleZ pre-recon). 

As mentioned above, in all of our tests 
we find consistency with $\Lambda$CDM model. 
We now turn to quantify 
the model selection 
compared to $\Lambda$CDM. 
For this purpose we use the Akaike information criterion, 
which incorporates trade-offs between the goodness of fits to the additional complexity  
of each model (\citealt{akaike74a}). 
For each model M we quantify $AIC_{\rm M} \equiv 2p-2\ln(L)$, 
where $p$ is the number of parameters and $L$ is the maximized value of the likelihood function. 
We then define $\Delta AIC \equiv AIC_{\Lambda {\rm CDM}}-AIC_{{\rm M}}$ as 
our indicator of the 
preferred model. 
A positive $\Delta AIC$ prefers model 
M over $\Lambda$CDM and vice versa. 
The relative likelihood of the models can be quantified as 
$\exp (\Delta AIC/2)$. 

In Table \ref {table:cosmo_constraints} we list the 
$\Delta AIC$ of the models $w$CDM, $o\Lambda$CDM, $ow$CDM, 
which should be read by column (for each data set combination). 
We find non-positive values of $\Delta AIC$ values for all the data sets 
which include BAO 
in all models, meaning that the model that is 
preferred given the data 
(CMB, WiggleZ, 6dFGS) is $\Lambda$CDM. 
E.g, when comparing the $w$CDM model to $\Lambda$CDM 
and using the CMB+(WiggleZ w/recon) we obtain $\Delta AIC=-2.8$, 
i.e, 
the relative likelihood of the $w$CDM model is 0.247 times that of 
$\Lambda$CDM according to the Akaike information criteria.

\section{Summary}\label{section:summary}
We present improved distance measurements 
in the redshift shift range 
$0.2<z<1$ 
using the WiggleZ Dark Energy Survey galaxies, 
by applying the reconstruction of the 
\baf technique, which utilizes additional information encoded in the 
density field. 

The constraints on \dvrsrsfid are 
\dvrsrsfidN$\pm$\upmdvrsrsfidN Mpc,  \dvrsrsfidM$\pm$\upmdvrsrsfidM Mpc, \dvrsrsfidF$\pm$\upmdvrsrsfidF Mpc ($68\%$ CL) 
for effective redshifts $z_{\rm eff}=0.44,\ 0.6,\ 0.73$, respectively. 
These results are model-independent as we focus on the 
geometrical information contained in the \bafii, 
and not the full shape of $\xi$. 

Figure \ref{plot:dv_comparison} shows a comparison 
of WiggleZ \dvrs measurements obtained by various methods 
with other data sets 
and cosmological predictions. 
The \dvrs measurements obtained 
by analysis of the \baf position 
when using pre- and post-reconstruction 
data are shown to be consistent. 
Furthermore, 
these results also agree 
with those obtained by \cite{blake11c}, 
who used the full shape of the pre-reconstruction 
$\xi$ as a standard ruler. 

Interestingly, although we use 
a cosmology as predicted by  
WMAP  
as our fiducial, when converting 
redshifts to co-moving distances before counting 
the pairs of galaxies, 
the post-reconstruction \dvrs results show a preference 
for the distance-redshift predictions of 
the best-fit cosmologies measured by 
\cite{planck13xvi}  and BOSS (e.g, \citealt{sanchez13a}).  

These $3.4-4.8\%$ accuracy post-reconstruction \dvrs measurements 
represent a  
significant improvement from the pre-reconstruction case, 
and from the analysis of the full shape of $\xi$.\footnote{This statement is true for our analysis in the context of constraining \dvrsii; 
the full shape of $\xi$ contains more information, e.g $\Omega_{\rm m}h^2$ and $n_{\rm s}$, which is not investigated here.}
These measurement improvements 
are effectively equivalent 
to those expected from surveys with up to 
2.5 times the volume of WiggleZ.\footnote{
The calculation is based on squaring the uncertainty ratio, 
where we assume $\sigma_\alpha^2 \propto$ 1/Volume.}  
To be conservative, here we assume a comparison between 
our  post-reconstruction BAO-only results to those 
of the pre-reconstruction $\xi$ full-shape analysis reported by \cite{blake11c}.\footnote{  
When comparing between BAO-only pre- and post-reconstruction the 
improvement is 
effectively equivalent to surveys with volumes up to 4.7 larger than WiggleZ.} 

We test for sample variance by analyzing 600 mock simulations 
and  
find that reconstruction of the density field 
should yield a sharpened \baf $65\%$ of the time, 
and our \dvrsrsfid results are within these expectations. 

The main limitations 
of the 
WiggleZ combined volumes are the edge effects, 
completeness and large shot-noise. 
Although we show that reconstruction successfully 
works on the data and most mocks, 
we find that it fails to yield an improved significance 
of detection of the \baf in $\sim 30\%-40\%$ of the cases, 
depending on the redshift range. 
We also find that $3\%-6\%$ of the mock realizations  
fail to detect a \baf post-reconstruction.  

In Table \ref{table:dv_rs_iCij} we provide the inverse 
covariance matrix of the \dvrsrsfid measurements 
between these overlapping $\Delta z$ volumes,  
which can be used to calculate cosmological implications. 
We combine our measurements with CMB temperature anisotropies from 
Planck and CMB polarization of WMAP9, 
as well as the \baf of the 6dF Galaxy Survey.

Using these post-reconstruction \dvrsrsfid measurements 
we obtain  consistent 
measurements of 
fundamental cosmological parameters 
compared with those obtained when using the \cite{blake11c} $A(z)$ results. 
Assuming a curved cold dark matter model while varying 
the equation of state of dark energy, 
we find consistency with the flat $\Lambda$CDM model. 
The significant improvement in measuring \dvrsrsfidii, 
obtained by applying reconstruction,  
yields moderate improvements on constraining $\Omega_{\rm K}$ ($o\Lambda$CDM, $ow$CDM), 
and  only slight improvement in $w$ ($w$CDM, $ow$CDM), 
and $H_{0}$, $\Omega_{\rm m}$ (when examining the flat $\Lambda$CDM model). 

Testing the $\Lambda$CDM model we obtain 
a marginalized constraint of $H_0=67.15\pm 0.98$ kms$^{-1}$Mpc$^{-1}$, 
which is in a $2.6\sigma$ tension with the SH0ES measurement of $H_0=73.8\pm 2.4$kms$^{-1}$Mpc 
(\citealt{riess11a}).\footnote{Calculation: (73.8-67.15)/$\sqrt{2.4^2+0.98^2}=2.6$}  
The density of matter is constrained in the range $\Omega_{\rm m}=0.317\pm 0.014$. 
Relaxing the assumption of flatness we constrain the curvature to $\Omega_{\rm K}=-0.0043\pm0.0047$. 
When assuming a flat $w$CDM model, the equation of state of dark energy 
is estimated to be $w_{\rm DE}=-1.08 \pm 0.135$. 

In the analysis of the cosmological constraints we do 
not compare results with those of the SDSS. 
Although the overlap between the surveys is small, 
current investigation is underway to quantify the covariance of 
the \dvrs measurements of the surveys 
(Beutler, Blake et al.; in prep). 

To summarize we find that, although 
the reconstruction procedure is most effective in contiguous 
surveys, it can be applied successfully in surveys that are patchy, 
that have high shot-noise and  
significant edge effects. This demonstrates the power of the 
technique in producing 
a sharper \baf from which we can obtain significantly improved unbiassed distance measurements. 

\section*{Acknowledgments}
We thank 
Florian Beutler, 
Daniel Eisenstein, 
Shahab Joudaki, 
Antony Lewis, 
Felipe Marin and 
Ariel Sanchez
for useful discussions. 
EK and JK are supported by the Australian Research Council Centre of Excellence for All-sky Astrophysics (CAASTRO), through project number CE110001020.
CB acknowledges the support of the Australian Research Council
through the award of a Future Fellowship. 
TMD acknowledges the support of the Australian Research Council through a Future Fellowship award, FT100100595.
The numerical simulation was supported by the SwinSTAR supercomputer
at Swinburne University of Technology and the Raijin supercomputer
through the Flagship Allocation Scheme of the NCI National Facility at
the ANU.


\end{document}